\documentclass[12pt, a4paper]{article}
 \usepackage[font=small,format=plain,labelfont=bf,up,textfont=normal,up,justification=justified,singlelinecheck=false]{caption}
\usepackage{subcaption}
\usepackage{mathrsfs}
\usepackage[a4paper, left=2cm,right=2cm]{geometry}
\usepackage[colorlinks=true,linkcolor=black,citecolor=teal,urlcolor=MidnightBlue,filecolor=black]{hyperref}
\usepackage{amsfonts}
\usepackage{amsmath,amssymb}
\usepackage{setspace}
\usepackage{slashed}
\usepackage{braket}
\usepackage[dvipsnames]{xcolor}
\definecolor{SchoolColor}{rgb}{0.6471, 0.1098, 0.1882} % Crimson
\usepackage{subfloat}
\usepackage[utf8,applemac]{inputenc}
\usepackage{tensor}
\usepackage{cite}
\usepackage{tikz}
\usetikzlibrary{calc}
\usetikzlibrary{patterns}
\usetikzlibrary{arrows.meta}
\usetikzlibrary{decorations.text}
\usetikzlibrary{decorations.pathmorphing}
\usepackage[pdftex]{pict2e}
\usepackage{caption}

\usepackage{graphicx}% Include figure files
\usepackage{bm} % For bold math symbols
\allowdisplaybreaks[4]
\usepackage[utf8,applemac]{inputenc}
\usepackage{tensor}
\usepackage{cite}
\usepackage{tikz}
\graphicspath{{figure/}}
\usepackage{array}
\usepackage{booktabs}

\usepackage{multirow}

\usepackage{dcolumn}% Align table columns on decimal point
\usepackage{bm}% bold math

\usepackage{verbatim}

\usepackage{textcomp} % \textless, \textgreater, \textbrokenbar macros

\numberwithin{equation}{section}
\newcommand{\xhy}{\color{red}}
\newcommand{\bea}{\begin{eqnarray}}
\newcommand{\eea}{\end{eqnarray}}
\newcommand{\be}{\begin{equation}}
\newcommand{\ee}{\end{equation}}
\newcommand{\bs}{\begin{subequations}}
\newcommand{\es}{\end{subequations}}
\def\nn{\nonumber}

\def\p{\partial}

\newcommand{\beqs}{\begin{eqnarray}}
\newcommand{\eeqs}{\end{eqnarray}}

\newcommand{\dd}{D^>}
\newcommand{\dx}{D^<}
\newcommand{\scri}{\mathcal{I}}
\newcommand{\mbx}{\mathbf{x}}

\numberwithin{equation}{section}

\newcommand{\Rmnum}[1]{\uppercase\expandafter{\romannumeral #1\relax}}

\def\c.c.{\mathrm{c.c.}}

\def\de{\delta}
\def\ep{\epsilon}

\def\lam{\lambda}

\def\om{\omega}
\def\De{\Delta}\def\Om{\Omega}

\newcommand{\lon}{\color{blue}}

\begin{document}
\setlength{\unitlength}{1cm}
\begin{titlepage}

\begin{flushright}\vspace{-3cm}
{\small
%{\tt arXiv:yymm.nnnn} \\
\today }\end{flushright}
\vspace{0.5cm}
\begin{center}
	{{ \LARGE{\bf{Thermal correlator at null infinity}}}}\vspace{5mm}%\vspace{8pt}\\in higher dimensional CFT}}}} \vspace{5mm}

	\centerline{ Jiang Long\footnote{longjiang@hust.edu.cn}\ \& Hong-Yang Xiao\footnote{xiaohongyang@hust.edu.cn}}
	\vspace{2mm}
	\normalsize
	\bigskip\medskip
%	\textit{Asia Pacific Center for Theoretical Physics,\\ Pohang 37673, Korea}\\\
%\vspace{2mm}

	\textit{School of Physics, Huazhong University of Science and Technology, \\ Luoyu Road 1037, Wuhan, Hubei 430074, China
	}
	%\vfil
	%\pacs{04.70.Dy}
	
	\vspace{25mm}
	
	\begin{abstract}
		\noindent
		{We study the thermal Carrollian correlators at null infinity in the real-time formalism. We derive the Feynman rules to calculate these correlators in the position space. We compute the bulk-to-bulk, bulk-to-boundary and boundary-to-boundary propagators for massless scalar theory. Due to the doubling of the fields degrees of freedom, the number of each propagator is quadrupled. The bulk-to-boundary propagators have the form of (extended) Bose-Einstein distribution in the position space. Utilizing the contour integral  of the propagators, we can transform the Feynman rules to momentum space. Interestingly, while the external lines and amplitudes in momentum space depend on the contour, Carrollian correlators in position space are independent of it. We show how to compute four-point correlators at finite temperature. The tree level correlators can be written as the summation of Barnes zeta functions and reduce to the  ones in the zero temperature limit.  }\end{abstract}
	
	%\pacs{04.65.+e,04.70.-s,11.30.-j,12.10.-g}

\end{center}
%%%%%%%%%%%%%%%%%%%%%%%%%%%%%%%%%%%%%%%%%%%%%%%%%%%%%%%%%%%%%%%%%%%%%%%%%%%%%%%%%%%%%%%%

\end{titlepage}
\tableofcontents

\section{Introduction}
A half century ago, Bekenstein and Hawking found the relation between the entropy of a black hole and the area of its event horizon \cite{Bekenstein:1973ur,Hawking:1975vcx}. Many important achievements, including the black hole thermodynamics \cite{Bardeen:1973gs} and the Unruh effect in Rindler spacetime \cite{Unruh:1976db}, %and the  holographic entanglement entropy \cite{Ryu:2006bv}, 
are ultimately connected or motivated by this discovery. 

In recent years, motivated by the holographic principle \cite{1993gr.qc....10026T,Susskind:1994vu} and its explicit realization of AdS/CFT \cite{Maldacena:1997re}, researchers have become increasingly interested in searching for flat holography \cite{Polchinski:1999ry,Susskind:1998vk,Giddings:1999jq,deBoer:2003vf,Arcioni:2003xx,Arcioni:2003td,Mann:2005yr} which is the key to understanding gravitational physics in the real world. So far, two scenarios, the celestial \cite{Pasterski:2016qvg,Pasterski:2017kqt,Pasterski:2017ylz} and Carrollian holography \cite{Donnay:2022aba,Bagchi:2022emh},  have been proposed to  explore this topic. We will focus on Carrollian holography since it is based on geometric properties of the Carrollian manifold \cite{Une, Gupta1966OnAA,Duval:2014uva,Duval:2014lpa,Duval:2014uoa}, matches perfectly with asymptotic symmetries \cite{Barnich:2010eb,Sachs:1962wk,Campiglia:2014yka, Campiglia:2015yka,Duval:2014uva,Duval:2014lpa}, field quantization \cite{1978JMP....19.1542A,Ashtekar:1981sf,Ashtekar:1981bq,Ashtekar:1987tt} and provides fruitful algebras  \cite{Liu:2022mne,Liu:2023qtr,Liu:2023gwa,Li:2023xrr,Liu:2024nkc,Liu:2024rvz,Guo:2024qzv}, superduality transformations \cite{Liu:2023qtr,Liu:2023gwa,Liu:2023jnc} and unexpected observable quantities such as helicity flux density\cite{long2025gravitational}.

Based on holographic principle,  the symmetries at the null boundary of an
asymptotically flat spacetime are expected to be  Carrollian conformal symmetries in
one lower dimension. Aspects of Carrollian conformal field theories have been investigated in \cite{Bagchi:2019xfx,Baiguera:2022lsw,saha2022intrinsic,Duval:2014lpa,Duval:2014uva,Ciambelli:2019lap}. Moreover, various Carrollian field theories  have been introduced in the literature. These include Carrollian scalars \cite{Baiguera:2022lsw,Bagchi:2022eav,de2023carroll,gupta2021constructing,chen2023constructing,Bagchi:2019xfx,sharma2025studies,bekaert2024holographic,banerjee2023one,ciambelli2024dynamics}, fermions \cite{bagchi2023magic,banerjee2023carroll,bergshoeff2024carroll,hao2024bms,yu2023free,ara2024flat,ekiz2025quantization}, Yang-Mills \cite{islam2023carrollian}, and  supersymmetric \cite{barducci2019vector} theories. There are several ways of constructing Carrollian field theories. Firstly, one can use symmetry principle to constrain the theory. Based on Carroll covariance, actions \cite{Bagchi:2022eav} and dynamics \cite{Rivera-Betancour:2022lkc} of scalar fields on a Carrollian manifold are derived. The second way to construct Carrollian field theories is called contraction, which means taking the ultra-relativistic limit $c\rightarrow0$\cite{Duval:2014uoa,Henneaux:2021yzg,Bagchi:2016bcd,Bagchi:2019clu,Bagchi:2019xfx} where $c$ is the speed of light. By imposing this limit on the equations of motion, one can  obtain two distinct Carrollian field theories from two different Carroll contractions\cite{Bagchi:2016bcd,Bagchi:2019xfx,Duval:2014uoa,basu2018dynamical}, which are conventionally called electric and magnetic branch, respectively. While the construction of the electric branch is quite straightforward, there exist some difficulties in the construction of the magnetic branch\cite{Bagchi:2019clu}. An alternate way to construct the Carrollian field theories by contraction is based on the Hamiltonian action principle\cite{Henneaux:2021yzg}. Within this formalism, the electric branch can be obtained by discarding all the spatial derivatives in the Hamiltonian density, while only the time derivatives remain. Conversely, the magnetic branch emerges when the spatial derivatives are kept only. Other methods of constructing Carrollian field theories include taking the flat limit of AdS\cite{kraus2025carrollian,Bagchi:2023fbj} and null-reduction of the Bargmann invariant actions in one higher dimension\cite{chen2023constructing}. \iffalse The quantisation of Carrollian scalar field theories can be found in \cite{banerjee2023one,de2023carroll,chen2024quantization}.\fi 

In Carrollian conformal field theories, the most important quantities are Carrollian correlators that are  the analogs of  those in conformal field theories. Basically, they are correlation functions of certain primary fields inserted at the null boundary that satisfy Ward identities associated with the Carrollian conformal symmetries. 
 In the framework of Carrollian holography, the standard scattering amplitude in momentum space is mapped to the so-called Carrollian amplitude \cite{Liu:2022mne, Donnay:2022wvx, Salzer:2023jqv,Nguyen:2023miw, Liu:2024nfc, Mason:2023mti,Ruzziconi:2024zkr} for massless scattering.  Recently, the concept of Carrollian amplitude has been  generalized to higher dimensions \cite{Liu:2024llk} and general Carrollian manifolds \cite{Li:2024kbo}. The Carrollian amplitude  can be  identified as the Carrollian correlator by relating  bulk fields to boundary operators. For this reason, we will use Carrollian correlator and Carrollian amplitude interchangeably throughout this work.

\begin{figure}
    \centering
    \usetikzlibrary{decorations.text}
    \begin{tikzpicture} [scale=0.8]
        \draw[dashed,thick] (-3,-3) node[below]{\footnotesize $i^-$}  -- (3,3) node[above]{\footnotesize $i^+$};
        \draw[draw,thick] (3,3) -- (6,0) node[right]{\footnotesize $i^0$};
        \node at (4.8,1.8) {\footnotesize $\mathcal{I}^+$};
        \draw[dashed,thick] (-3,3) node[above] {\footnotesize $i^+$} -- (3,-3) node[below]{\footnotesize $i^-$};
        \draw[draw, thick](3,-3) -- (6,0);
        \draw[draw, thick](-3,-3) -- (-6,0) node[left]{\footnotesize $i^0$};
        \draw[draw,thick](-6,0) -- (-3,3);
        \node at(-4.8,1.8) {\footnotesize $\mathcal{I}^+$};
        \node at (-4.8,-1.8){\footnotesize $\mathcal{I}^-$};
        \node at (4.8,-1.8) {\footnotesize $\mathcal{I}^-$};
        \node at (3,0){\text{I}};
        \node at (-3,0){\text{II}};
       \draw[decorate, decoration={snake, amplitude=0.4mm, segment length=1mm}] (-3,3) -- (3,3);
       \draw[draw,thick](-3,3) -- (3,3);
       \draw[decorate, decoration={snake, amplitude=0.4mm, segment length=1mm}] (-3,-3) -- (3,-3);
       \draw[draw,thick](-3,-3) -- (3,-3);
       \fill (1.7,0) circle (2pt);
       \draw[decorate, decoration={snake, amplitude=0.8mm,segment length=2mm}](1.7,0) node[below, xshift=-0.1cm,yshift=-0.1cm]{\footnotesize $x$} -- (-3.8,2.2) node[above,xshift=-0.4cm] {\footnotesize $$};
       \draw[decorate, decoration={snake, amplitude=0.8mm,segment length=2mm}](1.7,0) -- (-3.5,-2.5) node[left,yshift=-0.1cm]{\footnotesize $$};
       \draw[decorate, decoration={snake, amplitude=0.8mm,segment length=2mm}](1.7,0) -- (3.6,2.4) node[right]{\footnotesize $$};
       \draw[decorate, decoration={snake, amplitude=0.8mm,segment length=2mm}] (1.7,0) -- (3.5,-2.5) node[right] {\footnotesize $$};
    \end{tikzpicture}
    \caption{A Feynman diagram for four graviton scattering in a maximally extended Schwarzschild black hole. The dashed lines are event horizons and the wavy lines are bulk-to-boundary propagators for gravitons. The wavy line with a horizontal line represents the singularity. One should integrate out the bulk points, including the black hole and white hole as well as the two asymptotic flat regions I and II to obtain the Carrollian amplitude.} 
    \label{fourgravitons}
\end{figure}
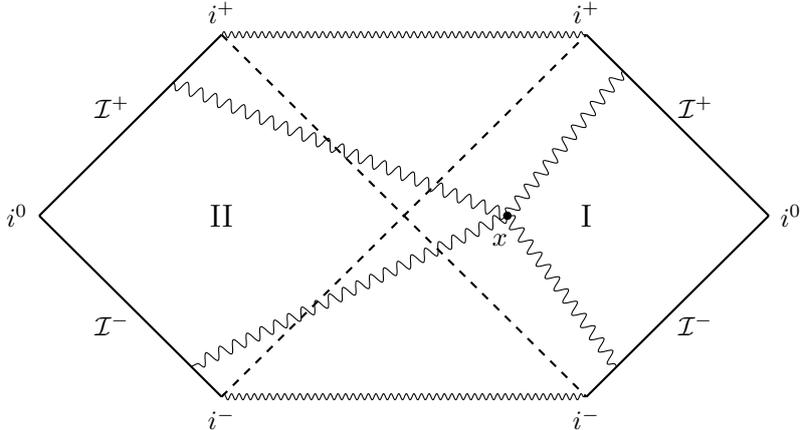

Based on these exciting developments, especially the Feynman rules \cite{Liu:2024nfc} of Carrollian amplitude and the scattering in Rindler spacetime \cite{Li:2024kbo}, there appears to be no conceptual difficulty in dealing with black hole scattering problems using the technologies of Carrollian amplitude. For any globally hyperbolic spacetime $\mathbb M$, one can always find future and past  null hypersurfaces that determine its causal development. By constructing the bulk-to-boundary and bulk-to-bulk propagators in $\mathbb M$ and taking into account the interaction vertices, one can always draw the Feynman diagrams and write down the associated Feynman integrals for scattering processes. In particular, this is possible for Schwarzschild black holes. Figure \ref{fourgravitons} is the Penrose diagram of a maximally extended Schwarzschild black hole and we have drawn a Feynman diagram that represents four graviton interactions in the bulk. The gravitons are connected to the future/past null infinity $(\mathcal{I}^{\pm})$ by retarded/advanced bulk-to-boundary propagators. Unfortunately, the technical difficulties in constructing the analytic propagators in Schwarzschild black hole and the messy integrals on the bulk spacetime prevent us from obtaining concrete conclusion, at least at this moment.

 In AdS/CFT,  an AdS black hole has been conjectured to be dual to a putative thermal conformal field theory located at the boundary  whose temperature is exactly the Hawking temperature of the AdS black hole \cite{Witten:1998qj}. When the cosmological constant tends to zero, the AdS black hole reduces to  a black hole in an asymptotically flat spacetime. It would be interesting to understand the corresponding limit of the dual conformal field theory at finite temperature along the line of \cite{Alday:2024yyj}. This indicates a concrete dual description of black holes in asymptotically flat spacetime. It seems that it should be a certain thermal Carrollian field theory. Moreover, the previous exploration on the Rindler spacetime amplitude also supports the existence of a thermal Carrollian field theory  at the null boundary \cite{Li:2024kbo}. 
 
In this paper, we will turn to a description of a thermal Carrollian field theory at the null boundary of Minkowski spacetime whose bulk cousin is the usual thermal quantum field theory. We develop a real-time formalism to construct the bulk-to-bulk, bulk-to-boundary and boundary-to-boundary propagators as well as thermal correlators at the null boundary. The Feynman rules, both in  position  and  momentum space, have been presented to compute Carrollian correlators. Interestingly, the bulk-to-boundary propagators take the form of an (extended) Bose-Einstein distribution in the position space. The Carrollian correlators reduce to the Carrollian amplitudes in the zero temperature limit. 

The layout of this  paper is  as follows. In section \ref{real}, we review the minimal aspects of the real-time formalism  relevant for this work, including the Schwinger-Keldysh contour and the doubling of the degrees of freedom of the fields. In section \ref{fen}, we  explore the method  of extracting the Carrollian correlators from bulk Green's functions and show the associated Feynman rules, In section \ref{propaga}, a complete set of propagators for Carrollian correlators is given explicitly and the KMS symmetry is verified. Then we turn to the calculation of the Carrollian correlators in the following section. We  discuss open questions in section \ref{dis}. Technical details are  relegated to several appendices.  In appendix A we review some aspects of Carrollian holography including Carrollian symmetries and amplitudes. In Appendix B we discuss the integral representation of the propagators. Appendix C  lists some properties of the step and the sign functions and Appendix D is an introduction of Barnes zeta function.

\section{Real-time formalism}\label{real}
The real-time formalism has been reviewed in the reports \cite{Chou:1984es,landsman1987real}. In AdS/CFT, the  thermal propagators in real-time formalism have been discussed by \cite{Son:2002sd,Herzog:2002pc}.  We will follow the book \cite{le2000thermal} to present the formalism and 
 work with the real scalar field theory at finite temperature $T=\beta^{-1}$ with zero chemical potential. The field operator $\Phi(x)$ in the Heisenberg picture is
\begin{align}
   \Phi(x)=e^{it H}\Phi(0,\bm x)e^{-it H}
\end{align}
where the time coordinate $t=x^0$ is allowed to be complex and $H$ is the Hamiltonian. The thermal Green's functions are defined as
\begin{align}
    \mathcal G_C(x_1,\cdots,x_n)\equiv \frac{\text{tr}e^{-\beta H}T_C(\Phi(x_1)\cdots\Phi(x_n))}{\text{tr}e^{-\beta H}} =\braket{T_C(\Phi(x_1)\cdots\Phi(x_n))}_\beta\label{green}
\end{align}
where the time-ordering operator $T_C$ is taken along a complex time path $C$. To be more precise, one may choose a parametric definition $t=\tt f(\lambda)$ of the path, with $\lambda$ real and monotonically increasing, namely the ordering along the path will correspond to the ordering in $\lambda$. One can also introduce the path $\theta$- and $\delta$-functions
\begin{align}
    \theta_C(t-t')=\theta(\lambda-\lambda'),\quad \delta_C(t-t')=\Big|\frac{\partial \tt f}{\partial \lambda}\Big|^{-1} \delta(\lambda-\lambda'),
\end{align}
such that one can write the path-ordered Green's functions, for example,
%\begin{subequations}
\begin{align}
    T_C(\Phi(x)\Phi(x'))=\theta_C(t-t')\Phi(x)\Phi(x')+\theta_C(t'-t)\Phi(x')\Phi(x).
\end{align}
%\begin{align}
%    \partial_tT_C(\hat{\fai}(x)\hat{\fai}(x'))=\delta_C(t-t')[\hat{\fai}(x),\hat{\fai}(x')]+T_C(\partial_t\hat{\fai}(x)\hat{\fai}(x'))
%\end{align}
%\end{subequations}
One can also extend functional differentiation
\begin{align}
    \frac{\delta J(x)}{\delta J(x')}=\delta_C(t-t')\delta^{(3)}(\textbf{x}-\textbf{x}')\label{varationJ}
\end{align}
for $c$-number functions $J(x)$ living on the path $C$. There is a generating functional $Z_C(\beta;J)$ 
\begin{align}
    Z_C(\beta;J)=\text{tr}\left[e^{-\beta H}T_Ce^{i\int_Cd^4xJ(x)\Phi(x)}\right]
\end{align}
which allows us to obtain Green's functions from functional differentiation w.r.t. sources $J(x)$
\begin{align}
    \mathcal G_C(x_1,\cdots,x_n)=\frac{1}{Z(\beta)}\frac{\delta^nZ_C(\beta;J)}{i\delta J(x_1)\cdots i\delta J(x_n)}\Big{|}_{J=0},\label{GCx}
\end{align}
where the path $C$ must go through all the arguments of the Green's function we are interested in. Note that $Z_C(\beta;J=0)=Z(\beta)=\text{tr} e^{-\beta H}$ is the partition function without source.

For $n=2$, the two-point Green's function $\mathcal G_C(x,x')$ is defined through the equation 
\begin{align}\label{DCxx'}
   \mathcal G_C(x,x')=\theta_C(t-t')\mathcal G_C^{>}(x,x')+\theta_C(t'-t)\mathcal G_C^{<}(x,x')
\end{align}
where 
\bea 
\mathcal G_C^{>}(x,x')=\braket{\Phi(x)\Phi(x')}_\beta,\quad \mathcal G_C^{<}(x,x')=\braket{\Phi(x')\Phi(x)}_\beta
\eea   are properly defined in the strips $-\beta\leq \text{Im}(t-t')\leq0$ and $0\leq \text{Im}(t-t')\leq\beta$ respectively. The propagator (\ref{DCxx'}) is well defined provided that we take path $C$ such that the imaginary part of $t$ is non-increasing when the parameter $\lambda$ increases.
\iffalse 
Adding space components and using complex values of $t$ instead of real ones, taking translation invariance into account, we write (\ref{DCxx'}) as
\begin{align}
    D_C(x,x')=\theta_C(t-t')\left[\dd_C(x-x')-\dx_C(x-x')\right]+\dx_C(x-x')
\end{align}
Instead of a spectral function $\rho(k_0)$, we now have a generalized spectral function $\rho(k)=\rho(k_0,\mathbf{k})$ in 4 dimensions. From the relation between $\dd(k)$, $\dx(k)$ and $\rho(k)$, we obtain a very useful representation of the propagator:
\begin{align}\label{DCxx'=intrhok}
    D_C(x-x')=\int\frac{d^4k}{(2\pi)^4}e^{-ik\cdot(x-x')}\left[\theta_C(t-t')+f(k_0)\right]\rho(k)
\end{align}
The free propagator $D_C^F$ follows from (\ref{DCxx'=intrhok}) by inserting the free spectral function:
\begin{align}
    \rho_F(k)=2\pi sgn(k_0)\delta(k^2-m^2)
\end{align}\fi

\begin{figure}
    \centering
    \usetikzlibrary{decorations.text}
    \begin{tikzpicture} [scale=0.8]
  % \label{sigmatimepath}
        \draw[->,thick] (-6,0) -- (-2,0) node[above]{\footnotesize $C_1$};
        \draw[->,thick] (-2,0)-- (6,0) node[right]{\footnotesize \text{Re}\ $t$};
        \draw[->,thick] (0,-4) -- (0,2) node[above]{\footnotesize \text{Im}\ $t$};
        \node at (-0.2,0.2) {\footnotesize $0$};
        \filldraw[black] (-5,0) circle (2pt) node[above]{\footnotesize $t_i$};
        \filldraw[black] (5,0) circle (2pt) node[above]{\footnotesize $-t_i$};
        \draw[->,thick] (5,0) -- (5,-1) node[right]{\footnotesize $C_3$};
        \draw[draw,thick] (5,-1) -- (5,-2) node[below]{\footnotesize $-t_i-i\sigma$}; 
        \filldraw[black] (5,-2) circle (2pt);
        \draw[->,thick] (5,-2) -- (-2,-2) node[below]{\footnotesize $C_2$};
        \draw[draw,thick] (-2,-2) -- (-5,-2) node[above]{\footnotesize $t_i-i\sigma$}; 
        \filldraw[black] (-5,-2) circle (2pt);
        \draw[->,thick] (-5,-2) -- (-5,-3) node[left]{\footnotesize $C_4$}; 
        \draw[draw,thick] (-5,-3) -- (-5,-4) node[below]{\footnotesize $t_i-i\beta$}; 
        \filldraw[black] (-5,-4) circle (2pt);
%\caption{The time path $C$.}
    \end{tikzpicture}
    \caption{\centering{The time path $C$ in the real-time formalism.}}
    \label{sigmatimepath}
\end{figure}
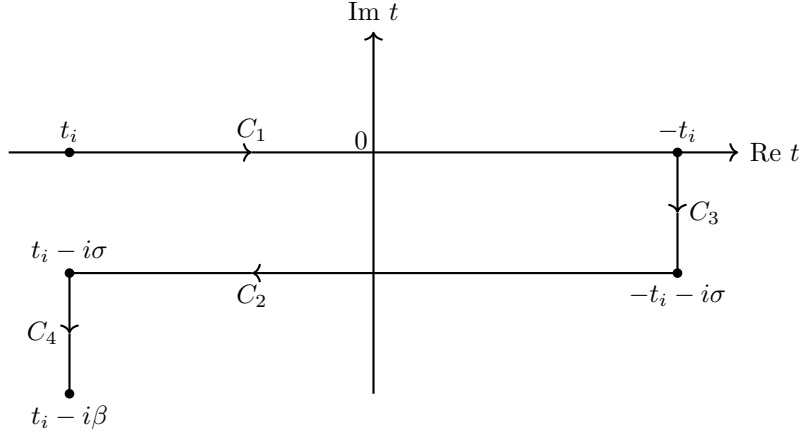

We now turn to the derivation for the generating functional $Z_C(\beta;J)$ in a path integral representation. Let $\Phi(x)=\Phi(t,\bm{x})$ be the field operator in the Heisenberg picture and  $\ket{\Phi(\textbf{x});t}$ be the state vector  at time $t$ which is an eigenstate of $\Phi(x)$ with eigenvalue $\Phi(\textbf{x})$
\begin{align}
    \Phi(x)\ket{\Phi(\textbf{x});t}=\Phi(\textbf{x})\ket{\Phi(\textbf{x});t}.
\end{align}
We recall that
\begin{align}
    \ket{\Phi(\textbf{x});t}=e^{iHt}\ket{\Phi(\textbf{x});t=0}
\end{align}
and write the thermal average of an operator $O$ as
\begin{align}\nonumber
    \braket{O}_{\beta}=\frac{1}{Z(\beta)}\text{tr}\left(e^{-\beta\hat{H}}O\right)=\frac{1}{Z(\beta)}\int\mathcal D\Phi \braket{\Phi(\textbf{x});t|e^{-\beta H}O|\Phi(\textbf{x});t}
    =\frac{1}{Z(\beta)}\int \mathcal D\Phi\braket{\Phi(\textbf{x});t-i\beta|O|\Phi(\textbf{x});t}
\end{align}
where $\mathcal D\Phi$ indicates a sum over all possible field configurations $\Phi(\textbf{x})$. Then we write $Z(\beta;J)$ in the form
\begin{align}
    Z_C(\beta;J)=\int\mathcal D\Phi\braket{\Phi(\mathbf{x});t_i-i\beta|T_Ce^{i\int_Cd^4xJ(x)\Phi(x)}|\Phi(\mathbf{x});t_i}
\end{align}
where we have chosen for time $t$ the initial time $t_i$ of the path $C$ and then the final time is $t_f=t_i-i\beta$. Then we cast $Z_C(\beta;J)$ into the form of a path integral:
\begin{align}
    Z_C(\beta;J)=\int\mathcal{D}\Phi e^{i\int_Cd^4x(\mathcal{L}(\Phi)+J(x)\Phi(x))}
\end{align}
with the boundary condition $\Phi(t;\mathbf{x})=\Phi(t-i\beta;\mathbf{x})$. The Lagrangian $\mathcal L(\Phi)$ is the kinematic term  minus the potential in which the kinematic term is  quadratic in $\Phi$ while the potential term $V(\Phi)$ is responsible for the interactions
\be 
\mathcal{L}(\Phi)=-\frac{1}{2}(\partial_\mu\Phi)^2-V(\Phi).
\ee By using the standard trick to replace $\Phi$ to $\frac{\delta}{i\delta J}$ in the potential term \cite{hollik2014quantum}, we find the partition function 
\begin{align}
    Z_C(\beta;J)=e^{-i\int_Cd^4xV\left(\frac{\delta}{i\delta J(x)}\right)}Z_C^{(0)}(\beta;J),
\end{align}
where the free generating functional $Z_C^{(0)}(\beta;J)$ is computed by a Gaussian integration
\begin{align}\label{ZCFCC}
    Z^{(0)}_C(\beta;J)=\mathcal{N}e^{-\frac{1}{2}\int_Cd^4x\int_Cd^4x'J(x)G_C(x-x')J(x')},
\end{align} where $\mathcal N$ is a normalization constant and $G_C(x-x')$ is the extended Feynman propagator with the fields inserted in the path $C$ \begin{align}
    G_C(x-x')=\theta_C(t-t')G_C^{>}(x-x')+\theta_C(t'-t)G_C^{<}(x-x').\label{feynmanD}
\end{align} The G-greater and G-lesser are defined as
\bea 
G_C^{>}(x-x')=\braket{\Phi(x)\Phi(x')}^{(0)}_\beta,\quad G_C^{<}(x-x')=\braket{\Phi(x')\Phi(x)}^{(0)}_\beta.
\eea The propagator \eqref{feynmanD} can be compared with the Green's function \eqref{DCxx'}. They share the same form. The former can be computed in the free theory while the latter should include the interactions. Thus we have used a superscript $(0)$ to denote the free theory. To simplify notation, we will omit the superscript $(0)$ from now on. 
%\subsection{Integration path and propagators}
Up to now the restrictions on the time path $C$ are that it starts from an initial time $t_i$, ends at a final time $t_i-i\beta$, and between these times the imaginary part of $t$ must be a non-increasing function of the path parameter $\lambda$. Furthermore, $C$ must contain the real axis, since we are ultimately interested in Green's functions whose time arguments take real values. These restrictions still leave open many possibilities for the path $C$. We shall describe the standard choice:
\begin{itemize}
    \item  $C$ starts from a real value $t_i$, large and negative.
\item  $C$ follows the real axis up to a large positive value $-t_i$. This part of $C$ is denoted by $C_1$.
\item Then the path goes from $-t_i$ to $-t_i-i\sigma$, with $0<\sigma<\beta$, along a vertical straight line denoted by $C_3$.
\item There is a second horizontal straight line $C_2$ going from $-t_i-i\sigma$ to $t_i-i\sigma$.
\item Finally, the path follows a vertical straight line $C_4$ from $t_i-i\sigma$ to $t_i-i\beta$.
\end{itemize} The choice of the time path $C$ is 
\be 
C=\cup_{i=1}^4 C_i
\ee and it has been shown in Figure \ref{sigmatimepath}. 

In the limit $t_i\rightarrow-\infty$, the two vertical segments $C_3$ and $C_4$ are moved to infinity and  their contributions to the partition function vanish\cite{rammer2011quantum}. It is convenient to choose $t$ to be real variables running from $-\infty$ to $\infty$ and to label the source $J(x)$ with an index $a$, $a=1,2$,  according to the part $C_a$ of the path on which it lives
\begin{align}
    J_1(x)=J(t,\mathbf{x}),\quad J_2(x)=J(t-i\sigma,\mathbf{x}).
\end{align}
At the same time, the functional differentiation \eqref{varationJ} is replaced by 
\begin{align}
    \frac{\delta J_a(x)}{\delta J_b(x')}=\delta_{ab}\delta^{(4)}(x-x').
\end{align}
With these conventions, the partition function \eqref{ZCFCC} becomes 
\begin{align}\label{ZCF=jaDabjb}
    Z_C^{(0)}(\beta;J)=\mathcal{N}e^{-\frac{1}{2}\int_{-\infty}^{\infty}d^4x\int_{-\infty}^{\infty}d^4x'J_a(x)G_{ab}(x-x')J_b(x')}
\end{align}
where the real-time propagators are
\begin{subequations}\label{DFAB}
\begin{align}
    G_{11}(x-x')&=G_F(x-x'),\\
    G_{22}(x-x')&=G_F^*(x-x'),\\
    G_{12}(x-x')&=G^{<}(t-t'+i\sigma,\bm x-\bm x'),\\
    G_{21}(x-x')&=G^{>}(t-t'-i\sigma,\bm x-\bm x').
\end{align}
\end{subequations}
The second equation stems from $\theta_C(t)=\theta(-t)$ on $C_2$, while last two equations follow by noting that ``times'' on $C_2$ are always later than ``times" on $C_1$. Taking the change of sign on $C_2$ due to our convention (\ref{ZCF=jaDabjb}) into account, we arrive at the final form of the  generating functional
\begin{align}\label{ZC}
    Z_C(\beta;J)=\mathcal{N}e^{-i\int_{-\infty}^{\infty}d^4x\left[V\left(\frac{\delta}{i\delta J_1(x)}\right)-V\left(\frac{\delta}{i\delta J_2(x)}\right)\right]}e^{-\frac{1}{2}\int_{-\infty}^{\infty}d^4x\int_{-\infty}^{\infty}d^4x'J_a(x)G_{ab}(x-x')J_b(x')}
\end{align}
which is also equivalent to the path integral
\begin{align}\label{ZCpathintegral}
    Z_C(\beta;J)=\int\left(\prod_{a=1}^2\mathcal{D}\Phi_a\right) e^{i\int d^4 x \left(\mathcal{L}(\Phi_1)-\mathcal{L}(\Phi_2)\right)+i\int_{-\infty}^{\infty}d^4xJ_a(x)\Phi_a(x)}.
\end{align}
One notes that (\ref{ZCpathintegral}) may be interpreted by identifying $\Phi_2$ as a ghost field living on $C_2$. We thus arrive at a doubling of the field degrees of freedom. Of course only the ``physical" fields $\Phi_1(x)$ appear on the external lines of Green's functions, which are obtained from functional differentiation w.r.t. $J_1(x)$. However, the ghost field induces a modification of the naive Feynman rules, since the propagators in (\ref{DFAB}) have off-diagonal elements.

By a Fourier transform
\be 
G_{ab}(x-y)=\int \frac{d^4p}{(2\pi)^4}G_{ab}(k)e^{ik\cdot(x-y)},\label{FourierDab}
\ee 
one can derive the explicit expression of the free propagator (\ref{DFAB}) in the momentum-space
\begin{subequations}\label{Dab}
\begin{align}
    G_{11}(k)&=\frac{i}{-k^2+i\epsilon}+n(|k^0|)2\pi\delta(k^2)=(G_{22}(k))^*,\\
    G_{12}(k)&=e^{\sigma k^0}[n(|k^0|)+\theta(-k^0)]2\pi\delta(k^2),\\
    G_{21}(k)&=e^{-\sigma k^0}[n(|k^0|)+\theta(k^0)]2\pi\delta(k^2).
\end{align}
\end{subequations}
\iffalse 
\begin{subequations}
\begin{align}
    D_{11}^F(k)=\frac{i}{k^2-m^2+i\eta}+n(k_0)2\pi\delta(k^2-m^2)=(D^F_{22}(k))^*
\end{align}
\begin{align}
    D^F_{12}(k)=e^{\sigma k_0}f(k_0)\rho_F(k)=e^{\sigma k_0}[n(k_0)+\theta(-k_0)]2\pi\delta(k^2-m^2)
\end{align}
\begin{align}
    D^F_{21}(k)=e^{-\sigma k_0}[1+f(k_0)]\rho_F(k)=e^{-\sigma k_0}[n(k_0)+\theta(k_0)]2\pi\delta(k^2-m^2)
\end{align}
\end{subequations}\fi In the expressions, the occupation number is the form of Bose-Einstein distribution
\be 
n(\omega)=\frac{1}{e^{\beta\omega}-1}
\ee where $\omega$ is assumed to be positive. However, one can always extend it to the whole complex plane. A useful identity for $n(\omega)$ is 
\be 
n(\omega)+n(-\omega)=-1.\label{sumn}
\ee 
We notice that the off-diagonal elements of the extended  Feynman propagators depend on $\sigma$. However, it could be shown that the physical results are independent of the choice of $\sigma$ \cite{landsman1987real}. In the literature, there are two useful choices for $\sigma$  as follows:
\begin{itemize}
    \item  Thermo-field dynamics(TFD) \cite{Umezawa:1982nv}. This is equivalent to the choice $\sigma=\frac{\beta}{2}$, leading to a symmetric propagator
\begin{align}\label{beta/2DF12}
    G_{12}(k)=G_{21}(k)=e^{\frac{\beta|k^0|}{2}}n(|k^0|)2\pi\delta(k^2).
\end{align}
\item Schwinger-Keldysh formalism (SKF)\cite{Schwinger:1960qe,Keldysh:1964ud}. This is equivalent to the choice $\sigma=0$, leading to
\begin{subequations}
\begin{align}
    G_{12}(k)&=[n(|k^0|)+\theta(-k^0)]2\pi\delta(k^2),\\
    G_{21}(k)&=[n(|k^0|)+\theta(k^0)]2\pi\delta(k^2).
\end{align}
\end{subequations}
 
\end{itemize}
The TFD and SKF are in many ways the same in form. In particular, the two approaches are identical in stationary situations. However, TFD and SKF are quite different in time-dependent non-equilibrium systems. The main source of the difference is that the time evolution of the density matrix itself is ignored in SKF while in TFD it is replaced by a time-dependent Bogoliubov transformation. In this sense TFD is a better candidate for time-dependent quantum field theory. Even in equilibrium situations, TFD has some remarkable advantages over SKF, the most notable feature being the Feynman diagram recipes\cite{chu1994unified}.
In the following, we will write down the general propagators for arbitrary choice of $\sigma$ for completeness.

\section{Feynman rules}\label{fen}
Taking the functional differentiation in \eqref{GCx}, one can easily obtain the Feynman rules for the Green's functions. 
We have on the one hand fields linked to external positions, which are of type 1, and on the other hand internal vertices which are of type 1 or type 2. Note that there could be off-diagonal propagator which connects vertices that mixes the fields of type 1 and type 2. Given a configuration of internal vertices, we have to join them by the corresponding propagators: $G_{11}$ links two vertices of type 1, $G_{12}$ a vertex of type 1 with a vertex of type 2, etc, and we must sum over all possibilities. One can find more details on the functional methods to derive the Feynman rules in \cite{Jordan:1986ug}.

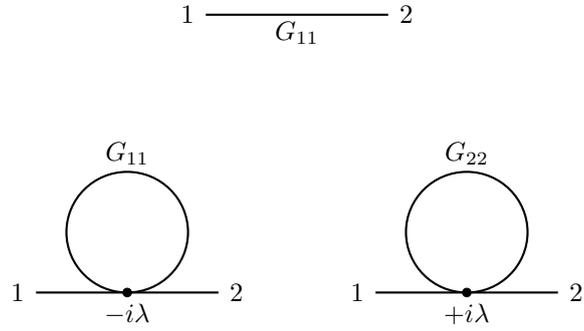
\begin{figure}
    \centering
    \usetikzlibrary{decorations.text}
    \begin{tikzpicture} [scale=0.8]
        \draw[draw,thick] (0,0) node[left]{\footnotesize $1$} -- (3,0) node[right]{\footnotesize $2$};
        \node at (1.5,-0.3) {\footnotesize $G_{11}$};
    \end{tikzpicture}\vspace{1cm}
    
    \begin{tikzpicture} [scale=0.8]
        \draw[draw,thick] (0,0) node[left]{\footnotesize $1$} -- (3,0) node[right]{\footnotesize $2$};
        \draw[draw,thick] (1.5,0) arc (270:-90:1);
        \node at (1.5,2.3) {\footnotesize $G_{11}$};
        \filldraw[black] (1.5,0) circle (2pt) node[below]{\footnotesize $-i\lambda$};
    \end{tikzpicture}\hspace{1cm}
    \begin{tikzpicture} [scale=0.8]
        \draw[draw,thick] (0,0) node[left]{\footnotesize $1$} -- (3,0) node[right]{\footnotesize $2$};
        \draw[draw,thick] (1.5,0) arc (270:-90:1);
        \node at (1.5,2.3) {\footnotesize $G_{22}$};
        \filldraw[black] (1.5,0) circle (2pt) node[below]{\footnotesize $+i\lambda$};
    \end{tikzpicture}
    \caption{\centering{Feynman diagrams for two-point Green's function up to $\mathcal{O}(\lambda)$.}}
    \label{twopoint}
\end{figure}

\begin{figure}
    \centering
    \usetikzlibrary{decorations.text}
    \begin{tikzpicture} [scale=0.8]
        \draw[draw,thick] (0,3) node[above]{\footnotesize $1$} -- (3,3) node[above]{\footnotesize $2$};
        \node at (1.5,3.3) {\footnotesize $G_{11}$};
        \draw[draw,thick] (0,0) node[below]{\footnotesize $3$} -- (3,0) node[below]{\footnotesize $4$};
        \node at (1.5,-0.3) {\footnotesize $G_{11}$};
    \end{tikzpicture}\hspace{1cm}
    \begin{tikzpicture} [scale=0.8]
        \draw[draw,thick] (0,3) node[above]{\footnotesize $1$} -- (0,0) node[below]{\footnotesize $3$};
        \node at (-0.4,1.5) {\footnotesize $G_{11}$};
        \draw[draw,thick] (3,3) node[above]{\footnotesize $2$} -- (3,0) node[below]{\footnotesize $4$};
        \node at (3.4,1.5) {\footnotesize $G_{11}$};
    \end{tikzpicture}\hspace{1cm}
    \begin{tikzpicture} [scale=0.8]
        \draw[draw,thick] (0,3) node[above]{\footnotesize $1$} -- (3,0) node[below]{\footnotesize $4$};
        \node at (1.1,2.5) {\footnotesize $G_{11}$};
        \draw[draw,thick] (0,0) node[below]{\footnotesize $3$} -- (1.4,1.4);
        \draw[draw,thick] (1.6,1.6) -- (3,3) node[above]{\footnotesize $2$};
        \node at (1.1,0.5) {\footnotesize $G_{11}$};
    \end{tikzpicture}\vspace{1cm}
    
    \begin{tikzpicture} [scale=0.8]
        \draw[draw,thick] (0,3) node[above]{\footnotesize $1$} -- (3,0) node[below]{\footnotesize $4$};
        \draw[draw,thick] (0,0) node[below]{\footnotesize $3$} -- (3,3) node[above]{\footnotesize $2$};
       \filldraw[black] (1.5,1.5) circle (2pt) node[below,yshift=-0.15cm]{\footnotesize $-i\lambda$};
       \node at (0.3,2.2) {\footnotesize $G_{11}$};
       \node at (0.3,0.8) {\footnotesize $G_{11}$};
       \node at (2.7,2.2) {\footnotesize $G_{11}$};
       \node at (2.7,0.8) {\footnotesize $G_{11}$};
    \end{tikzpicture}\hspace{1cm}
    \begin{tikzpicture} [scale=0.8]
        \draw[draw,thick] (0,3) node[above]{\footnotesize $1$} -- (3,0) node[below]{\footnotesize $4$};
        \draw[draw,thick] (0,0) node[below]{\footnotesize $3$} -- (3,3) node[above]{\footnotesize $2$};
       \filldraw[black] (1.5,1.5) circle (2pt) node[below,yshift=-0.15cm]{\footnotesize $+i\lambda$};
       \node at (0.3,2.2) {\footnotesize $G_{21}$};
       \node at (0.3,0.8) {\footnotesize $G_{21}$};
       \node at (2.7,2.2) {\footnotesize $G_{21}$};
       \node at (2.7,0.8) {\footnotesize $G_{21}$};
    \end{tikzpicture}
    \caption{\centering{Feynman diagrams for four-point  Green's function up to $\mathcal{O}(\lambda)$.}}
    \label{connectphi4}
\end{figure}
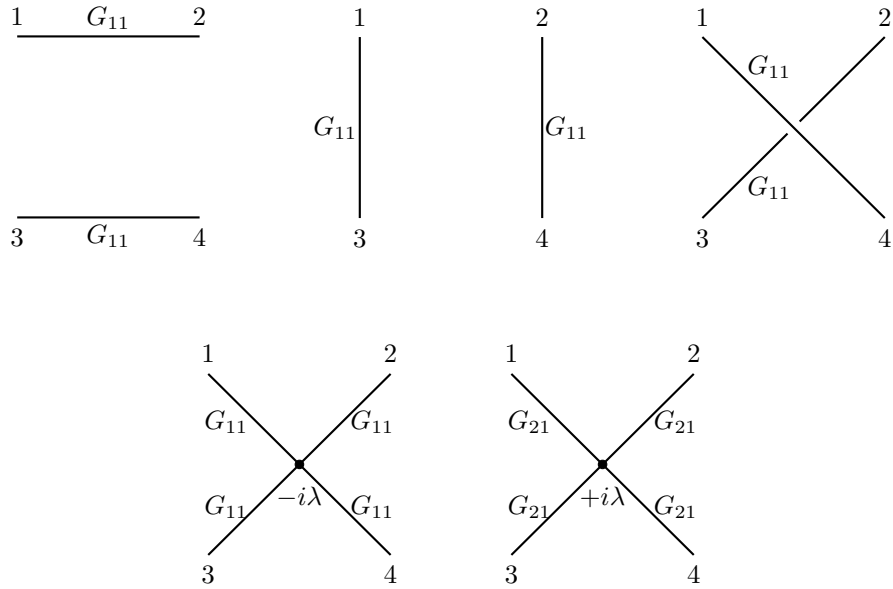

To be more precise, we take  $V(\Phi)=\frac{\lambda\Phi^4}{4!}$ as an example.  To each vertex of type 1 or type 2, we should associate a factor $-i\lambda$ or $+i\lambda$, respectively. For each line that connects internal vertex of type $a$ at $x$ and type $b$ at $x'$, we should  associate it with a propagator $G_{ab}(x-x')$. Finally, as the Feynman rules at zero temperature, we must integrate over all internal vertices with the measure $\int d^4x$, sum over all possible types of vertices and divide by a symmetry factor. The previous discussion can be checked by computing the generating function explicitly. As an illustration, the two-point Green's function reads
\bea 
   && \mathcal G(x_1,x_2)\nn\\&=&\frac{1}{Z_0}\frac{\delta^2}{i\delta J_1(x_1)i\delta J_1(x_2)}\left[\left(1-\frac{i\lambda}{4!}\int d^4x[(\frac{\delta}{\delta J_1(x)})^4-(\frac{\delta}{\delta J_2(x)})^4]\right)e^{-\frac{1}{2}\int d^4y\int d^4zJ_a(y)G_{ab}(y-z)J_b(z)}\right]\Big{|}_{J=0}\nn\\&=&G_{11}(x_1-x_2)-\frac{i\lambda}{2}\int d^4x[G_{11}(x-x_1)G_{11}(x-x_2)G_{11}(x-x)-G_{21}(x-x_1)G_{21}(x-x_2)G_{22}(x-x)],\nn\\
\eea  whose Feynman diagrams are shown in Figure \ref{twopoint}. In Figure \ref{connectphi4}, we show the Feynman diagrams for the four-point Green's function 
\be 
\mathcal G(x_1,x_2,x_3,x_4)=\langle \Phi(x_1)\Phi(x_2)\Phi(x_3)\Phi(x_4)\rangle_\beta
\ee up to $\mathcal{O}(\lambda)$. The first three diagrams are disconnected which can be obtained in free theory
\bea 
G_{11}(x_1-x_2)G_{11}(x_3-x_4)+G_{11}(x_1-x_3)G_{11}(x_2-x_4)+G_{11}(x_1-x_4)G_{11}(x_2-x_3).\label{tree4}
\eea 
There is no vertex of type 2 in above expression since the external positions are always of type 1.
The last {two diagrams encode} the leading order interaction 
\bea 
&&-i\lambda\int d^4x G_{11}(x_1-x)G_{11}(x_2-x)G_{11}(x_3-x)G_{11}(x_4-x)\nn\\&&+i\lambda\int d^4x G_{21}(x_1-x)G_{21}(x_2-x)G_{21}(x_3 -x)G_{21}(x_4-x).\label{loop4}
\eea The first and second line correspond to the vertex of type 1 and type 2, respectively.  In the following, we will always consider the connected Green's function since any disconnected diagrams can be built from the connected ones. 
The four-point correlator can also be derived from the generating functional. 
 Remember that the external points of Green's  functions can only be type 1 field,  and the four-point function is
\begin{align}
    \mathcal G(x_1,x_2,x_3,x_4)=\frac{1}{Z_0}(-i)^4\frac{\delta}{\delta J_1(x_1)}\frac{\delta}{\delta J_1(x_2)}\frac{\delta}{\delta J_1(x_3)}\frac{\delta}{\delta J_1(x_4)}Z_C(\beta;J)|_{J=0}.
\end{align}
Expanding up to order $\mathcal{O}(\lambda)$, we find
\bea 
    \mathcal G(x_1,x_2,x_3,x_4)&=&\frac{1}{Z_0}\frac{\delta^4}{\delta J_1(x_1)\delta J_1(x_2)\delta J_1(x_3)\delta J_1(x_4)}\nn\\&&\left[\left(1-\frac{i\lambda}{4!}\int d^4x(\frac{\delta}{\delta J_1(x)})^4+\frac{i\lambda}{4!}\int d^4x(\frac{\delta}{\delta J_2(x)})^4\right)e^{-\frac{1}{2}\int d^4y\int d^4zJ_a(y)G_{ab}(y-z)J_b(z)}\right]\Big{|}_{J=0}\nn\\&=&G_{11}(x_1-x_2)G_{11}(x_3-x_4)+G_{11}(x_1-x_3)G_{11}(x_2-x_4)+G_{11}(x_1-x_4)G_{11}(x_2-x_3)\nn\\&&-i\lambda\int d^4xG_{11}(x-x_1)G_{11}(x-x_2)G_{11}(x-x_3)G_{11}(x-x_4)\nn\\&&+i\lambda\int d^4xG_{12}(x-x_1)G_{12}(x-x_2)G_{12}(x-x_3)G_{12}(x-x_4),
\eea  which is exactly the summation of \eqref{tree4} and \eqref{loop4}.

Now we can consider the boundary field $\Sigma(u,\Omega)$ which is inserted at future null infinity $\mathcal{I}^+$ and related to the bulk field $\Phi(x)$ through the fall-off condition
\be 
\Phi(x)=\frac{\Sigma(u,\Omega)}{r}+o(r^{-1}).
\ee The Cartesian coordinates $x^\mu$ and the retarded coordinates $(u,r,\Omega)$ are related through 
\be 
x^\mu=u \bar{m}^\mu+r \ell^\mu,
\ee where $\ell^\mu$ is a null vector and $\bar m^\mu$ is a unit timelike vector
\bea \label{ell}
\ell^\mu=(1,\ell^i),\quad \bar m^\mu=(1,0,0,0).
\eea The unit normal vector of the sphere is 
\be 
\ell^i=(\sin\theta\cos\phi,\sin\theta\sin\phi,\cos\theta).
\ee 
Further details on conventions and notations for future/past null infinity are provided in Appendix \ref{carrollianamplitude}.

\begin{figure}
	    \centering
	    \begin{tikzpicture}[scale=1.3]
	     \draw[line width=1pt] (-1.299,0.75) node[above,yshift=0.25cm]{$y_{3}$}  -- (-2.5,1.5) node[right]{$x_{3}$};
	    \draw[line width=1pt] (-1.5,0)  -- (-2.8,0) node[left] {$x_{2}$};
	    \node[above] at (-1.7,0) {$y_{2}$};
	    \node[above] at (1.9,0) {$y_{n-1}$};
	     \draw[line width=1pt] (-1.2,-0.9) node[below,yshift=-0.15cm]{$y_{1}$}  -- (-2.5,-1.5) node[left]{$x_{1}$};
	     \draw[line width=1pt] (1.299,0.75) node[above,yshift=0.25cm]{$y_{{n-2}}$}  -- (2.5,1.5) node[right]{$x_{n-2}$};
	     \draw[line width=1pt] (1.5,0)  -- (2.8,0) node[right] {$x_{n-1}$};
	     \draw[line width=1pt] (1.299,-0.75) node[below,yshift=-0.25cm]{$y_{n}$}  -- (2.5,-1.5) node[right]{$x_{n}$};
	     \node at (-1.9,0.9) {\footnotesize $G_{a_31}$};
	     \node at (-2,-0.18) {\footnotesize $G_{a_21}$};
	     \node at (-1.9,-1.5) {\footnotesize $G_{a_11}$};
	     \node at (1.9,0.7) {\footnotesize $G_{a_{n-2}1}$};
	     \node at (2,-0.18) {\footnotesize $G_{a_{n-1}1}$};
	     \node at (1.8,-1.3) {\footnotesize $G_{a_n1}$};
	      \fill[gray!50] (0,0) circle (1.5);
	     \fill (-1.299,0.75) circle (1.5pt);
	       \fill (-2.5,-1.5) circle (1.5pt);
	    \fill (-2.5,1.5) circle (1.5pt);
	    \fill (-2.8,0) circle (1.5pt);
	    \fill (-1.5,0) circle (1.5pt);
	    \fill (-1.2,-0.9) circle (1.5pt);
	    \fill (1.299,0.75) circle (1.5pt);
	   \fill (1.299,-0.75) circle (1.5pt);
	   \fill (1.5,0) circle (1.5pt);
	   \fill (2.5,1.5) circle (1.5pt);
	   \fill (2.8,0) circle (1.5pt);
	   \fill (2.5,-1.5) circle (1.5pt);
	     \node at (1,-0.65) { $a_{n}$};
	     \node at (1.1,0) { $a_{n-1}$};
	    \node at (0.9,0.7) { $a_{n-2}$};
	    \node at (-1,0.7) { $a_3$};
	    \node at (-1.2,0) { $a_2$};
	    \node at (-1,-0.8) { $a_1$};
	    \node at (0,0) {\Large $ \mathcal{G}_{a_1a_2\cdots a_n}$};
	    \node at (0,1.8) {$\cdots$};
	    \end{tikzpicture}
	    \caption{The $n$ point connected Green's function. The  black lines  are connected to external points. Each external point  $x_j$ is connected to a vertex $y_j$ through Feynman propagators. The internal vertices should be integrated out. The shaded part is the connected and amputated correlation function $\mathcal G_{a_1a_2\cdots a_n}$ which could be constructed by Feynman rules in the position space.}
	    \label{bulkamputated}
	\end{figure}
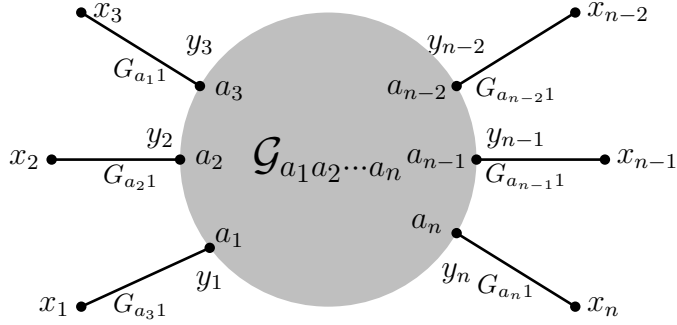

For a general connected Feynman diagram that contributes to the $n$ point Green's function $\mathcal G(x_1,x_2,\cdots,x_n)$, we collect the external positions in a set $E=\{x_1,x_2,\cdots,x_n\}$.  For each external position $x_i\in E$, we subtract a propagator $G_{a_i1}(x_i-y_i)$ from the Feynman diagram. The subscript  $a_i=1$ or $2$ corresponds to the internal vertices of type 1 or 2. The second subscript of the Feynman propagator is always 1 because the point $x_i$ is always type 1 for physical Green's functions. The point $y_i$ is an internal vertex that can be either type 1 or type 2 which should be integrated out.  Therefore, the $n$ point Green's function can be factorized as \footnote{The factorization is correct except that two external points $x_{i_1}$ and $x_{i_2}$ are linked by a propagator directly. Since we are considering connected Feynman diagrams, the exceptional case is only possible for two-point Green's function. The corresponding boundary-to-boundary correlators will be discussed later. }
\be 
\mathcal G(x_1,x_2,\cdots,x_n)=\sum_{a_1,a_2,\cdots,a_n}\left(\int\prod_{j=1}^n d^4y_j \right) \left(\prod_{i=1}^n G_{a_i1}(x_i-y_{i})\right)\mathcal G_{a_1a_2\cdots a_n}(y_1,y_2,\cdots,y_n),
\ee where the connected and amputated Green's function $\mathcal G_{a_1a_2\cdots a_n}(y_1,y_2,\cdots,y_n)$ is independent of the external points. We have shown the formula in Figure \ref{bulkamputated}.
Now one can take the limit $r_i\to\infty$ while keeping $u_i$ finite to extract the $n$ point correlator of the fields $\Sigma$ at finite temperature
\bea 
\langle \prod_{j=1}^n\Sigma(u_j,\Omega_j)\rangle_{\beta}&=&\left(\prod_{j=1}^n \lim_{r_j\to\infty,\ u_j\ \text{finite}} r_j\right) \mathcal G(x_1,x_2,\cdots,x_n)\nn\\&=&\sum_{a_1,a_2,\cdots,a_n}\left(\int\prod_{j=1}^n d^4y_j \right) \left(\prod_{i=1}^n D_{a_i1}(u_i,\Omega_i;y_{i})\right)\mathcal G_{a_1a_2\cdots a_n}(y_1,y_2,\cdots,y_n),\nn\\\label{Sigmanpoint}
\eea
where we have defined the retarded bulk-to-boundary propagator\footnote{We call it the retarded bulk-to-boundary propagator since the boundary field is located at $\mathcal{I}^+$ which is described by a retarded time $u$. Correspondingly, we will call $D_{ab}^{(-)}$ from bulk to $\mathcal{I}^-$ the advanced bulk-to-boundary propagator.}
\bea 
D_{ab}(u,\Omega;y)=\lim_{r\to\infty,\ u\ \text{finite}}r\  G_{ab}(x-y).\label{limitDab}
\eea 
One can read out the Feynman rules for the $n$ point correlator $\langle \prod_{j=1}^n\Sigma(u_j,\Omega_j)\rangle_{\beta}$ as follows. The external points are of type 1 and the bulk vertices are of type 1 or type 2. For each line that connects the vertex of type $a$ at $x$ and another vertex of type $b$ at $x'$, we join a bulk-to-bulk propagator $G_{ab}(x-x')$. For each line that connects the external point $(u,\Omega)$ and the bulk vertex of type $a$ at $x$, we should associate it with a bulk-to-boundary propagator $D_{a1}(u,\Omega;x)$. Certainly, one should attach a factor $-i\lambda$ or $+i\lambda$ to each bulk vertex of type 1 or type 2, respectively. Finally, we still need to integrate over all vertices, sum over all possible types of vertices and divide by a symmetry factor. 

Note that the formula \eqref{Sigmanpoint} and the associated Feynman rules are similar to the ones in \cite{Liu:2024nfc}, except that one should sum over all possible diagrams with different types of internal vertices. Actually, in the limit of zero temperature, we can show that the off-diagonal propagators vanish. Therefore, the Feynman rules reduce to the ones of \cite{Liu:2024nfc} in zero temperature limit. 

\begin{figure}
    \centering
    \usetikzlibrary{decorations.text}
    \begin{tikzpicture} [scale=0.8]
        \draw[draw,thick] (-1,5) node[above]{\footnotesize $i^+$} -- (4,0) node[right]{\footnotesize $i^0$};
        \draw[draw,thick] (4,0) -- (-1,-5) node[below]{\footnotesize $i^-$};
        \draw[draw,thick](-1,5) -- (-1,-5);
         \fill[gray!80] (0.5,0) circle (1);
         \draw[draw,thick] (0.5,1) node[below]{\footnotesize $a_1$} node[above left]{\footnotesize $y_1$} -- (0.7,3.3) node[above, xshift=0.33cm]{\footnotesize $(u_1,\Omega_1)$} ;
         \node at (1,1.5) {\rotatebox{-45}{\footnotesize $\cdots$}};
          \draw[draw,thick] (1.3,0.6) node[below]{\footnotesize $a_p$} node[right]{\footnotesize $y_p$} -- (2,2) node[above, xshift=0.33cm]{\footnotesize $(u_p,\Omega_p)$} ;
           \draw[draw,thick] (1.3,-0.6) node[left,xshift=0.2cm,yshift=0.25cm]{\footnotesize $a_{p+1}$} node[right]{\footnotesize $y_{p+1}$} -- (1.8,-2.2) node[right]{\footnotesize $(v_{p+1},\Omega_{p+1})$} ;
           \draw[draw,thick] (0.5,-1) node[above]{\footnotesize $a_{n}$} node[left, yshift=-0.05cm]{\footnotesize $y_{n}$} -- (0.3,-3.7) node[right]{\footnotesize $(v_{n},\Omega_{n})$} ;
           \node at (1,-1.5) {\rotatebox{45}{\footnotesize $\cdots$}};
        \node at (3.3,1.3) {\footnotesize $\mathcal{I}^+$};
        \node at (3.3,-1.3) {\footnotesize $\mathcal{I}^-$};
        \fill (0.5,1) circle (2pt);
        \fill (0.5,-1) circle (2pt);
        \fill (1.3,0.6) circle (2pt);
        \fill (1.3,-0.6) circle (2pt);
    \end{tikzpicture}\hspace{1cm}
    \caption{\centering{$n$-point correlator of mixed type.}}
    \label{npoint}
\end{figure}
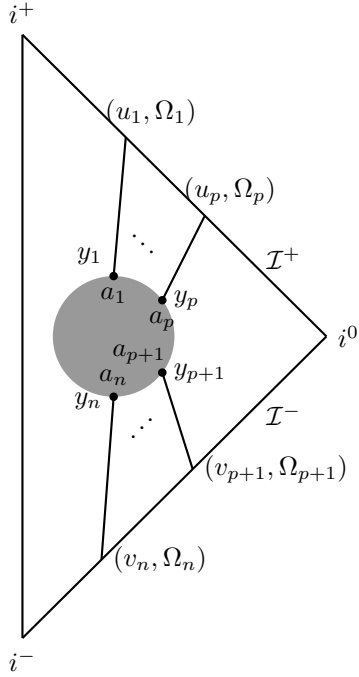

Near past null infinity $\mathcal{I}^-$, the fall-off condition of the bulk field is
\be 
\Phi(x)=\frac{\Sigma^{(-)}(v,\Omega)}{r}+o(r^{-1})
\ee where $(v,r,\Omega)$ are advanced coordinates. There should be another bulk-to-boundary propagator 
\be 
D_{ab}^{(-)}(v,\Omega;y)=\lim_{r\to\infty,\ v\ \text{finite}} r \ G_{ab}(x-y).
\ee 
 The previous discussion can be extended to the $n$ point correlator of mixed type 
\bea 
&&\langle \prod_{j=1}^m \Sigma(u_j,\Omega_j)\prod_{j=m+1}^n \Sigma^{(-)}(v_j,\Omega_j)\rangle_\beta\nn\\&=&\left(\prod_{j=m+1}^{n}\lim_{r_j\to\infty,\ v_j\ \text{finite}} r_j\right)\left(\prod_{j=1}^m \lim_{r_j\to\infty,\ u_j\ \text{finite}} r_j\right) \mathcal G(x_1,x_2,\cdots,x_n)\nn\\&=&\sum_{a_1,a_2,\cdots,a_n}\left(\int\prod_{j=1}^n d^4y_j \right) \left(\prod_{i=1}^m D_{a_i1}(u_i,\Omega_i;y_{i})\right)\left(\prod_{i=m+1}^n D^{(-)}_{a_i1}(v_i,\Omega_i;y_{i})\right)\mathcal G_{a_1a_2\cdots a_n}(y_1,y_2,\cdots,y_n),\nn\\\label{Sigmanpoint2}
\eea and the Feynman rule can be read out from the formula which is similar to the previous one. In Figure  \ref{npoint}, we have converted it into a Feynman diagram in the Penrose diagram.
\section{Propagators}\label{propaga}
We have derived the Feynman rules in previous section. In this section, we will work out the bulk-to-bulk, bulk-to-boundary and boundary-to-boundary propagators. %The contour integral representation 
\subsection{Bulk-to-bulk propagator}
The bulk-to-bulk propagator is the extended Feynman propagator whose momentum space  form is given by \eqref{Dab}. These expressions can be found by mode expansion or solving the Green's function in the bulk. Utilizing the Fourier transform \eqref{FourierDab}, we find the position space Feynman propagator 
\bs\begin{align}
    G_{11}(x-y)&=\frac{\theta(x^0-y^0)}{8\pi\beta |\bm x-\bm y|}\{\coth\frac{\pi}{\beta}[|\bm x-\bm y|-(x^0-y^0-i\epsilon)]+\coth\frac{\pi}{\beta}[|\bm x-\bm y|+(x^0-y^0-i\epsilon)]\}\nn\\& +\frac{\theta(y^0-x^0)}{8\pi\beta|\bm x-\bm y|}\{\coth\frac{\pi}{\beta}[|\bm x-\bm y|-(x^0-y^0+i\epsilon)]+\coth\frac{\pi}{\beta}[|\bm x-\bm y|+(x^0-y^0+i\epsilon)]\},\\
    G_{12}(x-y)&=-\frac{1}{4\pi\beta|\bm x-\bm y|}\frac{\sinh \frac{2\pi}{\beta}|\bm x-\bm y|}{\cosh\frac{2\pi}{\beta}((x^0-y^0)+i\sigma)-\cosh\frac{2\pi}{\beta}|\bm x-\bm y|},\\
    G_{21}(x-y)&=-\frac{1}{4\pi\beta|\bm x-\bm y|}\frac{\sinh \frac{2\pi}{\beta}|\bm x-\bm y|}{\cosh\frac{2\pi}{\beta}((x^0-y^0)-i\sigma)-\cosh\frac{2\pi}{\beta}|\bm x-\bm y|},\\
    G_{22}(x-y)&=\frac{\theta(x^0-y^0)}{8\pi\beta |\bm x-\bm y|}\{\coth\frac{\pi}{\beta}[|\bm x-\bm y|-(x^0-y^0+i\epsilon)]+\coth\frac{\pi}{\beta}[|\bm x-\bm y|+(x^0-y^0+i\epsilon)]\}\nn\\& +\frac{\theta(y^0-x^0)}{8\pi\beta|\bm x-\bm y|}\{\coth\frac{\pi}{\beta}[|\bm x-\bm y|-(x^0-y^0-i\epsilon)]+\coth\frac{\pi}{\beta}[|\bm x-\bm y|+(x^0-y^0-i\epsilon)]\}.
\end{align}\es We may set $\sigma=\frac{\beta}{2}$ in the above expressions and then the bulk-to-bulk propagator matrix is symmetric. In the zero temperature limit, $\beta\to\infty$, we find 
\bs\begin{align}
    D_{11}(x-y)&=\frac{1}{4\pi^2[(x-y)^2+i\epsilon]},\\
    D_{12}(x-y)&=0,\\
    D_{21}(x-y)&=0,\\
    D_{22}(x-y)&=\frac{1}{4\pi^2[(x-y)^2-i\epsilon]}.
\end{align}\es  The first one is the Feynman propagator at zero temperature while the last one is the complex conjugate of the first one. The second and the third propagators vanish in the zero temperature limit.

One can also obtain the following integral representation of the bulk-to-bulk propagator 
\bs\label{bulktobulkint}\begin{align}
    G_{11}(x-y)&=\frac{\theta(x^0-y^0)}{4\pi^2|\bm x-\bm y|}\int_0^\infty d\omega [(1+n(\omega)) e^{-i\omega(x^0-y^0)}+n(\omega) e^{i\omega(x^0-y^0)}]\sin \omega|\bm x-\bm y| \nn\\&+\frac{\theta(y^0-x^0)}{4\pi^2|\bm x-\bm y|}\int_0^\infty d\omega [n(\omega) e^{-i\omega(x^0-y^0)}+(1+n(\omega))e^{i\omega(x^0-y^0)}]\sin \omega|\bm x-\bm y|\nn\\&=\frac{\theta(x^0-y^0)}{4\pi^2|\bm x-\bm y|}\int_{-\infty}^\infty d\omega n(\omega) e^{i\omega(x^0-y^0)}\sin\omega|\bm x-\bm y| \nn\\&+\frac{\theta(y^0-x^0)}{4\pi^2|\bm x-\bm y|}\int_{-\infty}^\infty d\omega n(\omega) e^{-i\omega(x^0-y^0)}\sin\omega|\bm x-\bm y|,\\%\nn\\&=\frac{\theta(x^0-y^0)}{8\pi\beta |\bm x-\bm y|}\{\coth\frac{\pi}{\beta}[|\bm x-\bm y|-(x^0-y^0-i\epsilon)]+\coth\frac{\pi}{\beta}[|\bm x-\bm y|+(x^0-y^0-i\epsilon)]\}\nn\\& +\frac{\theta(y^0-x^0)}{8\pi\beta|\bm x-\bm y|}\{\coth\frac{\pi}{\beta}[|\bm x-\bm y|-(x^0-y^0+i\epsilon)]+\coth\frac{\pi}{\beta}[|\bm x-\bm y|+(x^0-y^0+i\epsilon)]\}
   % \nn\\&=-i\frac{1}{4\pi}\delta((x-y)^2)+\frac{1}{4\pi\beta|\bm x-\bm y|}[\frac{1}{e^{\frac{2\pi}{\beta}(x^0-y^0+|\bm x-\bm y|)}-1}-\frac{1}{e^{\frac{2\pi}{\beta}(x^0-y^0-|\bm x-\bm y|)}-1}],\\
    G_{12}(x-y)&=\frac{1}{4\pi^2|\bm x-\bm y|}\int_0^\infty d\omega n(\omega)[e^{\sigma \omega} e^{-i\omega(x^0-y^0)}+e^{(\beta-\sigma)\omega}e^{i\omega(x^0-y^0)}]\sin \omega|\bm x-\bm y|\nn\\&=\frac{1}{4\pi^2|\bm x-\bm y|}\int_{-\infty}^\infty d\omega n(\omega)e^{\sigma \omega}e^{-i\omega(x^0-y^0)} \sin\omega|\bm x-\bm y|,\\%\nn\\&=-\frac{1}{4\pi\beta|\bm x-\bm y|}\frac{\sinh \frac{2\pi}{\beta}|\bm x-\bm y|}{\cosh\frac{2\pi}{\beta}((x^0-y^0)+i\sigma)-\cosh\frac{2\pi}{\beta}|\bm x-\bm y|},\\
    G_{21}(x-y)&=\frac{1}{4\pi^2|\bm x-\bm y|}\int_0^\infty d\omega n(\omega)[e^{\sigma \omega} e^{i\omega(x^0-y^0)}+e^{(\beta-\sigma)\omega}e^{-i\omega(x^0-y^0)}]\sin \omega|\bm x-\bm y|\nn\\&=\frac{1}{4\pi^2|\bm x-\bm y|}\int_{-\infty}^\infty d\omega n(\omega)e^{\sigma \omega}e^{i\omega(x^0-y^0)} \sin\omega|\bm x-\bm y|,\\%\nn\\&=-\frac{1}{4\pi\beta|\bm x-\bm y|}\frac{\sinh \frac{2\pi}{\beta}|\bm x-\bm y|}{\cosh\frac{2\pi}{\beta}((x^0-y^0)-i\sigma)-\cosh\frac{2\pi}{\beta}|\bm x-\bm y|},\\
    G_{22}(x-y)&=\frac{\theta(x^0-y^0)}{4\pi^2|\bm x-\bm y|}\int_0^\infty d\omega [(1+n(\omega)) e^{i\omega(x^0-y^0)}+n(\omega) e^{-i\omega(x^0-y^0)}]\sin \omega|\bm x-\bm y| \nn\\&+\frac{\theta(y^0-x^0)}{4\pi^2|\bm x-\bm y|}\int_0^\infty d\omega [n(\omega) e^{i\omega(x^0-y^0)}+(1+n(\omega))e^{-i\omega(x^0-y^0)}]\sin \omega|\bm x-\bm y|\nn\\&=\frac{\theta(x^0-y^0)}{4\pi^2|\bm x-\bm y|}\int_{-\infty}^\infty d\omega n(\omega) e^{-i\omega(x^0-y^0)}\sin\omega|\bm x-\bm y| \nn\\&+\frac{\theta(y^0-x^0)}{4\pi^2|\bm x-\bm y|}\int_{-\infty}^\infty d\omega n(\omega) e^{i\omega(x^0-y^0)}\sin\omega|\bm x-\bm y|.%\frac{\theta(x^0-y^0)}{8\pi\beta |\bm x-\bm y|}\{\coth\frac{\pi}{\beta}[|\bm x-\bm y|-(x^0-y^0+i\epsilon)]+\coth\frac{\pi}{\beta}[|\bm x-\bm y|+(x^0-y^0+i\epsilon)]\}\nn\\& +\frac{\theta(y^0-x^0)}{8\pi\beta|\bm x-\bm y|}\{\coth\frac{\pi}{\beta}[|\bm x-\bm y|-(x^0-y^0-i\epsilon)]+\coth\frac{\pi}{\beta}[|\bm x-\bm y|+(x^0-y^0-i\epsilon)]\}.%\nn\\&=i\frac{1}{4\pi}\delta((x-y)^2)+\frac{1}{4\pi\beta|\bm x-\bm y|}[\frac{1}{e^{\frac{2\pi}{\beta}(x^0-y^0+|\bm x-\bm y|)}-1}-\frac{1}{e^{\frac{2\pi}{\beta}(x^0-y^0-|\bm x-\bm y|)}-1}].
\end{align}\es \iffalse We have used the integration 
\bea  
\int_{-\infty}^\infty d\omega \frac{e^{-i\omega t}\sin\omega x}{e^{\omega}-1}&=&\frac{\pi}{2}[\coth \pi(x-t)+\coth\pi(x+t)],\quad 0<\text{Im}(t)<1\\
\int_{-\infty}^\infty d\omega \frac{e^{\omega a}\sin\omega x}{e^{\omega}-1}&=&-\frac{\pi  \sinh (2 \pi  x)}{\cos (2 \pi  a)-\cosh (2 \pi  x)},\quad |\text{Im}(x)|<\text{min}\{\text{Re}(a),1-\text{Re}(a)\}.
\eea \fi
\subsection{Bulk-to-boundary propagator}
\paragraph{Retarded bulk-to-boundary propagator.}
We may write the retarded bulk-to-boundary propagator more explicitly as \be 
D_{ab}(u,\Omega;x)=\langle T'_C(\Phi_a(x)\Sigma_b(u,\Omega))\rangle_\beta,
\ee where we have defined a time-ordered product $T'_{C}$ through bulk reduction
\bea 
    T'_C(\Phi_a(x)\Sigma_b(u,\Omega))=\left\{\begin{array}{cc}\Sigma(u,\Omega)\Phi(t,\bm x), &a=1,b=1,\\
    \Sigma(u-i\sigma,\Omega)\Phi(t,\bm x),& a=1,b=2,\\
    \Phi(t-i\sigma,\bm x)\Sigma(u,\Omega),&a=2,b=1,\\
    \Phi(t-i\sigma,\bm x)\Sigma(u-i\sigma,\Omega),&a=2,b=2.\end{array}\right.\label{timeorderTp}
\eea In the first line, both of the boundary field $\Sigma$ and the bulk field $\Phi$ are in the path $C_1$. Since the time of $\Sigma$ approaches $+\infty$, we should put the boundary field $\Sigma$ before the bulk one. In the second line, the boundary field is inserted in the path $C_2$ while the bulk field is inserted in the path $C_1$. Therefore, the boundary field is always before the bulk field. In the third line, the boundary field is inserted in the path $C_1$ while the bulk field is inserted in $C_2$. Then the bulk field is always before the boundary field. In the last line, both of the boundary field and the bulk field are inserted in the path $C_2$, we should put the bulk field before the boundary field since the time of the boundary field approaches $+\infty$.

 We will write the bulk point $y^\mu$ in retarded coordinates 
\be 
y^\mu=u\bar m^\mu+r \ell^\mu.\label{labely}
\ee 
Using the formula \eqref{limitDab}, we find the 
retarded bulk-to-boundary propagators
\bs\label{retardedbb}\begin{align}
    D_{11}(u,\Omega;x)&=-\frac{1}{4\pi\beta}\frac{1}{e^{\frac{2\pi}{\beta}(u+\ell\cdot x-i\epsilon)}-1},\\%\quad x^0\in C_1,\quad u\in C_1\\
    D_{12}(u,\Omega;x)&=-\frac{1}{4\pi\beta}\frac{1}{e^{\frac{2\pi}{\beta}(u+\ell\cdot x-i\sigma)}-1},\\%\quad x^0\in C_1,\quad u\in C_2\\
    D_{21}(u,\Omega;x)&=-\frac{1}{4\pi\beta}\frac{1}{e^{\frac{2\pi}{\beta}(u+\ell\cdot x+i\sigma)}-1},\\%\quad x^0\in C_2,\quad u\in C_1\\
    D_{22}(u,\Omega;x)&=-\frac{1}{4\pi\beta}\frac{1}{e^{\frac{2\pi}{\beta}(u+\ell\cdot x+i\epsilon)}-1}.%\quad x^0\in C_2,\quad u\in C_2.
\end{align}\es The propagators $D_{11}$ and $D_{22}$ are the form of extended Bose-Einstein distribution, albeit in the position space.  They satisfy the relation 
\be 
D_{22}^*(u,\Omega;x)=D_{11}(u,\Omega;x),\quad D^*_{12}(u,\Omega;x)=D_{21}(u,\Omega;x).
\ee Setting $\sigma=\frac{\beta}{2}$, the retarded bulk-to-boundary propagators $D_{12}$ and $D_{21}$ become the form of extended Fermi-Dirac distribution in the position space
\begin{align}
    D_{12}(u,\Omega;x)= D_{21}(u,\Omega;x)=\frac{1}{4\pi\beta}\frac{1}{e^{\frac{2\pi}{\beta}(u+\ell\cdot x)}+1}.%\quad x^0\in C_1,\quad u\in C_2\\
\end{align}

A more interesting property is the discontinuity of the propagator $D_{11}$(and $D_{22}$) when crosses the hyperplane 
\be\label{bulkboundarygeodesic} 
u+\ell\cdot x=0,
\ee which is composed by the poles of the propagator. We compute the imaginary part through 
\bea 
D_{11}(u,\Omega;x)-D_{11}^*(u,\Omega;x)=-\frac{1}{4\pi\beta}\Big[\frac{1}{e^{\frac{2\pi}{\beta}(u+\ell\cdot x-i\epsilon)}-1}-\frac{1}{e^{\frac{2\pi}{\beta}(u+\ell\cdot x+i\epsilon)}-1}\Big]=-\frac{i}{4\pi}\delta(u+\ell\cdot x),
\eea where we have used the expansion in \eqref{bernoulli2} and the formula
\be 
\frac{1}{x+i\epsilon}-\frac{1}{x-i\epsilon}=-2\pi i\delta(x).
\ee 

The integral representation of  the retarded bulk-to-boundary propagators are
\bs\label{bulktoboundary}\begin{align}
    D_{11}(u,\Omega;x)&=-\frac{1}{8\pi^2 i}\int_{\mathcal C} d\omega \frac{e^{i\omega(u+\ell\cdot x-i\epsilon)}}{e^{\beta\omega}-1}=-\frac{1}{8\pi^2 i}\int_{\mathcal C} d\omega n(\omega)e^{i\omega(u+\ell\cdot x-i\epsilon)},\\
    D_{12}(u,\Omega;x)&=-\frac{1}{8\pi^2 i}\int_{\mathcal C} d\omega \frac{e^{i\omega(u+\ell\cdot x-i\sigma)}}{e^{\beta\omega}-1}=-\frac{1}{8\pi^2 i}\int_{\mathcal C} d\omega n(\omega) e^{i\omega(u+\ell\cdot x-i\sigma)},\\
    D_{21}(u,\Omega;x)&=-\frac{1}{8\pi^2 i}\int_{\mathcal C} d\omega \frac{e^{i\omega(u+\ell\cdot x-i(\beta-\sigma))}}{e^{\beta\omega}-1}=-\frac{1}{8\pi^2 i}\int_{\mathcal C} d\omega n(\omega)e^{i\omega(u+\ell\cdot x-i(\beta-\sigma))},\\
    D_{22}(u,\Omega;x)&=-\frac{1}{8\pi^2 i}\int_{\mathcal C} d\omega \frac{e^{i\omega(u+\ell\cdot x-i(\beta-\epsilon))}}{e^{\beta\omega}-1}=-\frac{1}{8\pi^2 i}\int_{\mathcal C} d\omega n(\omega) e^{i\omega(u+\ell\cdot x-i(\beta-\epsilon))},
\end{align}\es where contour $\mathcal C$ is from $-\infty$ and wraps around $\omega=0$ in a clockwise {way} to the positive $\omega $ axis and then goes to $+\infty$. This has been shown in Figure  \ref{contourC} and we have
\be 
\mathcal C=\mathcal C_1\cup \mathcal C_2\cup \mathcal C_3.
\ee 
\begin{figure}
    \centering
    \usetikzlibrary{decorations.text}
    \begin{tikzpicture} [scale=0.8]
  % \label{sigmatimepath}
        \draw[->,thick] (-6,0) -- (-3,0) node[above]{\footnotesize $\mathcal{C}_1$};
        \draw[draw,thick] (-3,0)-- (-0.5,0) ;
        \node at (-0.8,-0.3){\footnotesize $-\eta$};
        \draw[->,thick] (0.5,0) -- (3,0) node[above]{\footnotesize $\mathcal{C}_2$};
        \draw[->,thick] (3,0) -- (6,0) node[right]{\footnotesize \text{Re}$\ \omega$};
        \draw[thick,->] (-0.5,0) arc (180:90:0.5) node[above]{\footnotesize $\mathcal{C}_3$};
        \draw[draw,thick] (0,0.5) arc (90:0:0.5) ;
        \node at (0.8,-0.3){\footnotesize $+\eta$};
        \filldraw[black] (0,0) circle (2pt) node[below]{\footnotesize $0$};
%\caption{The time path $C$.}
    \end{tikzpicture}
    \caption{\centering{The contour $\mathcal C$.}}
    \label{contourC}
\end{figure}
\begin{figure}
    \centering
    \usetikzlibrary{decorations.text}
    \begin{tikzpicture} [scale=0.8]
        \draw[->,thick] (-6,0) -- (-3,0) node[above]{\footnotesize $\mathcal{C}_1$};
        \draw[draw,thick] (-3,0)-- (-0.5,0) ;
        \node at(-0.8,0.3){\footnotesize $-\eta$};
        \draw[->,thick] (0.5,0) -- (3,0) node[above]{\footnotesize $\mathcal{C}_2$};
        \draw[->,thick] (3,0) -- (6,0) node[right]{\footnotesize \text{Re}$\ \omega$};
        \draw[thick,->] (-0.5,0) arc (180:270:0.5) node[below]{\footnotesize $\mathcal{C}'_3$};
        \draw[draw,thick] (0,-0.5) arc (270:360:0.5);
        \node at(0.8,0.3){\footnotesize $+\eta$};
        \filldraw[black] (0,0) circle (2pt) node[above]{\footnotesize $0$};
    \end{tikzpicture}
    \caption{\centering{The contour $\mathcal C'$.}}
    \label{ContourCp}
\end{figure}
To prove this point, we assume $u+\ell\cdot x>0$ at first. Then using the residue theorem, 
\bea 
-\frac{1}{8\pi^2 i}\int_{\mathcal C} d\omega \frac{e^{i\omega(u+\ell\cdot x-i\epsilon)}}{e^{\beta\omega}-1}=2\pi i \sum_{k=1}^\infty \text{Res}_{\omega=\frac{2\pi i k}{\beta}}\left(-\frac{1}{8\pi^2 i}\int_{\mathcal C_{11}} d\omega \frac{e^{i\omega(u+\ell\cdot x-i\epsilon)}}{e^{\beta\omega}-1}\right)=D_{11}.
\eea When $u+n\cdot x<0$, we can also use the residue theorem 
\bea 
-\frac{1}{8\pi^2 i}\int_{\mathcal C} d\omega \frac{e^{i\omega(u+\ell\cdot x-i\epsilon)}}{e^{\beta\omega}-1}=-2\pi i \sum_{k=0}^\infty \text{Res}_{\omega=-\frac{2\pi i k}{\beta}}\left(-\frac{1}{8\pi^2 i}\int_{\mathcal C_{11}} d\omega \frac{e^{i\omega(u+\ell\cdot x-i\epsilon)}}{e^{\beta\omega}-1}\right)=D_{11}.
\eea Introducing the notation 
\be 
\epsilon_{ab}=\left\{\begin{array}{cc}\epsilon,&a=1,b=1,\\
\sigma,&a=1,b=2,\\
\beta-\sigma,&a=2,b=1,\\
\beta-\epsilon,&a=2,b=2,\end{array}\right.
\ee the retarded bulk-to-boundary propagator can be unified as 
\be 
D_{ab}(u,\Omega;x)=-\frac{1}{8\pi^2 i}\int_{\mathcal C} d\omega n(\omega) e^{i\omega(u+\ell\cdot x-i\epsilon_{ab})}=-\frac{1}{4\pi\beta}\frac{1}{e^{\frac{2\pi}{\beta}(u+\ell\cdot x-i\epsilon_{ab})}-1}.
\ee 
Another integral representation of the retarded bulk-to-boundary propagator is
\bea 
D_{ab}(u,\Omega;x)&=&\frac{1}{8\pi^2 i}\int_{\mathcal C'} d\omega \frac{e^{\beta\omega-i\omega(u+\ell\cdot x-i\epsilon_{ab})}}{e^{\beta\omega}-1}=\frac{1}{8\pi^2 i}\int_{\mathcal C'} d\omega(1+n(\omega)) e^{-i\omega(u+\ell\cdot x-i\epsilon_{ab})}\nn\\&=&\frac{1}{8\pi^2 i}\int_{\mathcal C'} d\omega n(\omega) e^{-i\omega(u+\ell\cdot x+i(\beta-\epsilon_{ab}))} .\label{bulktoboundary2}
\eea 
\iffalse 
\bs\label{bulktoboundary2}\begin{align}
     D_{11}(u,\Omega;x)&=\frac{1}{8\pi^2 i}\int_{\mathcal C'} d\omega \frac{e^{\beta\omega-i\omega(u+n\cdot x-i\epsilon)}}{e^{\beta\omega}-1}=\frac{1}{8\pi^2 i}\int_{\mathcal C'} d\omega(1+n(\omega)) e^{-i\omega(u+n\cdot x-i\epsilon)},\label{contourCpD11}\\
    D_{12}(u,\Omega;x)&=\frac{1}{8\pi^2 i}\int_{\mathcal C'} d\omega \frac{e^{\beta\omega-i\omega(u+n\cdot x-i\sigma)}}{e^{\beta\omega}-1}=\frac{1}{8\pi^2 i}\int_{\mathcal C'} d\omega(1+n(\omega)) e^{-i\omega(u+n\cdot x-i\sigma)},\\
    D_{21}(u,\Omega;x)&=\frac{1}{8\pi^2 i}\int_{\mathcal C'} d\omega \frac{e^{\beta\omega-i\omega(u+n\cdot x+i\sigma)}}{e^{\beta\omega}-1}=\frac{1}{8\pi^2 i}\int_{\mathcal C'} d\omega(1+n(\omega)) e^{-i\omega(u+n\cdot x+i\sigma)},\\
    D_{22}(u,\Omega;x)&=\frac{1}{8\pi^2 i}\int_{\mathcal C'} d\omega \frac{e^{\beta\omega-i\omega(u+n\cdot x+i\epsilon)}}{e^{\beta\omega}-1}=\frac{1}{8\pi^2 i}\int_{\mathcal C'} d\omega(1+n(\omega)) e^{-i\omega(u+n\cdot x+i\epsilon)}.
\end{align}\es \fi 
As  shown in Figure \ref{ContourCp}, the path $\mathcal{C}'$ is from $-\infty$ and wraps $\omega=0$ in an anti-clockwise way to the positive axis and then goes to $+\infty$ along the real axis. More precisely, 
\be 
\mathcal C'=\mathcal C_1\cup \mathcal C_2\cup \mathcal C'_3.
\ee 
There are two ways to relate the contour $\mathcal C$ to $\mathcal C'$. 
\begin{itemize}
    \item Complex conjugate. From the Figure \ref{contourC} and \ref{ContourCp}, it is clear that the complex conjugate of $\mathcal C$ is exactly $\mathcal C'$
    \be 
    \mathcal C'=\mathcal C^*.
    \ee More precisely, we change variable $\omega$ to its complex conjugate $\omega^*$  and then the contour $\mathcal C$ for $\omega$ integration becomes the contour $\mathcal C'$ for $\omega^*$.
     \item Inverse. This is realized by  changing $\omega$ to its inversion $-\omega$. Then the contour $\mathcal C$ will change to $-\mathcal C'$ which is clear from the Figure \ref{contourC} and  \ref{ContourCp}. We denote the inversion of $\mathcal C$ briefly as 
     \be 
     \mathcal C'=-\mathcal C.
     \ee 
\end{itemize}
At zero temperature, we have 
\be 
\lim_{\beta\to\infty}1+n(\omega)=\theta(\omega).
\ee Therefore we reproduce the zero temperature bulk-to-boundary propagator \cite{Liu:2024nfc}
\be 
D_{11}(u,\Omega;x)=\frac{1}{8\pi^2 i}\int_0^\infty d\omega e^{-i\omega(u+\ell\cdot x-i\epsilon)},
\ee 
where the integral domain is restricted to positive real axis such that the boundary field is composed of positive frequency modes (outgoing modes) at $\mathcal{I}^+$. However, at finite temperature, the contour $\mathcal C$ or $\mathcal C'$ is deformed to the region with negative frequency modes, indicating that both incoming and outgoing modes of the boundary field contribute to the bulk-to-boundary propagator. For later convenience, we  define a generalized occupation number in the frequency space
\bea 
n_{ab}(\omega;\mathcal C)=n(\omega)e^{\omega\epsilon_{ab}}=\left\{\begin{array}{cc}n(\omega)e^{ \omega\epsilon},&a=1,b=1,\\
n(\omega)e^{\omega\sigma},&a=1,b=2,\\
n(\omega)e^{\omega(\beta-\sigma)}=(1+n(\omega))e^{-\omega\sigma},&a=2,b=1,\\
n(\omega)e^{\omega(\beta-\epsilon)}=(1+n(\omega))e^{-\omega\epsilon},&a=2,b=2.\end{array}\right.
\eea Then the retarded bulk-to-boundary propagator becomes
\be 
D_{ab}(u,\Omega;x)=-\frac{1}{8\pi^2 i}\int_{\mathcal C}d\omega n_{ab}(\omega;\mathcal C) e^{i\omega(u+\ell\cdot x)}.
\ee Note that the occupation number  depends on the contour $\mathcal C$. When we choose the contour $\mathcal C'$, the occupation number would be
\bea 
n_{ab}(\omega;\mathcal C')=(1+n(\omega))e^{-\omega\epsilon_{ab}}=\left\{\begin{array}{cc} (1+n(\omega))e^{-\omega\epsilon},&a=1,b=1,\\
(1+n(\omega))e^{-\omega\sigma},&a=1,b=2,\\
(1+n(\omega))e^{-\omega(\beta-\sigma)}=n(\omega)e^{\omega\sigma},&a=2,b=1,\\
(1+n(\omega))e^{-\omega(\beta-\epsilon)}=n(\omega)e^{\omega\epsilon},&a=2,b=2\end{array}\right.
\eea and the  retarded bulk-to-boundary propagator is 
\be 
D_{ab}(u,\Omega;x)=\frac{1}{8\pi^2 i}\int_{\mathcal C'}d\omega n_{ab}(\omega;\mathcal C')e^{-i\omega(u+\ell\cdot x)}.
\ee We will omit the dependence on the contour in the occupation number when it is clear from the context of the article.

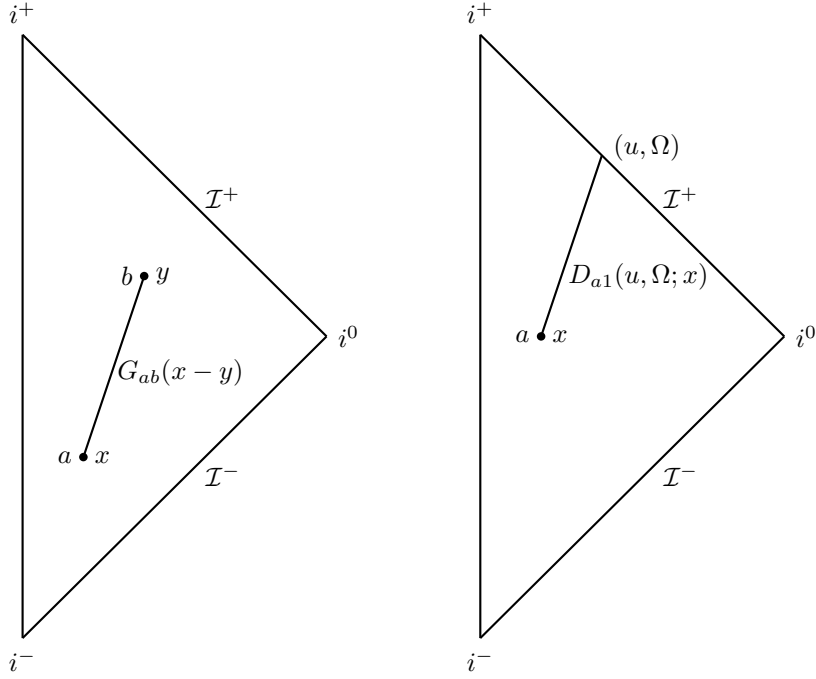
\begin{figure}
    \centering
    \usetikzlibrary{decorations.text}
    \begin{tikzpicture} [scale=0.8]
        \draw[draw,thick] (-1,5) node[above]{\footnotesize $i^+$} -- (4,0) node[right]{\footnotesize $i^0$};
        \draw[draw,thick] (4,0) -- (-1,-5) node[below]{\footnotesize $i^-$};
        \draw[draw,thick](-1,5) -- (-1,-5);
        \node at (2.3,2.3) {\footnotesize $\mathcal{I}^+$};
        \node at (2.3,-2.3) {\footnotesize $\mathcal{I}^-$};
        \draw[draw,thick] (0,-2) node[left]{\footnotesize $a$} node[right]{\footnotesize $x$}-- (1,1) node[left]{\footnotesize $b$} node[right]{\footnotesize $y$};
        \node at (1.6,-0.6){\footnotesize $G_{ab}(x-y)$};
        \fill (0,-2) circle (2pt);
        \fill (1,1) circle (2pt);
    \end{tikzpicture}\hspace{1cm}
    \begin{tikzpicture} [scale=0.8]
        \draw[draw,thick] (-1,5) node[above]{\footnotesize $i^+$} -- (4,0) node[right]{\footnotesize $i^0$};
        \draw[draw,thick] (4,0) -- (-1,-5) node[below]{\footnotesize $i^-$};
        \draw[draw,thick](-1,5) -- (-1,-5);
        \node at (2.3,2.3) {\footnotesize $\mathcal{I}^+$};
        \node at (2.3,-2.3) {\footnotesize $\mathcal{I}^-$};
        \draw[draw,thick] (0,0) node[left]{\footnotesize $a$} node[right]{\footnotesize $x$} -- (1,3) node[right,yshift=0.1cm]{\footnotesize $(u,\Omega)$};
        \node at (1.6,1){\footnotesize $D_{a1}(u,\Omega;x)$};
        \fill (0,0) circle (2pt);
    \end{tikzpicture}\hspace{1cm}
    \caption{\centering{Bulk-to-bulk and bulk-to-boundary propagator in Penrose diagram.}}
    \label{bktobkbktobd}
\end{figure}

In Figure \ref{bktobkbktobd}, we have shown the bulk-to-bulk propagator and bulk-to-boundary propagator in Penrose diagram.

\paragraph{Feynman rules in momentum space.}
Given the bulk-to-boundary propagator, we can relate the Carrollian correlator to the momentum space one
\bea 
&&\langle \prod_{j=1}^n\Sigma(u_j,\Omega_j)\rangle_{\beta}\nn\\&=&\sum_{a_1,a_2,\cdots,a_n}\left(\int\prod_{j=1}^n d^4y_j \right) \left(\prod_{i=1}^n D_{a_i1}(u_i,\Omega_i;y_{i})\right)\mathcal G_{a_1a_2\cdots a_n}(y_1,y_2,\cdots,y_n)\nn\\&=& \left(\frac{1}{8\pi^2 i}\right)^n \sum_{a_1,a_2,\cdots,a_n}\left(\int \prod_{j=1}^n dy_j\right)\left(\int_{\mathcal C'}\prod_{i=1}^n d\omega_i n_{a_i1}(\omega_i) e^{-i\omega_i(u_i+\ell_i\cdot y_i)}\right)\mathcal G_{a_1a_2\cdots a_n}(y_1,y_2,\cdots,y_n)\nn\\&=& \left(\frac{1}{8\pi^2 i}\right)^n\sum_{a_1,a_2,\cdots,a_n}\left(\int_{\mathcal C'}\prod_{j=1}^n d\omega_j n_{a_j1}(\omega_j) e^{-i\omega_ju_j}\right)(2\pi)^4 \delta^{(4)}(\sum_{j=1}^n p_j)i \mathcal{M}_{a_1a_2\cdots a_n}(p_1,p_2,\cdots,p_n).\nn\\\label{fernmanmomentum}
\eea We have used the integral representation of the bulk-to-boundary propagator in the third line. By defining $p_j=\omega_j \ell_j$, we transform the connected and  amputated Green's function to momentum space one at the last step
\bea 
(2\pi)^4 \delta^{(4)}(\sum_{j=1}^n p_j) i\mathcal{M}_{a_1a_2\cdots a_n}(p_1,p_2,\cdots,p_n)=\left(\int \prod_{j=1}^n d^4 y_j e^{-ip_j y_j}\right) \mathcal G_{a_1a_2\cdots a_n}(y_1,y_2,\cdots,y_n).
\eea We have separated out a Dirac  delta function follows from the conservation of four-momentum. The generalized $\mathcal M$ matrix carries index of type 1 or type 2. At zero temperature, the generalized $\mathcal M$ matrix becomes the usual one 
\bea 
(2\pi)^4 \delta^{(4)}(\sum_{j=1}^n p_j) i\mathcal{M}(p_1,p_2,\cdots,p_n)= \left(\int \prod_{j=1}^n d^4 y_j e^{-ip_j y_j}\right) \mathcal G_{\text{connected and amputated}}(y_1,y_2,\cdots,y_n).
\eea 
To be more precise, we take the zero temperature limit. The occupation number $n_{ab}(\omega)$ on the path $\mathcal C'$ becomes
\bea 
\lim_{\beta\to\infty}n_{11}(\omega)=\theta(\omega),\quad \lim_{\beta\to\infty}n_{21}(\omega)=0,
\eea and only the index of type 1 contributes to the correlator. Therefore, the formula \eqref{fernmanmomentum} becomes exactly the one in \cite{Liu:2024nfc}
\bea 
&&\lim_{\beta\to\infty}\langle \prod_{j=1}^n\Sigma(u_j,\Omega_j)\rangle_{\beta}\nn\\&=&\left(\frac{1}{8\pi^2 i}\right)^n\left(\int_{0}^\infty\prod_{j=1}^n d\omega_j  e^{-i\omega_ju_j}\right)(2\pi)^4 \delta^{(4)}(\sum_{j=1}^n p_j)i \lim_{\beta\to\infty}\mathcal{M}_{1,1\cdots 1}(p_1,p_2,\cdots,p_n)
\eea with the identification 
\be 
\mathcal{M}(p_1,p_2,\cdots,p_n)= \lim_{\beta\to\infty}\mathcal{M}_{1,1\cdots 1}(p_1,p_2,\cdots,p_n).
\ee In summary, the Carrollian correlator at  
finite temperature is still a modified Fourier transform of the generalized momentum space amplitude. We define a momentum space quantity 
\be 
i\mathcal C_{a_1a_2\cdots a_n}(p_1,p_2,\cdots,p_n)=\left(\frac{1}{8\pi^2 i}\right)^n \prod_{j=1}^n n_{a_j 1}(\omega_j) i\mathcal M_{a_1\cdots a_n}(p_1,p_2,\cdots,p_n).
\ee The Feynman rules for $i\mathcal C_{a_1a_2\cdots a_n}(p_1,p_2,\cdots,p_n)$ are as follows: 
The external points are of type 1 and the bulk points are of type 1 or type 2. For each external point at $(u,\Omega)$, there is an associated external line  with  momentum $p=\omega\ell$ that connects a vertex of type $a$ and we should join a factor $n_{a1}(\omega)$. For each vertex of type 1  or type 2, we joint a factor $-i\lambda$ or $+i\lambda$ respectively. For each internal line that connects two  vertices of type $a$ and type $b$, there is an associated momentum $p$ and we should join a Feynman propagator $G_{ab}(p)$. At each vertex, the four momentum is conserved and we should integrate out all the loop momentum $p$ with the measure $\int\frac{d^4p}{(2\pi)^4}$. Finally, we divide the symmetry factor and sum over all possible types of vertices. The Feynman rules are summarized below. 
\begin{itemize}
\item  One must assign types 1 and 2 to the vertices of a diagram in all the possible ways; The external points are always type 1.

\item  Each vertex of type 1 brings a factor $-i\lambda$ and of type 2 a $+i\lambda$
 \begin{eqnarray}	     
	    \begin{tikzpicture}[scale=0.7,baseline=(current bounding box.center)]
	    	\fill (0,0) circle (2pt);
	    	\draw (0,0) node[below] {$1$} -- (1,1);
	    	\draw (0,0) -- (1,-1);
	    	\draw (0,0) -- (-1,1);
	    	\draw (0,0) -- (-1,-1);
	    \end{tikzpicture} \ \ =-i\lambda,\hspace{1cm}
	     \begin{tikzpicture}[scale=0.7,baseline=(current bounding box.center)]
	    	\fill (0,0) circle (2pt);
	    	\draw (0,0) node[below] {$2$} -- (1,1);
	    	\draw (0,0) -- (1,-1);
	    	\draw (0,0) -- (-1,1);
	    	\draw (0,0) -- (-1,-1);
	    \end{tikzpicture} \ \ =+i\lambda.\nn
	\end{eqnarray}

\item  A vertex of type $a$ and a vertex of type $b$ are connected by the free propagator $G_{ab}(p)$ where $p$ is the associated momentum
	\begin{align}
			\begin{tikzpicture}[baseline=(current bounding box.center)]
				\fill (0,0) circle (1.5pt);
				\fill (2,0) circle (1.5pt);
				\draw[->,thick] (0,0) node[below] {$a$} -- (1,0) node[below] {$p$};
				\draw (1,0) -- (2,0) node[below] {$b$};
			\end{tikzpicture} \
			=G_{ab}(p).\nn
		\end{align}

\item  Each loop momentum $p$ must be integrated with the measure $\int\frac{d^4p}{(2\pi)^4}$.

\item Each external line between an external point $(u,\Omega)$ and a bulk vertex of type $a$ has an associated external momentum $p=\omega \ell$ and one should join an occupation number $n_{a1}(\omega)$ where $\omega$ is the dual energy of the retarded time $u$  
\begin{align}
	      \begin{tikzpicture}[baseline=(current bounding box.center)]
		           \fill (0,0) circle (1.5pt);
		           \draw[->, thick] (0,0) node[below] {$a$} -- (1,0) node[below]{\footnotesize $\omega$};
		           \draw[draw, thick] (1,0) -- (2,0);
   	         \end{tikzpicture} \
	       = n_{a1}(\omega).\label{externalpoint}
        \end{align} 
\item Divide by the symmetry factor and sum over all possible types of vertices.
\end{itemize} The first four rules and the last one are the same as the usual ones except that one should take care of different types of vertices. The fifth rule comes from the bulk-to-boundary propagator in the momentum space.

After obtaining the momentum space quantity $i\mathcal C_{a_1a_2\cdots a_n}(p_1,p_2,\cdots,p_n)$, we should add an overall factor that represents the momentum conservation $(2\pi)^4 \delta(\sum_{j=1}^n p_j)$ and Fourier transform it along the contour $\mathcal C'$ with the measure $\left(\frac{1}{8\pi^2 i}\right)^n \int_{\mathcal C'}\prod_{j=1}^n  d\omega_j e^{-i\omega_j u_j}$
\bea 
&&\langle \prod_{j=1}^n\Sigma(u_j,\Omega_j)\rangle_{\beta}=\left(\frac{1}{8\pi^2 i}\right)^n \int_{\mathcal C'}\prod_{j=1}^n  d\omega_j e^{-i\omega_j u_j}(2\pi)^4 \delta^{(4)}(\sum_{j=1}^n p_j)i\mathcal C_{a_1a_2\cdots a_n}(p_1,p_2,\cdots,p_n).\nn\\
\eea 
One can also choose the path $\mathcal C$, then one should change the corresponding occupation number and the Fourier transform becomes 
\be 
\left(-\frac{1}{8\pi^2 i}\right)^n \int_{\mathcal C}\prod_{j=1}^n  d\omega_j e^{i\omega_j u_j}[\cdots].
\ee 

\begin{figure}
    \centering
    \usetikzlibrary{decorations.text}
    \begin{tikzpicture} [scale=0.8]
  % \label{sigmatimepath}
        \draw[draw,thick] (-1,5) node[above]{\footnotesize $i^+$} -- (4,0) node[right]{\footnotesize $i^0$};
        \draw[draw,thick] (4,0) -- (-1,-5) node[below]{\footnotesize $i^-$};
        \draw[draw,thick](-1,5) -- (-1,-5);
        \node at (2.3,2.3) {\footnotesize $\mathcal{I}^+$};
        \node at (2.3,-2.3) {\footnotesize $\mathcal{I}^-$};
        \draw[draw,thick] (0,0) node[left]{\footnotesize $-i\lambda$} -- (0.25,1.75) node[left]{\footnotesize $D_{11}$} -- (0.5,3.5) node[above right]{\footnotesize $(u_1,\Omega_1)$};
        \draw[draw,thick] (0,0) -- (1.2,2) node[ left]{\footnotesize $D_{11}$} -- (1.5,2.5) node[above right]{\footnotesize $(u_2,\Omega_2)$};
        \draw[draw,thick] (0,0)  -- (2.5,1.5) node[above right]{\footnotesize $(u_3,\Omega_3)$};
         \node at (1.73,0.68) {\footnotesize $D_{11}$};
        \draw[draw,thick] (0,0) -- (1.75,0.25) node[below]{\footnotesize $D_{11}$} -- (3.5,0.5) node[above right]{\footnotesize $(u_4,\Omega_4)$};
        
    \end{tikzpicture}\hspace{1cm}
    \begin{tikzpicture} [scale=0.8]
        \draw[draw,thick] (-1,5) node[above]{\footnotesize $i^+$} -- (4,0) node[right]{\footnotesize $i^0$};
        \draw[draw,thick] (4,0) -- (-1,-5) node[below]{\footnotesize $i^-$};
        \draw[draw,thick] (-1,5) --  (-1,-5);
        \node at (2.3,2.3) {\footnotesize $\mathcal{I}^+$};
        \node at (2.3,-2.3) {\footnotesize $\mathcal{I}^-$};
        \draw[draw,thick] (0,0) node[left]{\footnotesize $+i\lambda$} -- (0.25,1.75) node[left]{\footnotesize $D_{21}$} -- (0.5,3.5) node[above right]{\footnotesize $(u_1,\Omega_1)$};
        \draw[draw,thick] (0,0) -- (1.2,2) node[ left]{\footnotesize $D_{21}$} -- (1.5,2.5) node[above right]{\footnotesize $(u_2,\Omega_2)$};
        \draw[draw,thick] (0,0) -- (2.5,1.5) node[above right]{\footnotesize $(u_3,\Omega_3)$};
        \node at (1.73,0.68) {\footnotesize $D_{21}$};
        \draw[draw,thick] (0,0) -- (1.75,0.25) node[below]{\footnotesize $D_{21}$} -- (3.5,0.5) node[above right]{\footnotesize $(u_4,\Omega_4)$};
        
    \end{tikzpicture}
    \caption{The tree level four-point Carrollian correlator at $\mathcal I^+$ in $\Phi^4$ theory. There are two types of vertices. }
    \label{fourpointCarrolliancorrelator}
\end{figure}
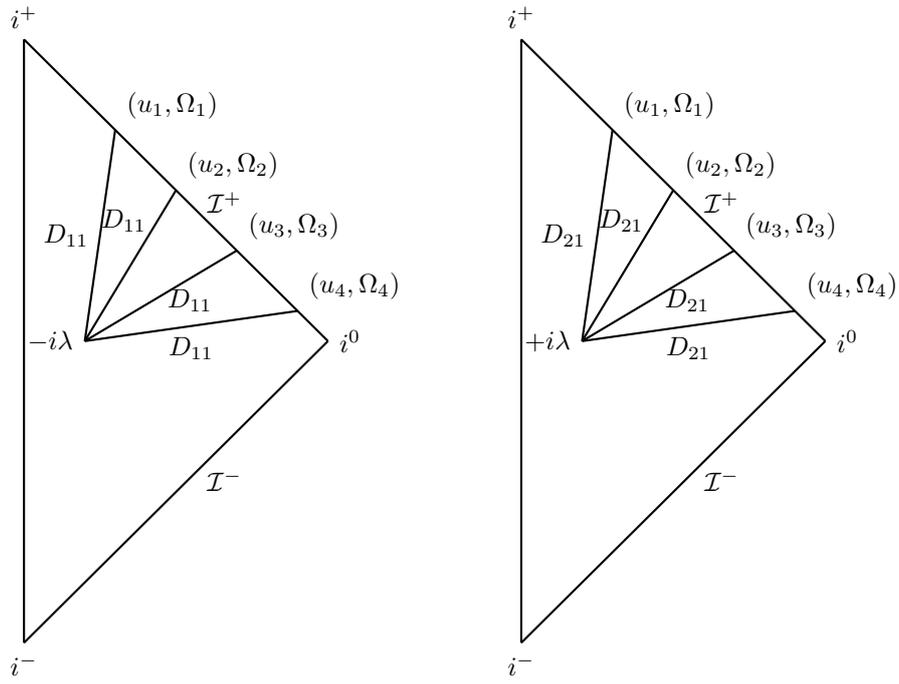

As an illustration, we consider the Feynman diagrams  that correspond to the four-point connected correlators in Figure \ref{fourpointCarrolliancorrelator}.

Using the Carrollian space Feynman rules, we find 
\bea 
&&\langle \Sigma(u_1,\Omega_1)\Sigma(u_2,\Omega_2)\Sigma(u_3,\Omega_3)\Sigma(u_4,\Omega_4)\rangle_\beta\nn\\&=&-i\lambda \int d^4x D_{11}(u_1,\Omega_1;x)D_{11}(u_2,\Omega_2;x)D_{11}(u_3,\Omega_3;x)D_{11}(u_4,\Omega_4;x)\nn\\&&+i\lambda \int d^4x D_{21}(u_1,\Omega_1;x)D_{21}(u_2,\Omega_2;x)D_{21}(u_3,\Omega_3;x)D_{21}(u_4,\Omega_4;x).
\eea The first line and the second line correspond to the diagram with vertex 1 and 2, respectively. Using the integral representation, we find \bea 
&&\langle \Sigma(u_1,\Omega_1)\Sigma(u_2,\Omega_2)\Sigma(u_3,\Omega_3)\Sigma(u_4,\Omega_4)\rangle_\beta\nn\\&=&-i\lambda\left(\frac{1}{8\pi^2}\right)^4\int d^4x \left(\prod_{j=1}^4 \int_{\mathcal C} d\omega_j n(\omega_j)\right)[e^{i\sum_{j=1}^4\omega_j(u_j+\ell_j\cdot x-i\epsilon)}-e^{i\sum_{j=1}^4 \omega_j(u_j+\ell_j\cdot x-i(\beta-\sigma))}]\nn\\&=&-i\lambda \left(\frac{1}{4\pi}\right)^4\left(\prod_{j=1}^4 \int_{\mathcal C} d\omega_j n(\omega_j)\right) \delta(\sum_{j=1}^4 \omega_j n_j)[e^{i\sum_{j=1}^4\omega_j(u_j-i\epsilon)}-e^{i\sum_{j=1}^4 \omega_j(u_j-i(\beta-\sigma))}].
\eea Due to the conservation of  energy, the exponential functions in the integrand are equal and then the connected  four-point correlator vanishes at $\mathcal{O}(\lambda)$. Note that the null result also appears at zero temperature where the conservation of  energy cannot be satisfied since the boundary fields are composed only by outgoing modes.

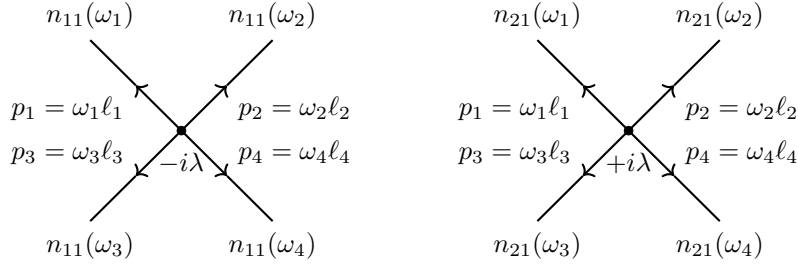
\begin{figure}
    \centering
    \usetikzlibrary{decorations.text}
    \begin{tikzpicture} [scale=0.8]
        \filldraw[black] (0,0) circle (2pt) node[below,yshift=-0.15cm]{\footnotesize $-i\lambda$};
        \draw[->,thick] (0,0) -- (-0.75,0.75) node[below left]{\footnotesize $p_1=\omega_1\ell_1$};
        \draw[draw,thick] (-0.75,0.75) --  (-1.5,1.5) node[above]{\footnotesize $n_{11}(\omega_1)$};
        \draw[->,thick] (0,0)  -- (0.75,0.75) node[below right]{\footnotesize $p_2=\omega_2\ell_2$};
        \draw[draw,thick] (0.75,0.75) --  (1.5,1.5) node[above]{\footnotesize $n_{11}(\omega_2)$};
        \draw[->,thick] (0,0) -- (-0.75,-0.75) node[above left]{\footnotesize $p_3=\omega_3\ell_3$};
        \draw[draw,thick] (-0.75,-0.75) --  (-1.5,-1.5) node[below]{\footnotesize $n_{11}(\omega_3)$};
        \draw[->,thick] (0,0)  -- (0.75,-0.75) node[above right]{\footnotesize $p_4=\omega_4\ell_4$};
        \draw[draw,thick] (0.75,-0.75) --  (1.5,-1.5) node[below]{\footnotesize $n_{11}(\omega_4)$};
    \end{tikzpicture}\hspace{1cm}
    \begin{tikzpicture} [scale=0.8]
        \filldraw[black] (0,0) circle (2pt) node[below,yshift=-0.15cm]{\footnotesize $+i\lambda$};
        \draw[->,thick] (0,0) -- (-0.75,0.75) node[below left]{\footnotesize $p_1=\omega_1\ell_1$};
        \draw[draw,thick] (-0.75,0.75) --  (-1.5,1.5) node[above]{\footnotesize $n_{21}(\omega_1)$};
        \draw[->,thick] (0,0)  -- (0.75,0.75) node[below right]{\footnotesize $p_2=\omega_2\ell_2$};
        \draw[draw,thick] (0.75,0.75) --  (1.5,1.5) node[above]{\footnotesize $n_{21}(\omega_2)$};
        \draw[->,thick] (0,0) -- (-0.75,-0.75) node[above left]{\footnotesize $p_3=\omega_3\ell_3$};
        \draw[draw,thick] (-0.75,-0.75) --  (-1.5,-1.5) node[below]{\footnotesize $n_{21}(\omega_3)$};
        \draw[->,thick] (0,0)  -- (0.75,-0.75) node[above right]{\footnotesize $p_4=\omega_4\ell_4$};
        \draw[draw,thick] (0.75,-0.75) --  (1.5,-1.5) node[below]{\footnotesize $n_{21}(\omega_4)$};
    \end{tikzpicture}\hspace{1cm}
    \caption{\centering{Feynman diagrams for  four-point connected correlator in momentum space.}}
    \label{fourpointmomentum}
\end{figure}

We can also compute the  four-point connected correlator in the momentum space at first. The Feynman diagrams are shown in Figure \ref{fourpointmomentum}. 

The momentum space Carrollian correlator is\footnote{We choose contour $\mathcal C$ here.}
\bs\begin{align}
i\mathcal C_{1,1,1,1}(p_1,p_2,p_3,p_4)&=-i\lambda n_{11}(\omega_1)n_{11}(\omega_2)n_{11}(\omega_3)n_{11}(\omega_4)=-i\lambda \prod_{j=1}^4  n(\omega_j),\\
i\mathcal C_{2,2,2,2}(p_1,p_2,p_3,p_4)&=+i\lambda n_{21}(\omega_1)n_{21}(\omega_2)n_{21}(\omega_3)n_{21}(\omega_4)=+i\lambda e^{\sigma(\sum_{j=1}^4 \omega_j)}\prod_{j=1}^4 n(\omega_j).
\end{align}\es 
The dependence on $\sigma$ can be dropped since the total energy is conserved and then we find 
\be 
\mathcal C_{1,1,1,1}(p_1,p_2,p_3,p_4)+\mathcal C_{2,2,2,2}(p_1,p_2,p_3,p_4)=0\quad \text{at tree level}.
\ee Therefore, its Fourier transform is also zero
\be 
\langle \Sigma(u_1,\Omega_1)\Sigma(u_2,\Omega_2)\Sigma(u_3,\Omega_3)\Sigma(u_4,\Omega_4)\rangle_\beta=0\quad\text{at tree level}
\ee which is consistent with the one from the Carrollian space Feynman rules. To get a non-trivial tree level  four-point Carrollian correlator, we should consider the boundary operators both at $\mathcal{I}^+$ and $\mathcal{I}^-$. We will derive the advanced bulk-to-boundary propagator at first.

\paragraph{Advanced bulk-to-boundary propagator.}
To approach $\mathcal I^-$, we can parameterize the bulk points
\be 
y^\mu=v \bar m^\mu+r \bar \ell^\mu
\ee with $\bar \ell^\mu=(-1,\ell^i)$. By taking the limit $r\to\infty$ with $v$ finite, we obtain the advanced bulk-to-boundary propagator
\be 
D_{ab}^{(-)}(v,\Omega;x)=-\frac{1}{4\pi\beta}\frac{1}{e^{\frac{2\pi}{\beta}(\bar \ell\cdot x-v-i\epsilon_{ba})}-1}.
\ee To be more precise, 
\bs\begin{align}
    D_{11}(v,\Omega;x)&=-\frac{1}{4\pi\beta}\frac{1}{e^{\frac{2\pi}{\beta}(\bar \ell\cdot x-v-i\epsilon)}-1},\\
    D_{12}(v,\Omega;x)&=-\frac{1}{4\pi\beta}\frac{1}{e^{\frac{2\pi}{\beta}(\bar \ell\cdot x-v+i\sigma)}-1},\\
    D_{21}(v,\Omega;x)&=-\frac{1}{4\pi\beta}\frac{1}{e^{\frac{2\pi}{\beta}(\bar \ell\cdot x-v-i\sigma)}-1},\\
    D_{22}(v,\Omega;x)&=-\frac{1}{4\pi\beta}\frac{1}{e^{\frac{2\pi}{\beta}(\bar \ell\cdot x-v+i\epsilon)}-1}.
\end{align}
\es 
The integral representation is 
\bea 
 D_{ab}^{(-)}(v,\Omega;x)=-\frac{1}{8\pi^2 i}\int_{\mathcal{C}} d\omega \frac{e^{i\omega(\bar \ell\cdot x-v-i\epsilon_{ba})}}{e^{\beta \omega}-1}=-\frac{1}{8\pi^2 i}\int_{\mathcal{C}} d\omega n(\omega)e^{i\omega(\bar \ell\cdot x-v-i\epsilon_{ba})}.
\eea 
\iffalse 
\bs\begin{align}
    D_{11}(v,\Omega;x)&=\frac{1}{8\pi^2 i}\int_{\mathcal{C}} d\omega \frac{e^{i\omega(\bar n\cdot x-v-i\epsilon)}}{e^{\beta \omega}-1}=\frac{1}{8\pi^2 i}\int_{\mathcal{C}} d\omega n(\omega)e^{i\omega(\bar n\cdot x-v-i\epsilon)},\\
     D_{12}(v,\Omega;x)&=\frac{1}{8\pi^2 i}\int_{\mathcal{C}} d\omega \frac{e^{i\omega(\bar n\cdot x-v+i\sigma)}}{e^{\beta \omega}-1}=\frac{1}{8\pi^2 i}\int_{\mathcal{C}} d\omega n(\omega) e^{i\omega(\bar n\cdot x-v+i\sigma)},\\   D_{21}(v,\Omega;x)&=\frac{1}{8\pi^2 i}\int_{\mathcal{C}} d\omega \frac{e^{i\omega(\bar n\cdot x-v-i\sigma)}}{e^{\beta \omega}-1}=\frac{1}{8\pi^2 i}\int_{\mathcal{C}} d\omega n(\omega) e^{i\omega(\bar n\cdot x-v-i\sigma)},\\
       D_{22}(v,\Omega;x)&=\frac{1}{8\pi^2 i}\int_{\mathcal{C}} d\omega \frac{ e^{i\omega(\bar n\cdot x-v+i\epsilon)}}{e^{\beta\omega}-1}=\frac{1}{8\pi^2 i}\int_{\mathcal{C}} d\omega n(\omega) e^{i\omega(\bar n\cdot x-v+i\epsilon)}.
\end{align}\es  \fi
One may also use the contour $\mathcal C'$ to obtain another integral representation 
\bea 
D^{(-)}_{ab}(v,\Omega;x)&=&\frac{1}{8\pi^2 i}\int_{\mathcal{C}'} d\omega \frac{e^{\beta\omega-i\omega(\bar \ell\cdot x-v-i\epsilon_{ba})}}{e^{\beta \omega}-1}=\frac{1}{8\pi^2 i}\int_{\mathcal{C}'} d\omega (1+n(\omega)) e^{-i\omega(\bar \ell\cdot x-v-i\epsilon_{ba})}\nn\\&=&\frac{1}{8\pi^2 i}\int_{\mathcal{C}'} d\omega n(\omega) e^{-i\omega(\bar \ell\cdot x-v+i(\beta-\epsilon_{ba}))}.
\eea 
Refer to the retarded bulk-to-boundary propagator, we define the generalized occupation number for the advanced bulk-to-boundary propagator 
\bs\begin{align}
    n^{(-)}_{ab}(\omega;\mathcal C)&=n(\omega)e^{\omega\epsilon_{ba}},\\
    n^{(-)}_{ab}(\omega;\mathcal C')&=(1+n(\omega))e^{-\omega\epsilon_{ba}}.
\end{align}\es 
Then we can derive the Carrollian correlator of the mixed type 
\bea 
&&\langle \prod_{j=1}^m \Sigma(u_j,\Omega_j)\prod_{j=m+1}^n \Sigma^{(-)}(v_j,\Omega_j)\rangle_\beta\nn\\&=&\sum_{a_1,a_2,\cdots,a_n}\left(\int\prod_{j=1}^n d^4y_j \right) \left(\prod_{i=1}^m D_{a_i1}(u_i,\Omega_i;y_{i})\right)\left(\prod_{i=m+1}^n D^{(-)}_{a_i1}(v_i,\Omega_i;y_i)\right)\mathcal G_{a_1a_2\cdots a_n}(y_1,y_2,\cdots,y_m),\nn\\
&=&\left(\frac{1}{8\pi^2 i}\right)^n \left(\int \prod_{j=1}^n d^4 y_j\right) \left(\int_{\mathcal C'} \prod_{i=1}^m d\omega_i n_{a_i 1}(\omega_i)e^{-i\omega_i(u_i+\ell_i\cdot y_i)}\right)\left(\int_{\mathcal C'} \prod_{i=m+1}^n d\omega_i n_{a_i1}^{(-)}(\omega_i)e^{-i\omega_i(\bar \ell_i\cdot y_i-v_i)}\right)\nn\\&&\times \mathcal G_{a_1a_2\cdots a_n}(y_1,y_2,\cdots,y_n)\nn\\&=&\left(\frac{1}{8\pi^2 i}\right)^n\left(\int_{\mathcal C'} \prod_{i=1}^m d\omega_i n_{a_i1}(\omega_i)e^{-i\omega_i u_i}\right)\left(\int_{\mathcal C'}\prod_{i=m+1}^n d\omega_i n_{a_i1}^{(-)}(\omega_i) e^{i\omega_i v_i}\right)\nn\\&&\times \left(\int \prod_{j=1}^n d^4 y_je^{-ip_j\cdot y_j}\right) \mathcal G_{a_1a_2\cdots a_n}(y_1,y_2,\cdots,y_n)\nn\\&=&\left(\frac{1}{8\pi^2 i}\right)^n\left(\int_{\mathcal C'} \prod_{i=1}^m d\omega_i n_{a_i1}(\omega_i)e^{-i\omega_i u_i}\right)\left(\int_{\mathcal C'}\prod_{i=m+1}^n d\omega_i n_{a_i1}^{(-)}(\omega_i) e^{i\omega_i v_i}\right)\nn\\&&\times (2\pi)^4 \delta^{(4)}(\sum_{j=1}^n p_j)i\mathcal M_{a_1a_2\cdots a_n}(p_1,\cdots,p_n).
\label{Sigmanpointmomentum}
\eea
We have defined 
\be 
p_j=\left\{\begin{array}{cc}\omega_j \ell_j,&j=1,2,\cdots,m,\\
\omega_j\bar \ell_j,&j=m+1,\cdots,n.\end{array}\right.
\ee 
Similar to the previous discussion, we may define 
\bea 
i\mathcal C_{a_1a_2\cdots a_n}(p_1,p_2,\cdots,p_n)=\left(\prod_{j=1}^m n_{a_j 1}(\omega_j)\right)\left(\prod_{j=m+1}^n n_{a_j 1}^{(-)}(\omega_j)\right) i\mathcal M_{a_1a_2\cdots a_n}(p_1,p_2,\cdots,p_n),\label{Carrollianmomentum}
\eea then the Carrollian amplitude at finite temperature becomes
\bea 
&&\langle \prod_{j=1}^m \Sigma(u_j,\Omega_j)\prod_{j=m+1}^n \Sigma^{(-)}(v_j,\Omega_j)\rangle_\beta\nn\\&=&\left(\frac{1}{8\pi^2 i}\right)^n\left(\int_{\mathcal C'} \prod_{j=1}^m d\omega_j e^{-i\omega_ju_j}\right)\left(\int_{\mathcal C'}\prod_{j=m+1}^n d\omega_j e^{i\omega_j v_j}\right)(2\pi)^4 \delta^{(4)}(\sum_{j=1}^n p_j)i\mathcal C_{a_1a_2\cdots a_n}(p_1,p_2,\cdots,p_n).\nn\\
\eea The  quantity \eqref{Carrollianmomentum} can be obtained from similar Feynman rules in the momentum space. One just needs to distinguish the retarded and advanced external lines. For each retarded(or advanced) external line that connects the external point $(u,\Omega)$(or $(v,\Omega)$), there is an outgoing(or incoming) momentum $p=\omega \ell$ (or $p=\omega\bar \ell$). For each retarded(or advanced) external line that connects to a bulk vertex of type $a$, we join an occupation number $n_{a1}(\omega)$ (or $n^{(-)}_{a1}(\omega)$). 
To be more precise, we should replace \eqref{externalpoint} to
	\begin{align}
	      \begin{tikzpicture}[baseline=(current bounding box.center)]
		           \fill (0,0) circle (1.5pt);
		           \draw[->, thick] (0,0) node[below] {\footnotesize $a$} -- (1,0) node[below]{\footnotesize $\omega$};
		           \draw[draw, thick] (1,0) -- (2,0);
   	         \end{tikzpicture} \
	       = n_{a1}(\omega)
        \end{align}   and 
        \begin{align}
	      \begin{tikzpicture}[baseline=(current bounding box.center)]
		           \fill (2,0) circle (1.5pt);
		           \draw[->,thick] (0,0) -- (1,0)  node[below] {$\omega$}; \draw[draw,thick] (1,0) -- (2,0) node[below]{\footnotesize $a$};
   	         \end{tikzpicture} \
	       = n^{(-)}_{a1}(\omega).
        \end{align} In these diagrams, we have used the arrow to distinguish the outgoing and incoming states. The arrow from bulk to boundary denotes the outgoing state while the one from boundary to bulk denotes the incoming state.
All the other Feynman rules remain the same form as before. 

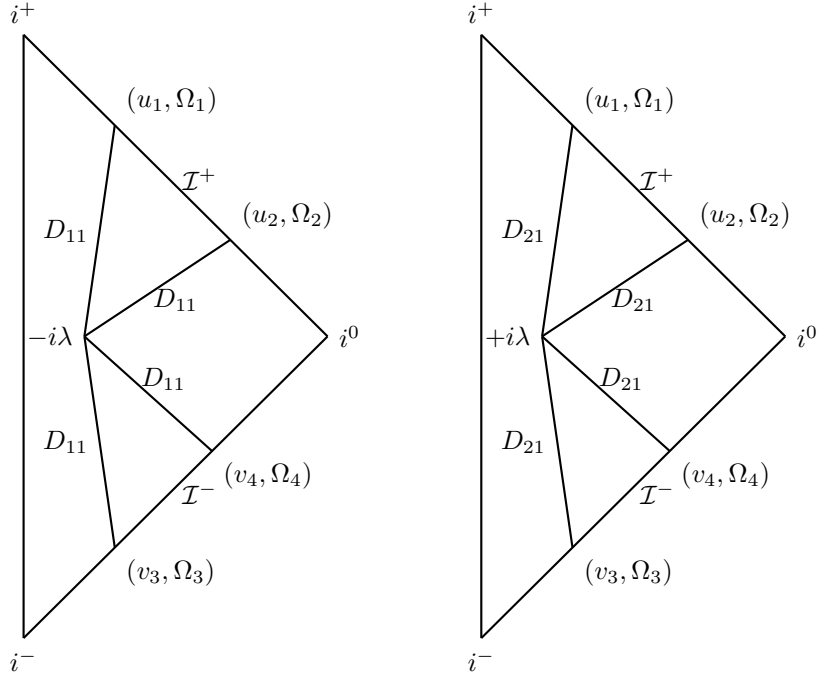
\begin{figure}
    \centering
    \usetikzlibrary{decorations.text}
    \begin{tikzpicture} [scale=0.8]
        \draw[draw,thick] (-1,5) node[above]{\footnotesize $i^+$} -- (4,0) node[right]{\footnotesize $i^0$};
        \draw[draw,thick] (4,0) -- (-1,-5) node[below]{\footnotesize $i^-$};
        \draw[draw,thick](-1,5) -- (-1,-5);
        \node at (1.9,2.6) {\footnotesize $\mathcal{I}^+$};
        \node at (1.9,-2.6) {\footnotesize $\mathcal{I}^-$};
        \draw[draw,thick] (0,0) node[left]{\footnotesize $-i\lambda$} -- (0.25,1.75) node[left]{\footnotesize $D_{11}$} -- (0.5,3.5) node[above right]{\footnotesize $(u_1,\Omega_1)$};
        \draw[draw,thick] (0,0)  -- (2.4,1.6) node[above right]{\footnotesize $(u_2,\Omega_2)$};
        \node at (1.5,0.6) {\footnotesize $D_{11}$};
        \draw[draw,thick] (0,0)  -- (0.25,-1.75) node[left]{\footnotesize $D_{11}$} -- (0.5,-3.5) node[below right]{\footnotesize $(v_3,\Omega_3)$};
        \draw[draw,thick] (0,0)  -- (2.1,-1.9) node[below right]{\footnotesize $(v_4,\Omega_4)$};
        \node at (1.3,-0.7) {\footnotesize $D_{11}$};
    \end{tikzpicture}\hspace{1cm}
    \begin{tikzpicture} [scale=0.8]
        \draw[draw,thick] (-1,5) node[above]{\footnotesize $i^+$} -- (4,0) node[right]{\footnotesize $i^0$};
        \draw[draw,thick] (4,0) -- (-1,-5) node[below]{\footnotesize $i^-$};
        \draw[draw,thick](-1,5) -- (-1,-5);
        \node at (1.9,2.6) {\footnotesize $\mathcal{I}^+$};
        \node at (1.9,-2.6) {\footnotesize $\mathcal{I}^-$};
        \draw[draw,thick] (0,0) node[left]{\footnotesize $+i\lambda$} -- (0.25,1.75) node[left]{\footnotesize $D_{21}$} -- (0.5,3.5) node[above right]{\footnotesize $(u_1,\Omega_1)$};
        \draw[draw,thick] (0,0)  -- (2.4,1.6) node[above right]{\footnotesize $(u_2,\Omega_2)$};
        \node at (1.5,0.6) {\footnotesize $D_{21}$};
        \draw[draw,thick] (0,0)  -- (0.25,-1.75) node[left]{\footnotesize $D_{21}$} -- (0.5,-3.5) node[below right]{\footnotesize $(v_3,\Omega_3)$};
        \draw[draw,thick] (0,0)  -- (2.1,-1.9) node[below right]{\footnotesize $(v_4,\Omega_4)$};
        \node at (1.3,-0.7) {\footnotesize $D_{21}$};
    \end{tikzpicture}
    \caption{\centering{Four-point Carrollian correlator of type (2,2) at tree level in $\Phi^4$ theory.}}
    \label{22fourpointCarrolliancorrelator}
\end{figure}

Now we consider the four-point Carrollian correlator as shown in Figure \ref{22fourpointCarrolliancorrelator},  this is an  alternative  four-point correlator with two fields inserted at $\mathcal{I}^+$ while the other two at $\mathcal I^-$. 

Using the Feynman rules in position space, 
\bea 
&&\langle \Sigma(u_1,\Omega_1)\Sigma(u_2,\Omega_2)\Sigma^{(-)}(v_3,\Omega_3)\Sigma^{(-)}(v_4,\Omega_4)\rangle_\beta\nn\\&=&-i\lambda \int d^4x D_{11}(u_1,\Omega_1;x)D_{11}(u_2,\Omega_2;x)D^{(-)}_{11}(v_3,\Omega_3;x)D^{(-)}_{11}(v_4,\Omega_4;x)\nn\\&&+i\lambda \int d^4x D_{21}(u_1,\Omega_1;x)D_{21}(u_2,\Omega_2;x)D^{(-)}_{21}(v_3,\Omega_3;x)D^{(-)}_{21}(v_4,\Omega_4;x)\nn\\&=&-i\lambda \left(\frac{1}{8\pi^2}\right)^4\int d^4x\left(\prod_{j=1}^4 \int_{\mathcal C} d\omega_j n(\omega_j)\right)\nn\\&&\times [e^{\sum_{j=1}^2 i\omega_j(u_j+\ell_j\cdot x-i\epsilon)+\sum_{j=3}^4 i \omega_j(\bar \ell_j\cdot x-v_j-i\epsilon)}-e^{\sum_{j=1}^2 i \omega_j(u_j+\ell_j\cdot x-i(\beta-\sigma))+\sum_{j=3}^4 i\omega_j(\bar \ell_j\cdot x-v_j-i\sigma)}]\nn\\&=&-i\lambda\left(\frac{1}{4\pi}\right)^{4}\left(\prod_{j=1}^4 \int_{\mathcal C} d\omega_j n(\omega_j)\right)\delta^{(4)}(\omega_1\ell_1+\omega_2\ell_2+\omega_3\bar \ell_3+\omega_4\bar \ell_4)\nn\\&&\times e^{i(\omega_1u_1+\omega_2u_2-\omega_3 v_3-\omega_4 v_4)}[1-e^{(\omega_1+\omega_2)(\beta-\sigma)+(\omega_3+\omega_4)\sigma}].
\eea 
The energy conservation leads to 
\be 
\omega_1+\omega_2=\omega_3+\omega_4,
\ee therefore, the result is independent of the choice of $\sigma$ as expected
\bea 
&&\langle \Sigma(u_1,\Omega_1)\Sigma(u_2,\Omega_2)\Sigma^{(-)}(v_3,\Omega_3)\Sigma^{(-)}(v_4,\Omega_4)\rangle_\beta\nn\\&=&-i\lambda\left(\frac{1}{4\pi}\right)^{4}\left(\prod_{j=1}^4 \int_{\mathcal C} d\omega_j n(\omega_j)\right)\delta^{(4)}(\omega_1\ell_1+\omega_2\ell_2+\omega_3\bar \ell_3+\omega_4\bar \ell_4) e^{i(\omega_1u_1+\omega_2u_2-\omega_3 v_3-\omega_4 v_4)}(1-e^{\beta(\omega_1+\omega_2)})\nn\\&=&i\lambda\left(\frac{1}{4\pi}\right)^{4}\left(\prod_{j=1}^4 \int_{\mathcal C} d\omega_j\right)\delta^{(4)}(q) e^{i(\omega_1u_1+\omega_2u_2-\omega_3 v_3-\omega_4 v_4)} n(\omega_3)n(\omega_4)(1+n(\omega_1)+n(\omega_2))\nn\\&=&i\lambda\left(\frac{1}{4\pi}\right)^{4}\left(\prod_{j=1}^4 \int_{\mathcal C} d\omega_j\right)\delta^{(4)}(q) e^{i(\omega_1u_1+\omega_2u_2-\omega_3 v_3-\omega_4 v_4)} n(\omega_1)n(\omega_2)(1+n(\omega_3)+n(\omega_4))\label{finitetem}
\eea where we have defined the four momentum
\be 
q^\mu=\omega_1\ell^\mu_1+\omega_2\ell^\mu_2+\omega_3\bar \ell^\mu_3+\omega_4\bar \ell^\mu_4.
\ee At zero temperature, we use the identity 
\be 
n(\omega)=-\theta(-\omega)
\ee and flip the sign of the frequencies in the integral, then the four-point correlator reduces to 
\bea 
&&\langle \Sigma(u_1,\Omega_1)\Sigma(u_2,\Omega_2)\Sigma^{(-)}(v_3,\Omega_3)\Sigma^{(-)}(v_4,\Omega_4)\rangle\nn\\&=&-i\lambda \left(\frac{1}{4\pi}\right)^{4}\left(\prod_{j=1}^4 \int_0^\infty d\omega_j\right)\delta^{(4)}(q) e^{-i(\omega_1u_1+\omega_2u_2-\omega_3 v_3-\omega_4 v_4)}
\eea which is exactly the  four-point Carrollian amplitude at zero temperature.

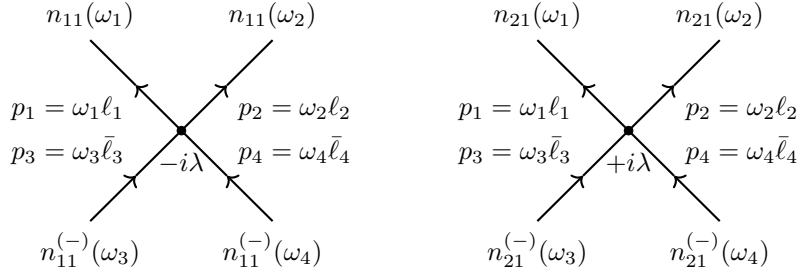
\begin{figure}
    \centering
    \usetikzlibrary{decorations.text}
    \begin{tikzpicture} [scale=0.8]
       \filldraw[black] (0,0) circle (2pt) node[below,yshift=-0.15cm]{\footnotesize $-i\lambda$};
        \draw[->,thick] (0,0) -- (-0.75,0.75) node[below left]{\footnotesize $p_1=\omega_1\ell_1$};
        \draw[draw,thick] (-0.75,0.75) --  (-1.5,1.5) node[above]{\footnotesize $n_{11}(\omega_1)$};
        \draw[->,thick] (0,0)  -- (0.75,0.75) node[below right]{\footnotesize $p_2=\omega_2\ell_2$};
        \draw[draw,thick] (0.75,0.75) --  (1.5,1.5) node[above]{\footnotesize $n_{11}(\omega_2)$};
        \draw[->,thick] (-1.5,-1.5) node[below]{\footnotesize $n^{(-)}_{11}(\omega_3)$} -- (-0.75,-0.75) ;
        \draw[draw,thick] (-0.75,-0.75) node[above left]{\footnotesize $p_3=\omega_3\bar{\ell}_3$} --  (0,0) ;
        \draw[->,thick] (1.5,-1.5) node[below]{\footnotesize $n^{(-)}_{11}(\omega_4)$} -- (0.75,-0.75) ;
        \draw[draw,thick] (0.75,-0.75) node[above right]{\footnotesize $p_4=\omega_4\bar{\ell}_4$} --  (0,0) ;
    \end{tikzpicture}\hspace{1cm}
    \begin{tikzpicture} [scale=0.8]
        \filldraw[black] (0,0) circle (2pt) node[below,yshift=-0.15cm]{\footnotesize $+i\lambda$};
        \draw[->,thick] (0,0) -- (-0.75,0.75) node[below left]{\footnotesize $p_1=\omega_1\ell_1$};
        \draw[draw,thick] (-0.75,0.75) --  (-1.5,1.5) node[above]{\footnotesize $n_{21}(\omega_1)$};
        \draw[->,thick] (0,0)  -- (0.75,0.75) node[below right]{\footnotesize $p_2=\omega_2\ell_2$};
        \draw[draw,thick] (0.75,0.75) --  (1.5,1.5) node[above]{\footnotesize $n_{21}(\omega_2)$};
        \draw[->,thick] (-1.5,-1.5) node[below]{\footnotesize $n^{(-)}_{21}(\omega_3)$} -- (-0.75,-0.75) ;
        \draw[draw,thick] (-0.75,-0.75) node[above left]{\footnotesize $p_3=\omega_3\bar{\ell}_3$} --  (0,0) ;
        \draw[->,thick] (1.5,-1.5) node[below]{\footnotesize $n^{(-)}_{21}(\omega_4)$} -- (0.75,-0.75) ;
        \draw[draw,thick] (0.75,-0.75) node[above right]{\footnotesize $p_4=\omega_4\bar{\ell}_4$} --  (0,0) ;
    \end{tikzpicture}\hspace{1cm}
    \caption{Feynman diagrams for four-point connected correlator of type (2,2) in momentum space.}
    \label{4ptmomentumspace}
\end{figure}
The process can be simplified dramatically in momentum space, and the Feynman diagrams are shown in Figure \ref{4ptmomentumspace} in which we choose the path $\mathcal C$ and then  
\bs\begin{align}
    i\mathcal C_{1,1,1,1}&=-i\lambda n_{11}(\omega_1)n_{11}(\omega_2)n_{11}^{(-)}(\omega_3)n_{11}^{(-)}(\omega_4)=-i\lambda\prod_{j=1}^4 n(\omega_j),\\
    i\mathcal C_{2,2,2,2}&=+i\lambda n_{21}(\omega_1)n_{21}(\omega_2)n_{21}^{(-)}(\omega_3)n_{21}^{(-)}(\omega_4)=+i\lambda \prod_{j=1}^2(1+ n(\omega_j))\prod_{j=3}^4 n(\omega_j).
\end{align}\es Now the conservation of the momentum leads to 
\be 
\omega_1+\omega_2=\omega_3+\omega_4, 
\ee therefore, we find
\bea 
i\mathcal C_{1,1,1,1}+i\mathcal C_{2,2,2,2}=i\lambda n(\omega_3)n(\omega_4)(1+n(\omega_1)+n(\omega_2)),
\eea which matches with \eqref{finitetem}.
\iffalse 
\bs\begin{align}
    D_{11}(v,\Omega;x)&=\frac{1}{8\pi^2 i}\int_{\mathcal{C}'} d\omega \frac{e^{\beta\omega-i\omega(\bar n\cdot x-v-i\epsilon)}}{e^{\beta \omega}-1}=\frac{1}{8\pi^2 i}\int_{\mathcal{C}'} d\omega (1+n(\omega)) e^{-i\omega(\bar n\cdot x-v-i\epsilon)},\\
     D_{12}(v,\Omega;x)&=\frac{1}{8\pi^2 i}\int_{\mathcal{C}'} d\omega \frac{e^{\beta\omega-i\omega(\bar n\cdot x-v+i\sigma)}}{e^{\beta \omega}-1}=\frac{1}{8\pi^2 i}\int_{\mathcal{C}'} d\omega (1+n(\omega)) e^{-i\omega(\bar n\cdot x-v+i\sigma)},\\   D_{21}(v,\Omega;x)&=\frac{1}{8\pi^2 i}\int_{\mathcal{C}'} d\omega \frac{e^{\beta\omega-i\omega(\bar n\cdot x-v-i\sigma)}}{e^{\beta \omega}-1}=\frac{1}{8\pi^2 i}\int_{\mathcal{C}'} d\omega (1+n(\omega)) e^{-i\omega(\bar n\cdot x-v-i\sigma)},\\
       D_{22}(v,\Omega;x)&=\frac{1}{8\pi^2 i}\int_{\mathcal{C}'} d\omega \frac{e^{\beta\omega-i\omega(\bar n\cdot x-v+i\epsilon)}}{e^{\beta \omega}-1}=\frac{1}{8\pi^2 i}\int_{\mathcal{C}'} d\omega (1+n(\omega))e^{-i\omega(\bar n\cdot x-v+i\epsilon)}.
\end{align}\es\fi 

Before we close this subsection, we just mention that the map $(v,\Omega)\to(u,\Omega^{\text{P}})$ will connect the retarded and advanced bulk-to-boundary propagators through the identity
\be
D^{(-)}_{ab}(u,\Omega^{\text{P}};x)+D_{3-a,3-b}(u,\Omega;x)=\frac{1}{4\pi\beta},\quad a,b=1,2.
\ee 
The superscript $\text{P}$ denotes the antipodal point of $\Omega=(\theta,\phi)$ with 
\be 
\Omega^{\text{P}}=(\pi-\theta,\pi+\phi).
\ee

\subsection{Boundary-to-boundary propagator}
As an analog of the limit \eqref{limitDab}, we may try to define the boundary-to-boundary propagator from $\mathcal I^-$ to $\mathcal I^+$ by 
\bea 
B_{ab}(u,\Omega;v',\Omega')=\lim_{r'\to\infty,\ v'\ \text{finite}} r'\ D_{ab}(u,\Omega;x')=\lim_{r'\to\infty,\ v'\ \text{finite}}\  \lim_{r\to\infty,\ u\ \text{finite}} r' r\ \ G_{ab}(x-x').\label{Babdefv}
\eea 
To be more precise, the boundary-to-boundary propagator can be reduced from \eqref{timeorderTp}
\bea  
B_{ab}(u,\Omega;v',\Omega')=\left\{\begin{array}{cc} \langle \Sigma(u,\Omega)\Sigma(v',\Omega')\rangle_\beta,&a=1,b=1,\\
\langle\Sigma(u-i\sigma,\Omega)\Sigma(v',\Omega')\rangle_\beta,&a=1,b=2,\\
\langle\Sigma(v'-i\sigma,\Omega')\Sigma(u,\Omega)\rangle_\beta,&a=2,b=1,\\
\langle\Sigma(v'-i\sigma,\Omega')\Sigma(u-i\sigma,\Omega)\rangle_\beta,&a=2,b=2.\end{array}\right.
\eea For the physical field inserted at the boundary, only the propagator $B_{11}(u,\Omega;v',\Omega')$ is important. We will abbreviate it as $B(u,\Omega;v',\Omega')$ in the following.
Note that the formal definition \eqref{Babdefv} may suffer divergence. To clarify this point, we consider the boundary-to-boundary propagator at zero temperature. In this case, the bulk-to-boundary propagator is 
\be 
D(u,\Omega;x')=-\frac{1}{8\pi^2(u+\ell\cdot x'-i\epsilon)}.
\ee Now we calculate  the limit 
\bea 
B(u,\Omega;v',\Omega')&=&-\frac{1}{8\pi^2}\lim_{r'\to\infty,\ v'\ \text{finite}}\frac{r'}{u-v'+r'+r'\cos\gamma(\Omega,\Omega')-i\epsilon}\nn\\&=&\frac{1}{4\pi}\log\frac{u-v'-i\epsilon}{-i\eta r'} \delta(\Omega-\Omega^{'\text{P}})-\frac{1}{8\pi^2\left(1+\cos\gamma(\Omega,\Omega')-i\eta\right)}.\label{limitB}
\eea The function $\gamma(\Omega,\Omega')$ is the angle between two normal vectors $\bm\ell$ and $\bm\ell'$.
The limit can be found as follows. At first, when $\Omega\not=\Omega^{'\text{P}}$, $1+\cos\gamma>0$ and then the limit is finite which is the second part of the above equation. When $\Omega=\Omega^{'\text{P}}$, the limit is divergent. It is reasonable to assume that it is proportional to $\delta(\Omega-\Omega^{'\text{P}})$ and propose the following identity in the large $r'$ limit
\bea 
-\frac{1}{8\pi^2}\frac{r'}{u-v'+r'+r'\cos\gamma(\Omega,\Omega')-i\epsilon}+\frac{1}{8\pi^2\left(1+\cos\gamma(\Omega,\Omega')-i\eta\right)}=\alpha\delta(\Omega-\Omega^{'\text{P}}).
\eea Integrate both sides on the unit sphere, and expand the result in the large $r'$ limit, we find 
\bea 
\frac{1}{4\pi}\log\frac{u-v'-i\epsilon}{2r'}+\frac{1}{4\pi}\log\frac{2}{-i\eta}=\alpha.\label{alpha}
\eea The second term on the left hand side is from 
\bea 
\int d\Omega'\frac{1}{8\pi^2(1+\cos\gamma(\Omega,\Omega')-i\eta)}=2\pi\int_{-1}^1 dx \frac{1}{8\pi^2(1+x-i\eta)}=\frac{1}{4\pi}\log\frac{2}{-i\eta},
\eea where the small positive parameter $\eta\to 0$ is important to regularize the integral. We find the constant $\alpha$
\be 
\alpha=\frac{1}{4\pi}\log\frac{u-v'-i\epsilon}{-i\eta r'}.
\ee Note that the $\log r'$ divergence is balanced by the $\log\eta$ divergence once we keep the product $\eta r'$ finite. More explicitly, the regularized cutoff $\eta$ may be identified as%{\xhy \cite{jorstad2024comment}} 
\be 
\eta=\frac{\epsilon}{r'} \label{ideneta}
\ee from \eqref{limitB}.  Therefore, the product $\eta r'=\epsilon$ is a natural cutoff for $\Omega=\Omega^{'\text{P}}$.   This proves the formula \eqref{limitB}. Interestingly, the first term is the electric part which depends on time while the second term is the magnetic part which is time independent \cite{chen2023higher}. We learn that the integral of the magnetic part on the sphere is divergent such that it cancels the divergence from the electric part. For completeness, we will also present the $\mathcal I^+$ to $\mathcal I^+$ propagator at zero temperature as follows:
\be 
B(u,\Omega;u',\Omega')=-\frac{1}{4\pi}\log \frac{u-u'-i\epsilon}{-i\eta r'}\delta(\Omega-\Omega')+\frac{1}{8\pi^2(1-\cos\gamma(\Omega,\Omega')+i\eta)}.\label{mtom}
\ee 

Now we turn to the boundary-to-boundary propagator at finite temperature. We calculate the limit 
\bea 
B_{ab}(u,\Omega;v',\Omega')&=&-\frac{1}{4\pi \beta}\lim_{r'\to\infty,\ v'\ \text{finite}} \frac{r'}{e^{\frac{2\pi}{\beta}(u+\ell\cdot x'-i\epsilon_{ab})}-1}\nn\\&=&-\frac{1}{4\pi\beta}\lim_{r'\to\infty,\ v'\ \text{finite}}\frac{r'}{e^{\frac{2\pi}{\beta}(u-v'+r'(1+\cos\gamma(\Omega,\Omega'))-i\epsilon_{ab})}-1}.
\eea When $\Omega\not=\Omega^{'\text{P}}$, we set $r'\to \infty$, then the limit is 0. Therefore, we find 
\bea 
B(u,\Omega;v',\Omega')\propto \delta(\Omega-\Omega^{'\text{P}}).
\eea The proportional coefficient can be fixed by integrating out both sides on the unit sphere, 
\bea
B_{ab}(u,\Omega;v',\Omega')=\frac{1}{4\pi}\log\left(1-e^{-\frac{2\pi}{\beta}(u-v'-i\epsilon_{ab})}\right)\delta(\Omega-\Omega^{'\text{P}}).\label{Bfiniteuv}
\eea Therefore, we conclude that the magnetic branch disappears at finite temperature while the electric branch is still non-vanishing.
We can now examine whether it satisfies the Ward identities (\ref{Ward2}) associated with the conformal Carroll symmetries in Appendix \ref{Carrollian}. We found that while (\ref{Bfiniteuv}) satisfies the Ward identities for translations \eqref{timetrans}-\eqref{spatial3} and rotations \eqref{rotation1}-\eqref{rotation3}, it fails to satisfy those for Lorentz boosts \eqref{Lorentz1}-\eqref{Lorentz3}. This is attributed to the fact that, in finite temperature theories, Lorentz invariance is explicitly broken by the heat bath\cite{le2000thermal,laine2016basics}. In fact, the Lorentz boosts are also broken for finite temperature CFTs \cite{marchetto2023broken}.

To convince ourselves, we  use another method to obtain the same propagator. Recall the integral representation of the retarded bulk-to-boundary propagator \eqref{bulktoboundary} and notice the formula for the expansion of the plane wave into spherical waves, we find 
\bea 
B_{ab}(u,\Omega;v',\Omega')=-\frac{1}{4\pi}\int_{\mathcal C}\frac{d\omega}{\omega}n(\omega)e^{i\omega(u-v'-i\epsilon_{ab})} \delta(\Omega-\Omega^{'\text{P}}).
\eea  There is an equivalent integral representation by changing the variable $\omega\to-\omega$
\bea 
B_{ab}(u,\Omega;v',\Omega')=-\frac{1}{4\pi}\int_{\mathcal C'}\frac{d\omega}{\omega}(1+n(\omega))e^{-i\omega(u-v'-i\epsilon_{ab})} \delta(\Omega-\Omega^{'\text{P}}).\label{Babint}
\eea 
Interested reader can find more details in Appendix \ref{prop}. For the physical boundary-to-boundary propagator, we  utilize the residue theorem, and then the boundary-to-boundary propagator from $\mathcal I^-$ to $\mathcal I^+$ becomes
\bea 
B(u,\Omega;v',\Omega')=\frac{1}{4\pi}\log(1-e^{-\frac{2\pi}{\beta}(u-v'-i\epsilon)})\delta(\Omega-\Omega^{'\text{P}})\label{Buvp}
\eea which is exactly the same as \eqref{Bfiniteuv}. We choose the path $\mathcal C$ to compute the propagator. Note that for $u>v'$, we should sum over the residues in the upper half plane 
\bea 
B(u,\Omega;v',\Omega')&=&-\frac{1}{4\pi}2\pi i \sum_{k=1}^\infty \text{Res}_{\omega=\frac{2\pi ki}{\beta}} \frac{n(\omega)}{\omega}e^{i\omega(u-v'-i\epsilon)}\delta(\Omega-\Omega^{'\text{P}})\nn\\&=&\frac{1}{4\pi}\log( 1-e^{-\frac{2\pi}{\beta}(u-v'-i\epsilon)})\delta(\Omega-\Omega^{'\text{P}}),\quad u>v'.
\eea On the other hand, for $u<v'$, we should sum over the residues in the lower half plane as well as the one at $\omega=0$
\bea 
B(u,\Omega;v',\Omega')&=&-\frac{1}{4\pi}(-2\pi i) \sum_{k=0}^\infty \text{Res}_{\omega=\frac{-2\pi ki}{\beta}} \frac{n(\omega)}{\omega}e^{i\omega(u-v'-i\epsilon)}\delta(\Omega-\Omega^{'\text{P}})\nn\\&=&-\left(\frac{i}{4}+\frac{u-v'-i\epsilon}{2\beta}+\sum_{k=1}^\infty \frac{1}{4\pi k}e^{2\pi k (u-v'-i\epsilon)/\beta}\right)\delta(\Omega-\Omega^{'\text{P}})\nn\\&=&\frac{1}{4\pi}\log( 1-e^{-\frac{2\pi}{\beta}(u-v'-i\epsilon)})
\delta(\Omega-\Omega^{'\text{P}}),\quad u<v'.\eea We confirm that the boundary-to-boundary propagator \eqref{Buvp} is valid both for $u>v'$ and $u<v'$. However, the result is asymmetric under the exchange of $(u,\Omega)$ and $(v',\Omega')$
\bea 
B(u,\Omega;v',\Omega')\not=B(v',\Omega';u,\Omega).
\eea 
The asymmetry of the boundary-to-boundary propagator becomes more  transparent in the limit of zero temperature 
\bea 
B(u,\Omega;v',\Omega')=\left\{\begin{array}{cc}-\frac{1}{4\pi}\log \frac{2\pi}{\beta}(u-v'-i\epsilon)\delta(\Omega-\Omega^{'\text{P}}),&u>v',\\
0,&u<v'.\end{array}\right.
\eea This is consistent with the boundary-to-boundary propagator at zero temperature by identifying $\beta^{-1}$ as an IR cutoff \cite{Liu:2022mne}. The limit is also the same form as the boundary-to-boundary propagator on the Rindler horizon \cite{Li:2024kbo}. %Interestingly, 

As an analog of the limit \eqref{limitDab}, we may define the boundary-to-boundary correlator at $\mathcal I^+$ by 
\bea 
B_{ab}(u,\Omega;u',\Omega')=\lim_{r'\to\infty,\ u'\ \text{finite}} r'\ D_{ab}(u,\Omega;x')=\lim_{r'\to\infty,\ u'\ \text{finite}}\  \lim_{r\to\infty,\ u\ \text{finite}} r' r\ \ G_{ab}(x-x').\label{Babdef}
\eea 
To be more precise, the boundary-to-boundary propagator can be reduced from \eqref{timeorderTp}
\bea  
B_{ab}(u,\Omega;u',\Omega')=\left\{\begin{array}{cc} \langle \Sigma(u,\Omega)\Sigma(u',\Omega')\rangle_\beta,&a=1,b=1,\\
\langle\Sigma(u-i\sigma,\Omega)\Sigma(u',\Omega')\rangle_\beta,&a=1,b=2,\\
\langle\Sigma(u'-i\sigma,\Omega')\Sigma(u,\Omega)\rangle_\beta,&a=2,b=1,\\
\langle\Sigma(u'-i\sigma,\Omega')\Sigma(u-i\sigma,\Omega)\rangle_\beta,&a=2,b=2.\end{array}\right.
\eea However, the double limit in the definition is non-commutative 
\be 
\lim_{r'\to\infty,\ u'\ \text{finite}}\  \lim_{r\to\infty,\ u\ \text{finite}} r' r\ \ G_{ab}(x-x')\not=\lim_{r\to\infty,\ u\ \text{finite}}\  \lim_{r'\to\infty,\ u'\ \text{finite}} r' r\ \ G_{ab}(x-x').
\ee In contrast, double limit of the boundary-to-boundary propagator $B(u,\Omega;v',\Omega')$ from $\mathcal{I}^-$ to $\mathcal I^+$ is commutative
\bea 
B(u,\Omega;v',\Omega')=\lim_{r'\to\infty,\ v'\ \text{finite}}\lim_{r\to\infty,\ u\ \text{finite}}G(x-y)=\lim_{r\to\infty,\ u\ \text{finite}}\lim_{r'\to\infty,\ v'\ \text{finite}}G(x-y).
\eea This is because the time tends to $+\infty$ for the boundary field $\Sigma(u,\Omega)$  and to $-\infty$ for $\Sigma^{(-)}(v,\Omega)$. One should always put $\Sigma(u,\Omega)$ before the field $\Sigma^{(-)}(v',\Omega')$. Although the boundary-to-boundary propagator $B(u,\Omega;v',\Omega')$ is finite, the alternative one $B(u,\Omega;u',\Omega')$ still suffers   a divergence  which is proportional to the large radius $r'$ in the magnetic branch
\bea 
B(u,\Omega;u',\Omega')&=&\frac{1}{4\pi}\int_{\mathcal C}\frac{d\omega}{\omega}(1+n(\omega))e^{-i\omega(u-u'-i\epsilon)}\delta(\Omega-\Omega')+\frac{r'}{4\pi\beta}\nn\\&=&-\frac{1}{4\pi}\log \left(1-e^{\frac{2\pi}{\beta}(u-u'-i\epsilon)}\right)\delta(\Omega-\Omega')+\frac{r'}{4\pi\beta}.
\eea 
Note that the magnetic branch should be divergent such that the electric branch remains finite. 
We will also derive this boundary-to-boundary correlator using contour integral representation in Appendix \ref{prop}. A puzzle is that the above propagator cannot reproduce the magnetic part of \eqref{mtom} in the zero temperature limit. The problem can be solved by noticing the limits $r'\to\infty$ and $T\to 0$ are not commutative. To zoom into the limit, we define a dimensionless parameter 
\be 
\bar\beta=\frac{\beta}{r'}
\ee and consider the limit $r'\to\infty, \ \beta\to\infty$ with $\bar\beta$ finite. The boundary-to-boundary propagators become
\bs\begin{align}
    B(u,\Omega;u',\Omega')&=-\frac{1}{4\pi}\log \frac{1-e^{\frac{2\pi}{\bar\beta r'}(u-u'-i\epsilon)}}{1-e^{-i
    \frac{2\pi}{\bar\beta}\eta}}\delta(\Omega-\Omega')-\frac{1}{4\pi\bar\beta}\frac{1}{e^{-\frac{2\pi}{\bar\beta}(1-\cos\gamma+i\eta)}-1},\\
    B(u,\Omega;v',\Omega')&=\frac{1}{4\pi}\log \frac{1-e^{\frac{2\pi}{\bar\beta r'}(u-v'-i\epsilon)}}{1-e^{i\frac{2\pi}{\bar\beta}\eta}}\delta(\Omega-\Omega^{'\text{P}})-\frac{1}{4\pi\bar\beta}\frac{1}{e^{\frac{2\pi}{\bar\beta}(1+\cos\gamma-i\eta)}-1}.
\end{align}\es We have treated $r'$ as a regulator and preserved the leading order correlator in the large $r'$ limit. Recall the identification \eqref{ideneta}, we take the limit $\bar\beta\to\infty$ and then 
\bs\begin{align}
    B(u,\Omega;u',\Omega')&=-\frac{1}{4\pi}\log \frac{u-u'-i\epsilon}{-i\epsilon}\delta(\Omega-\Omega')+\frac{1}{8\pi^2}\frac{1}{1-\cos\gamma+i\eta},\\
    B(u,\Omega;v',\Omega')&=\frac{1}{4\pi}\log \frac{u-v'-i\epsilon}{-i\epsilon}\delta(\Omega-\Omega^{'\text{P}})-\frac{1}{8\pi^2}\frac{1}{1+\cos\gamma-i\eta}.
\end{align}\es Both the electric and the magnetic branches match exactly the ones in  \eqref{mtom} and \eqref{limitB}  in the limit $\bar\beta\to\infty$.

To remove the magnetic branch, one should take the  derivative of the boundary-to-boundary propagators with respect to time 
\bs\begin{align}
    \partial_uB(u,\Omega;u',\Omega')&=-\frac{1}{2\beta}\frac{e^{\frac{2\pi}{\beta}(u-u'-i\epsilon)}}{e^{\frac{2\pi}{\beta}(u-u'-i\epsilon)}-1}\delta(\Omega-\Omega'),\\ \partial_uB(u,\Omega;v',\Omega')&=-\frac{1}{2\beta}\frac{1}{e^{\frac{2\pi}{\beta}(u-v'-i\epsilon)}-1}\delta(\Omega-\Omega^{'\text{P}}).
\end{align}\es 
The above result is obtained in the limit $r'\to\infty$ with the temperature finite. One can also find 
\bs\begin{align}
    \partial_u\partial_{u'}B(u,\Omega;u',\Omega')&=\langle \dot\Sigma(u,\Omega)\dot\Sigma(u',\Omega')\rangle_\beta=-\frac{\pi}{4\beta^2}\frac{1}{\sinh^2\frac{\pi(u-u'-i\epsilon)}{\beta}}\delta(\Omega-\Omega'),\\ \partial_u\partial_{v'}B(u,\Omega;v',\Omega')&=\langle \dot\Sigma(u,\Omega)\dot\Sigma^{(-)}(v',\Omega')\rangle_\beta=\frac{\pi}{4\beta^2}\frac{1}{\sinh^2\frac{\pi(u-v'-i\epsilon)}{\beta}}\delta(\Omega-\Omega^{'\text{P}}).
\end{align}\es 

\iffalse It should be distinguished from the alternative limit 
\bea 
r'\to\infty,\quad \bar\beta\quad\text{finite}
\eea since we can find 
\bs\begin{align}
    \partial_uB(u,\Omega;u',\Omega')&=,\\
    \partial_u B(u,\Omega;v',\Omega')&=.
\end{align}\es \fi
%Before we close this subsection, we emphasize that there are significant distinctions for the boundary-to-boundary propagators from different limits. 

\iffalse For example, 
\bea 
\partial_u\partial_{u'}B(u,\Omega;u',\Omega')=\langle\dot\Sigma(u,\Omega)\dot\Sigma(u',\Omega')\rangle_\beta=-\frac{\pi}{4\beta^2\sinh^2\frac{\pi(u-u'-i\epsilon)}{\beta}}\delta(\Omega-\Omega').
\eea\fi %We will not tackle the problem to remove the divergence despite 

\subsection{KMS symmetry}
The density matrix operator $e^{-\beta H}$ may be viewed as an evolution operator for a time shift in the imaginary direction, which implies the  formal identity
\begin{align}\label{eq4.1}
    e^{-\beta H}\Phi(x^0-i\beta,\bm x)e^{\beta H}=\Phi(x^0,\mathbf{x}).
\end{align}
Consider the following correlator\footnote{We don't include the normalization factor compared with \eqref{green}  to simplify notation.}
\begin{align}
    \mathcal G_C(t_i,\cdots)=\text{tr}(e^{-\beta H}T_C\Phi(t_i,\mathbf{x})\cdots),
\end{align}
which contains a field whose time $t_i$ is the ``smallest'' on the contour $C$ (the $\cdots$ represents the other unwritten fields). The field operator that carries it should be placed at the rightmost position by the path ordering. Thus we have
\begin{align}
    \mathcal G_C(t_i,\cdots)=\text{tr}{\left(e^{-\beta H}[T_C\cdots]\Phi(t_i,\bm x)\right)},
\end{align}
where the path ordering now applies only to the remaining unwritten fields. Using the cyclic invariance of the trace and equation (\ref{eq4.1}), we then get
\bea 
   \mathcal G_C(t_i,\cdots)&=&\text{tr}{\left(\Phi(t_i,\bm x)e^{-\beta H}[T_C\cdots]\right)}=\text{tr}{\left(e^{-\beta H}\Phi(t_i-i\beta,\mathbf{x})[T_C\cdots]\right)}\nn\\
    &=&\text{tr}{\left(e^{-\beta H}[T_C\Phi(t_i-i\beta,\mathbf{x})\cdots]\right)}=\mathcal G_C(t_i-i\beta,\cdots),
\eea 
where we have used the fact that $t_i-i\beta$ is the ``largest" time on the contour $C$ in order to insert back the operator carrying it inside the path ordering. This equality is one of the forms of the Kubo-Martin-Schwinger (KMS) symmetry \cite{Kubo:1957mj,Martin:1959jp}: all bosonic path-ordered correlators take identical values at the two endpoints of the contour\cite{le2000thermal}. Although we have singled out the first field in the correlator, this identity applies equally to all the fields. There is an analog KMS symmetry for fermionic field. For two-point Green's function, the KMS symmetry implies
\bea 
\langle T_C(\Phi_a(t,\bm x)\Phi_b(t',\bm x'))\rangle_\beta=\langle T_C(\Phi_a(t-i\beta,\bm x)\Phi_b(t',\bm x'))\rangle_\beta,
\eea which can be checked for propagators explicitly. By moving one of the points to $\mathcal{I}^+$, we can easily obtain the KMS symmetry for the bulk-to-boundary propagator
\be 
D_{ab}(u-i\beta,\Omega; x')=D_{ab}(u,\Omega;x'),
\ee which is satisfied by \eqref{retardedbb}. A further KMS symmetry for the boundary-to-boundary propagator is also checked
\be 
B_{ab}(u-i\beta,\Omega;v',\Omega')=B_{ab}(u,\Omega;v',\Omega').
\ee 
Given the formula \eqref{Sigmanpoint}, any $n+m$ point boundary correlator should also satisfy the KMS symmetry
\bea 
&&\langle\left( \prod_{i=1}^{j-1}\Sigma(u_i,\Omega_i)\right)\Sigma(u_j-i\beta,\Omega_j)\left(\prod_{i=j+1}^n \Sigma(u_j,\Omega_j)\right)\left(\prod_{k=1}^m\Sigma^{(-)}(v_k',\Omega_k')\right)\rangle_\beta\nn\\&=&\langle\left( \prod_{i=1}^{j-1}\Sigma(u_i,\Omega_i)\right)\Sigma(u_j,\Omega_j)\left(\prod_{i=j+1}^n \Sigma(u_j,\Omega_j)\right)\left(\prod_{k=1}^m\Sigma^{(-)}(v_k',\Omega_k')\right)\rangle_\beta.
\eea

\section{Correlators}
Given the Feynman rules and the propagators in the previous sections, we can compute the correlators at finite temperature.  In general, an $n$-point correlator is composed of $m$ fields at $\mathcal I^+$ and $n-m$ fields at $\mathcal I^-$, we will use the notation 
\bea 
i\mathcal C^{(m,n-m)}(u_1,\Omega_1;\cdots;u_m,\Omega_m;v_{m+1},\Omega_{m+1};\cdots;v_n,\Omega_n)=\langle \prod_{j=1}^m \Sigma(u_j,\Omega_j)\prod_{j=m+1}^n \Sigma^{(-)}(v_j,\Omega_j)\rangle_\beta
\eea and call it the Carrollian correlator of type $(m,n-m)$. Correspondingly, the momentum space correlator is denoted as 
\be 
i\mathcal C^{(m,n-m)}_{a_1a_2\cdots a_n}(p_1,p_2,\cdots,p_n),
\ee where $a_j$ denotes the type of the vertex that connects to the boundary point $(u_j,\Omega_j)$. For $n=4$, there are five kinds of correlators among which $\mathcal C^{(4,0)}$ and $\mathcal C^{(0,4)}$ vanish. Therefore, there are three kinds of non-trivial correlators. %We will focus on the three kinds of four point Carrollian correlators. 
\subsection{Type $(2,2)$}
At tree level, we already obtain one non-trivial propagator \eqref{finitetem}. The null vectors $\ell^\mu$ and $\bar \ell^\mu$ can be defined through the stereographic coordinates of $S^2$
\bs\begin{align}
    \ell_j^\mu&=(1,\frac{z_j+\bar z_j}{1+z_j\bar z_j},-i\frac{z_j-\bar z_j}{1+z_j\bar z_j},-\frac{1-z_j\bar z_j}{1+z_j\bar z_j}),\\
    \bar \ell_j^\mu&=(-1,\frac{z_j+\bar z_j}{1+z_j\bar z_j},-i\frac{z_j-\bar z_j}{1+z_j\bar z_j},-\frac{1-z_j\bar z_j}{1+z_j\bar z_j}).
\end{align}\es To simplify notation, we fix $z_1=0,\ z_2=z,\ z_3=-1,\ z_4=0$ and then\footnote{After changing to the antipodal coordinates for $z_3$ and $z_4$, this convention is actually equivalent to the one in \cite{Liu:2024nfc} with 
\be 
z_1=0,\quad z_2=z,\quad z_3=1,\quad z_4=\infty.\nn
\ee Another tricky point is that the Lorentz boost invariance of the amplitude is lost at finite temperature since the system is in  thermodynamic equilibrium with the heat bath. It is not easy to obtain the thermal correlator from the one in a special inertial frame. Therefore, in a more general treatment, one should keep  the coordinates $z_j$ arbitrary. In those cases, the computation is parallel. We will not present the results here.} 
\bs\begin{align}
    p_1^\mu&=\omega_1(1,0,0,-1),\\
    p_2^\mu&=\omega_2(1,\frac{z+\bar z}{1+z\bar z},-i\frac{z-\bar z}{1+z\bar z},-\frac{1-z\bar z}{1+z\bar z}),\\
    p_3^\mu&=\omega_3(-1,-1,0,0),\\
    p_4^\mu&=\omega_4(-1,0,0,-1).
\end{align}\es 
We can solve the constraint $q=0$ by
\bea 
\omega_1=\frac{z-1}{1+z^2}\omega_2,\quad
\omega_3=\frac{2z}{1+z^2}\omega_2,\quad
\omega_4=\frac{z(z-1)}{1+z^2}\omega_2,\quad
\bar z=z.
\eea  The last equation implies that $z$ is a real number. Therefore, the Dirac delta function becomes
\bea 
\delta^{(4)}(q)=\frac{1+z^2}{2\omega_2}\delta(\omega_1-\frac{z-1}{1+z^2}\omega_2)\delta(\omega_3-\frac{2z}{1+z^2}\omega_2)\delta(\omega_4-\frac{z(z-1)}{1+z^2}\omega_2)\delta(\bar z-z)
\eea where the previous factor on the right hand side is the Jacobian by changing the variables. 
Note the integral representation 
\bea 
&&\langle \Sigma(u_1,\Omega_1)\Sigma(u_2,\Omega_2)\Sigma^{(-)}(v_3,\Omega_3)\Sigma^{(-)}(v_4,\Omega_4)\rangle_\beta\nn\\&=&\frac{i\lambda}{256\pi^4}\left(\prod_{j=1}^4\int_{\mathcal C'} d\omega_j \right) e^{-i\omega_1 u_1-i\omega_2 u_2+i\omega_3 v_3+i\omega_4 v_4} \delta^{(4)}(q)(1+n(\omega_1))(1+n(\omega_2))[1+n(\omega_3)+n(\omega_4)].\nn\\
\eea %To integrate out the Dirac delta function along the path $\mathcal C$, one should be careful with the Dirac delta function with complex argument. 

\iffalse 
\bea 
\sum_{k=1}^\infty \frac{e^{2k\alpha}}{k(e^{2k\gamma_1}-1)(e^{2k\gamma_2}-1)(e^{2k\gamma_3}-1)}
\eea \fi
%Generating function of multiple variate Bernouli polynomials?1905.01914?
%Multivariate Bernoulli distribution?
Note that the occupation number $n(\omega)$ diverges around $\omega=0$
\be 
n(\omega)\sim \frac{1}{\beta\omega}+\cdots
\ee which is non-analytic in the complex $\omega$ plane. To avoid the subtlety\footnote{It is tricky to compute the integral along the contour $\mathcal C$ with Dirac delta function whose argument is complex. It is shown that  the argument $x_0$ may be complex in the integration \cite{Lindell1993DeltaFE,1996tah..book.....P}  
\bea 
\int_{-\infty}^\infty dx f(x)\delta(x-x_0)=f(x_0)
\eea where the function  $f(x)$ is analytic. However, for non-analytic functions, the formula may break down \cite{Brewster2016GeneralizedDF}.},  we will consider the following correlator
\bea 
&&\langle\dot\Sigma(u_1,\Omega_1)\dot\Sigma(u_2,\Omega_2)\dot\Sigma^{(-)}(v_3,\Omega_3)\dot\Sigma^{(-)}(v_4,\Omega_4)\rangle_\beta\nn\\&=&iF(\lambda,z)\int_{-\infty}^\infty d\omega \omega^3 e^{-i\omega\chi}[1+n(\alpha_1\omega)][1+n(\omega)][1+n(\alpha_3\omega)+n(\alpha_4\omega)].%\nn\\&=&iF(\lambda,z)\int_0^\infty d\omega \omega^3 [e^{-i\omega\chi}g(\omega)-e^{-i\omega\chi}h(\omega)-e^{i\omega \chi}g(-\omega)+e^{i\omega\chi}h(-\omega)].
\eea We have replaced the integration variable $\omega_2$ by $\omega$  and deformed the path $\mathcal C'$ to the real axis since there is no pole at the origin of the integrand. The function $F(\lambda,z)$ is 
\be 
F(\lambda,z)=\frac{\lambda}{256\pi^4}\frac{z^2(z-1)^2}{(1+z^2)^2}
\ee \iffalse and the function $g(\omega)$ is 
\bs\begin{align} 
g(\omega)&=[1+n(\alpha_1\omega)][1+n(\omega)][1+n(\alpha_3\omega)][1+n(\alpha_4\omega)],\\
h(\omega)&=[1+n(\alpha_1\omega)][1+n(\omega)]n(\alpha_3\omega)n(\alpha_4\omega).
\end{align}\es  \fi  and the quantity $\chi$ is defined as 
\be 
\chi=u_2+\alpha_1 u_1-\alpha_3 u_3-\alpha_4 u_4
\ee where constants $\alpha_1,\alpha_3,\alpha_4$ are
\bea 
\alpha_1=\frac{z-1}{1+z^2},\quad \alpha_3=\frac{2z}{1+z^2},\quad \alpha_4=\frac{z(z-1)}{1+z^2}.
\eea We will also set $\alpha_2=1$ for later convenience. The signs of these constants depend on the domain of $z$ and we have shown them in   Table \ref{sign}.
\begin{table}
\begin{center}
\renewcommand\arraystretch{1.5}
    \begin{tabular}{|c||c|c|c|}\hline
Domain of $z$&$z<0$&$0<z<1$&$z>1$\\\hline\hline
$\alpha_1$&$-$&$-$&$+$\\\hline
$\alpha_2$&$+$&$+$&$+$\\\hline
$\alpha_3$&$-$&$+$&$+$\\\hline
$\alpha_4$&$+$&$-$&$+$\\\hline
\end{tabular}
\caption{\centering{The signs of the constants $\alpha_j$. }}\label{sign}
\end{center}
\end{table}
They satisfy the following identities
\bs\begin{align}
    \theta(\alpha_1)&=\theta(z-1),\quad\theta(\alpha_2)=1,\quad \theta(\alpha_3)=\theta(z),\quad\theta(\alpha_4)=\theta(z-1)+\theta(-z),\\
    \theta(-\alpha_1)&=\theta(1-z),\quad\theta(-\alpha_2)=0,\quad \theta(-\alpha_3)=\theta(-z),\quad\theta(-\alpha_4)=\theta(z)\theta(1-z).
\end{align}\es 
More identities on the step function can be found in Appendix \ref{step}. \iffalse We need to study the integral \be 
I(z,\chi)=\int_0^\infty d\omega \omega^3 [e^{-i\omega\chi}g(\omega)-e^{-i\omega\chi}h(\omega)-e^{i\omega \chi}g(-\omega)+e^{i\omega\chi}h(-\omega)].
\ee 
\begin{itemize}
    \item $z>1$. In this case, we have $\alpha_1>0,\alpha_2>0,\alpha_3>0$ and then
    \bs\begin{align} 
    g(\omega)&=\frac{e^{(\alpha_1+\alpha_2+\alpha_3+\alpha_4)\beta\omega}}{(e^{\beta\alpha_1\omega}-1)(e^{\beta\alpha_2\omega}-1)(e^{\beta\alpha_3\omega}-1)(e^{\beta\alpha_4\omega}-1)},\\
    h(\omega)&=\frac{e^{(\alpha_1+\alpha_2)\beta\omega}}{(e^{\beta\alpha_1\omega}-1)(e^{\beta\alpha_2\omega}-1)(e^{\beta\alpha_3\omega}-1)(e^{\beta\alpha_4\omega}-1)}.
    \end{align}\es  
\end{itemize} One the other hand, we can also rewrite it as 
\bs\begin{align}
    g(-\omega)&=\frac{1}{(e^{\beta\alpha_1\omega}-1)(e^{\beta\alpha_2\omega}-1)(e^{\beta\alpha_3\omega}-1)(e^{\beta\alpha_4\omega}-1)},\\
    h(-\omega)&=\frac{e^{(\alpha_3+\alpha_4)\beta\omega}}{(e^{\beta\alpha_1\omega}-1)(e^{\beta\alpha_2\omega}-1)(e^{\beta\alpha_3\omega}-1)(e^{\beta\alpha_4\omega}-1)}.
\end{align}\es 
The integral becomes
\bea 
I(z,\chi)=6
\eea \fi

Notice that the occupation number satisfies the identities
\bea 
n(\omega)=-\theta(-\omega)+s(\omega)n(|\omega|),\quad 1+n(\omega)=\theta(\omega)+s(\omega)n(|\omega|),
\eea where $s(\omega)$ is the sign function 
\bea 
s(\omega)=\theta(\omega)-\theta(-\omega)=\left\{\begin{array}{cc}1,&\omega>0,\\
-1,&\omega<0.\end{array}\right.
\eea It is easy to check the identities
\bea 
s(\alpha_1)=s(z-1),\quad s(\alpha_2)=1,\quad s(\alpha_3)=s(z),\quad s(\alpha_4)=s(z)s(z-1).
\eea Then we can separate the integration of the frequency into positive and negative part to obtain 
\bea 
&&\langle\dot\Sigma(u_1,\Omega_1)\dot\Sigma(u_2,\Omega_2)\dot\Sigma^{(-)}(v_3,\Omega_3)\dot\Sigma^{(-)}(v_4,\Omega_4)\rangle_\beta\nn\\&=&iF(\lambda,z)\int_{0}^\infty d\omega \omega^3 e^{-i\omega\chi}\Big[f_0+\sum_{j=1}^4 f_j n(|\alpha_j|\omega)+\sum_{j_1<j_2}f_{j_1j_2}n(|\alpha_{j_1}|\omega)n(|\alpha_{j_2}|\omega)\nn\\&&+\sum_{j_1<j_2<j_3}f_{j_1j_2j_3}n(|\alpha_{j_1}|\omega)n(|\alpha_{j_2}|\omega)n(|\alpha_{j_3}|\omega)+f_{1234}n(|\alpha_1|\omega)n(|\alpha_2|\omega)n(|\alpha_3|\omega)n(|\alpha_4|\omega)\Big]\nn\\&&-iF(\lambda,z)\int_0^\infty d\omega \omega^3 e^{i\omega\chi} \Big[f^{(-)}_0+\sum_{j=1}^4 f^{(-)}_j n(|\alpha_j|\omega)+\sum_{j_1<j_2}f^{(-)}_{j_1j_2}n(|\alpha_{j_1}|\omega)n(|\alpha_{j_2}|\omega)\nn\\&&+\sum_{j_1<j_2<j_3}f^{(-)}_{j_1j_2j_3}n(|\alpha_{j_1}|\omega)n(|\alpha_{j_2}|\omega)n(|\alpha_{j_3}|\omega)+f^{(-)}_{1234}n(|\alpha_1|\omega)n(|\alpha_2|\omega)n(|\alpha_3|\omega)n(|\alpha_4|\omega)\Big].\nn\\
\eea We have changed the variable $\omega\to-\omega$ for the integral of negative frequency. The functions $f$ can be found in Appendix \ref{step} and $f^{(-)}$ 
is related to $f$ by flipping the sign of the argument
\be 
f_{\cdots}^{(-)}(\omega)=f_{\cdots}(-\omega).\label{feq}
\ee The subscript $\cdots$ on the left hand side should be the same as the one on the right hand side.
Since the frequency $\omega$ is always positive in the integral, the functions $f$ and $f^{(-)}$ are actually independent of $\omega$
\bs\label{f0s}\begin{align} 
f_0&=f_1=f_2=f_3=f_4=f_{12}=f_{23}=f_{24}=\theta(z-1),\\ f_{13}&=f_{123}=s(z)s(z-1)=\theta(z-1)-\theta(z)\theta(1-z)+\theta(-z),\\
f_{14}&=f_{124}=s(z)=\theta(z)-\theta(-z),\\ f_{34}&=f_{134}=f_{234}=f_{1234}=0.
\end{align}\es 
\bs\label{f0sm}\begin{align}
    f_0^{(-)}&=f_{1}^{(-)}=f_{2}^{(-)}=f_3^{(-)}=f_4^{(-)}=f_{13}^{(-)}=f_{14}^{(-)}=f_{34}^{(-)}=f_{134}^{(-)}=f_{234}^{(-)}=f_{1234}^{(-)}=0,\\
    f_{12}^{(-)}&=-\theta(z-1),\\
    f_{23}^{(-)}&=-f_{24}^{(-)}=\theta(z)\theta(1-z)-\theta(-z),\\
   f_{123}^{(-)}&=-\theta(z-1)+\theta(1-z)\theta(z)-\theta(-z),\\
   f_{124}^{(-)}&=-s(z)=\theta(-z)-\theta(z).
\end{align}\es 
Note that \eqref{f0s}, \eqref{f0sm}  are only valid for $\omega>0$ and they are not contradict with \eqref{feq}.

Now we can treat the $f's$
as constants and the integrals are of the form 
\bea 
I(c;\chi;b_1,b_2,\cdots,b_r)=\int_0^\infty d\omega \omega^c e^{-i\omega\chi} \prod_{j=1}^r n(b_j \omega),\quad b_1,b_2,\cdots,b_r>0,\quad c> r-1
\eea which can be factorized into  Barnes  zeta functions
\be 
I(c;\chi;b_1,b_2,\cdots,b_r)=\Gamma(1+c)\zeta_r(c+1;\beta \sum_{j=1}^r b_j;\beta b_1,\beta b_2,\cdots,\beta b_r).
\ee The Barnes zeta function  $\zeta_r(c;x;w_1,w_2,\cdots,w_r)$ can be defined as  a Dirichlet series of multiple variables
\bea 
\zeta_r(c;x;w_1,w_2,\cdots,w_r)=\sum_{m_1=0}^\infty \sum_{m_2=0}^\infty\cdots\sum_{m_r=0}^\infty (x+m_1w_1+m_2w_2+\cdots+m_rw_r)^{-c}
\eea with 
\be 
 \text{Re}\ x>0,\quad \text{Re}\ w_j>0,\quad \text{Re}\ c>r,\quad j=1,2,\cdots,r.
\ee 
Here we present the result as follows 
\bea 
&&\langle\dot\Sigma(u_1,\Omega_1)\dot\Sigma(u_2,\Omega_2)\dot\Sigma^{(-)}(v_3,\Omega_3)\dot\Sigma^{(-)}(v_4,\Omega_4)\rangle_\beta\nn\\&=&6iF(\lambda,z)[\frac{1}{\chi^4}f_0+\sum_{j=1}^4 f_j \zeta_1(4;\beta|\alpha_j|+i\chi;\beta|\alpha_j|)\nn\\&&+\sum_{j_1<j_2}f_{j_1j_2}\zeta_2(4;\beta|\alpha_{j_1}|+\beta|\alpha_{j_2}|+i\chi;\beta|\alpha_{j_1}|,\beta|\alpha_{j_2}|)-\sum_{j_1<j_2}f_{j_1j_2}^{(-)}\zeta_2(4;\beta|\alpha_{j_1}|+\beta|\alpha_{j_2}|-i\chi;\beta|\alpha_{j_1}|,\beta|\alpha_{j_2}|)\nn\\&&+\sum_{j_1<j_2<j_3}f_{j_1j_2j_3}\zeta_3(4;\beta|\alpha_{j_1}|+\beta|\alpha_{j_2}|+\beta|\alpha_{j_3}|+i\chi;\beta|\alpha_{j_1}|,\beta|\alpha_{j_2}|,\beta|\alpha_{j_3}|)\nn\\&&-\sum_{j_1<j_2<j_3}f_{j_1j_2j_3}^{(-)}\zeta_3(4;\beta|\alpha_{j_1}|+\beta|\alpha_{j_2}|+\beta|\alpha_{j_3}|-i\chi);\beta|\alpha_{j_1}|,\beta|\alpha_{j_2}|,\beta|\alpha_{j_3}|)].
\eea Interested reader can find more details in Appendix \ref{Barnes}. The result is exact albeit one should be familiar with the Barnes zeta functions. In the following, 
we turn to the low and high temperature expansion to extract useful information. 
\paragraph{Low temperature expansion.}
Note that the first term is the four-point correlator at zero temperature 
\be 
\langle\dot\Sigma(u_1,\Omega_1)\dot\Sigma(u_2,\Omega_2)\dot\Sigma^{(-)}(v_3,\Omega_3)\dot\Sigma^{(-)}(v_4,\Omega_4)\rangle_{\beta=\infty}=\frac{6if_0}{\chi^4}F(\lambda,z).
\ee  By subtracting the zero temperature result, we can find the deviation from the zero temperature correlator. In the low temperature limit, we can use \eqref{hightem}
and expand it around $\beta=\infty$
\bea 
I(c;\chi;b_1,b_2,\cdots,b_r)&=&\beta^{-1-c}\Gamma(c+1)\zeta_r(c+1;b_1+b_2+\cdots+b_r;b_1,b_2,\cdots,b_r)+o(\beta^{-1-c}),\nn\\
\eea where 
\bea 
\zeta_r(c+1;b_1+b_2+\cdots+b_r;b_1,b_2,\cdots,b_r)=\sum_{m_1=1}^\infty\sum_{m_2=1}^\infty\cdots\sum_{m_r=1}^\infty (m_1b_1+m_2b_2+\cdots+m_rb_r)^{-1-c}\nn\\
\eea is a higher dimensional generalization of the  Riemann zeta function. %The series is convergent 

The leading order correction in the low temperature limit is 
\bea 
&&\langle\dot\Sigma(u_1,\Omega_1)\dot\Sigma(u_2,\Omega_2)\dot\Sigma^{(-)}(v_3,\Omega_3)\dot\Sigma^{(-)}(v_4,\Omega_4)\rangle_\beta^{\text{low temperature correction}}\nn\\&=&6iF(\lambda,z)T^4[\sum_{j=1}^4 f_j \zeta_1(4;|\alpha_j|;|\alpha_j|)+\sum_{j_1<j_2}(f_{j_1j_2}-f_{j_1j_2}^{(-)})\zeta_2(4;|\alpha_{j_1}|+|\alpha_{j_2}|;|\alpha_{j_1}|,|\alpha_{j_2}|)\nn\\&&+\sum_{j_1<j_2<j_3}(f_{j_1j_2j_3}-f^{(-)}_{j_1j_2j_3})\zeta_3(4;|\alpha_{j_1}|+|\alpha_{j_2}|+|\alpha_{j_3}|;|\alpha_{j_1}|,|\alpha_{j_2}|,|\alpha_{j_3}|)].%\frac{i\lambda}{256\pi^4}\frac{z^2(z-1)^2}{(1+z^2)^2}\int_{-\infty}^\infty d\omega \omega^3e^{-i\omega\chi} [\theta(\omega)\theta(\alpha_1\omega)(s(\alpha_3\omega)e^{-\beta|\alpha_3\omega|}+s(\alpha_4\omega)e^{-\beta|\alpha_4\omega|})\nn\\&&+\theta(\omega)s(\alpha_1\omega)(\theta(\alpha_3\omega)-\theta(-\alpha_4\omega))e^{-\beta|\alpha_1\omega|}+s(\omega)\theta(\alpha_1\omega)(\theta(\alpha_3\omega)-\theta(-\alpha_4\omega))e^{-\beta|\omega|}+\cdots]\nn\\&=&iF(\lambda,z)\int_0^\infty d\omega \omega^3 e^{-i\omega\chi}[\theta(z-1)(e^{-\beta\alpha_3\omega}+e^{-\beta\alpha_4\omega})+\theta(z-1)e^{-\beta\alpha_3\omega}-\theta(1-z)\theta(z)e^{}+\theta(-z)]\nn\\&=&
\eea 
\iffalse In figure \ref{coe}, we have shown the absolute value of the constant $\alpha_j$ as a function of the cross ratio $z$. Obviously, there are five critical values of $z$
\bea 
z_c^{(1)}=-1,\quad z_c^{(2)}=0,\quad z_c^{(3)}=\frac{1}{3},\quad z_c^{(4)}=1,\quad z_c^{(5)}=3
\eea that divide the domain of $z$ into six intervals. In each interval, we can find the order of the constants in table \ref{order}. 

\begin{figure}
    \centering
    \includegraphics[width=5in]{coefficients.pdf}
    \caption{Absolute value of the constants $\alpha_j$.}
    \label{coe}
\end{figure}
\begin{table}
\begin{center}
\renewcommand\arraystretch{1.5}
    \begin{tabular}{|c||c|c|c|}\hline
Intervals &$z<-1$&$-1<z<0$&$0<z<\frac{1}{3}$\\\hline
Order of the constants&$|\alpha_1|<|\alpha_3|<|\alpha_2|<|\alpha_4|$&$|\alpha_4|<|\alpha_3|<|\alpha_2|<|\alpha_1|$&$|\alpha_4|<|\alpha_3|<|\alpha_1|<|\alpha_2|$\\\hline\hline 
Intervals&$\frac{1}{3}<z<1$&$1<z<3$&$z>3$\\\hline 
Order of the constants&$|\alpha_4|<|\alpha_1|<|\alpha_3|<|\alpha_2|$&$|\alpha_1|<|\alpha_4|<|\alpha_3|<|\alpha_2|$&$|\alpha_1|<|\alpha_3|<|\alpha_4|<|\alpha_2|$\\\hline\hline
%$\alpha_4$&$+$&$-$&$+$\\\hline
\end{tabular}
\caption{\centering{The order of the constants $\alpha_j$. }}\label{order}
\end{center}
\end{table}\fi
This is independent of the function $\chi$.

\paragraph{High temperature expansion.}
In the high temperature limit,  we can use the expansion \eqref{expansionn} to obtain %expand the occupation number using Bernoulli numbers 
%\bea 
%n(\omega)=\sum_{n=0}^\infty \frac{B_n(0)}{n!}(\beta\omega)^{n-1},\quad 1+n(\omega)=\sum_{n=0}^\infty \frac{B_n(1)}{n!}(\beta \omega)^{n-1}.
%\eea Therefore 
\iffalse 
We find the high temperature expansion of the occupation number
\bs\begin{align}
n(\omega)&=\sum_{m=0}^\infty B_m \frac{(\beta\omega)^{m-1}}{m!}=(\beta\omega)^{-1}-\frac{1}{2}+\frac{1}{24}(\beta\omega)+\cdots,\\
1+n(\omega)&=(\beta\omega)^{-1}+\frac{1}{2}+\frac{1}{24}(\beta\omega)+\cdots.
\end{align}\es  \fi 
\bea 
&&\langle\dot\Sigma(u_1,\Omega_1)\dot\Sigma(u_2,\Omega_2)\dot\Sigma^{(-)}(v_3,\Omega_3)\dot\Sigma^{(-)}(v_4,\Omega_4)\rangle_\beta^{\text{high temperature expansion}}\nn\\&=&iF(\lambda,z) \int_{-\infty}^\infty d\omega \omega^3 e^{-i\omega \chi }[(1+n(\alpha_1\omega))(1+n(\alpha_2\omega))+(1+n(\alpha_1\omega))(1+n(\alpha_2\omega))n(\alpha_3\omega)\nn\\&&+(1+n(\alpha_1\omega))(1+n(\alpha_2\omega)) n(\alpha_4\omega)]\nn\\&=&iF(\lambda,z)\int_{-\infty}^\infty d\omega \omega^3 e^{-i\omega \chi}[\frac{1}{\beta^2\omega^2 \alpha_1\alpha_2}\sum_{n=0}^\infty \frac{P_{2,0,n}(\alpha_1,\alpha_2)}{n!}(\beta\omega)^n\nn\\&&+\frac{1}{\beta^3\omega^3\alpha_1\alpha_2\alpha_3}\sum_{n=0}^\infty \frac{P_{2,1,n}(\alpha_1,\alpha_2,\alpha_3)}{n!}(\beta\omega)^n+\frac{1}{\beta^3\omega^3\alpha_1\alpha_2\alpha_4}\sum_{n=0}^\infty \frac{P_{2,1,n}(\alpha_1,\alpha_2,\alpha_4)}{n!}(\beta\omega)^n]
\nn\\&=&iF(\lambda,z)\int_{-\infty}^\infty d\omega e^{-i\omega\chi}[\frac{\alpha_3+\alpha_4}{\alpha_1\alpha_3\alpha_4\beta^3}+\frac{(1+\alpha_1)(\alpha_3+\alpha_4)}{2\alpha_1\alpha_3\alpha_4\beta^2}\omega+\cdots]\nn\\&=&2\pi iF(\lambda,z)[\frac{\alpha_3+\alpha_4}{\alpha_1\alpha_3\alpha_4}T^3 \delta(\chi)+i\frac{(1+\alpha_1)(\alpha_3+\alpha_4)}{2\alpha_1\alpha_3\alpha_4}T^2\delta'(\chi)+\cdots].
\eea At high temperature, the  correlator is proportional to $T^3$ and it is non-vanishing only for 
\be 
\chi=0.
\ee 

\subsection{Type $(3,1)$}
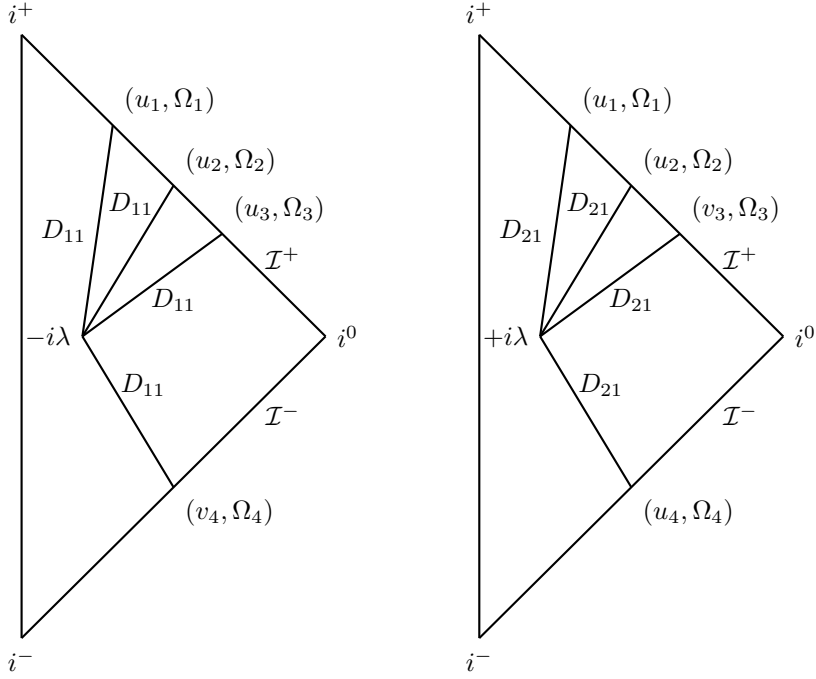
\begin{figure}
    \centering
    \usetikzlibrary{decorations.text}
    \begin{tikzpicture} [scale=0.8]
        \draw[draw,thick] (-1,5) node[above]{\footnotesize $i^+$} -- (4,0) node[right]{\footnotesize $i^0$};
        \draw[draw,thick] (4,0) -- (-1,-5) node[below]{\footnotesize $i^-$};
        \draw[draw,thick] (-1,5) -- (-1,-5);
        \node at (3.3,1.3) {\footnotesize $\mathcal{I}^+$};
        \node at (3.3,-1.3) {\footnotesize $\mathcal{I}^-$};
        \draw[draw,thick] (0,0) node[left]{\footnotesize $-i\lambda$} -- (0.25,1.75) node[left]{\footnotesize $D_{11}$} -- (0.5,3.5) node[above right]{\footnotesize $(u_1,\Omega_1)$};
        \draw[draw,thick] (0,0)  -- (1.5,2.5) node[above right]{\footnotesize $(u_2,\Omega_2)$};
        \node at (0.8,2.2) {\footnotesize $D_{11}$};
        \draw[draw,thick] (0,0) -- (2.3,1.7) node[above right]{\footnotesize $(u_3,\Omega_3)$};
        \node at (1.5,0.6) {\footnotesize $D_{11}$};
        \draw[draw,thick] (0,0)  -- (1.5,-2.5) node[below right]{\footnotesize $(v_4,\Omega_4)$};
        \node at (1,-0.8) {\footnotesize $D_{11}$};
    \end{tikzpicture}\hspace{1cm}
    \begin{tikzpicture} [scale=0.8]
        \draw[draw,thick] (-1,5) node[above]{\footnotesize $i^+$} -- (4,0) node[right]{\footnotesize $i^0$};
        \draw[draw,thick] (4,0) -- (-1,-5) node[below]{\footnotesize $i^-$};
        \draw[draw,thick] (-1,5) -- (-1,-5);
        \node at (3.3,1.3) {\footnotesize $\mathcal{I}^+$};
        \node at (3.3,-1.3) {\footnotesize $\mathcal{I}^-$};
        \draw[draw,thick] (0,0) node[left]{\footnotesize $+i\lambda$} -- (0.25,1.75) node[left]{\footnotesize $D_{21}$} -- (0.5,3.5) node[above right]{\footnotesize $(u_1,\Omega_1)$};
        \draw[draw,thick] (0,0)  -- (1.5,2.5) node[above right]{\footnotesize $(u_2,\Omega_2)$};
        \node at (0.8,2.2) {\footnotesize $D_{21}$};
        \draw[draw,thick] (0,0) -- (2.3,1.7) node[above right]{\footnotesize $(v_3,\Omega_3)$};
        \node at (1.5,0.6) {\footnotesize $D_{21}$};
        \draw[draw,thick] (0,0)  -- (1.5,-2.5) node[below right]{\footnotesize $(u_4,\Omega_4)$};
        \node at (1,-0.8) {\footnotesize $D_{21}$};
    \end{tikzpicture}
    \caption{\centering{Four-point Carrollian correlator  of type (3,1) at tree level in $\Phi^4$ theory.}}
    \label{31fourpointCarrolliancorrelator}
\end{figure}

It would be interesting to consider an alternative  four-point Carrollian correlator in which three fields are inserted at $\mathcal I^+$ while the fourth one at $\mathcal I^-$ 
\bea 
\langle \Sigma(u_1,\Omega_1)\Sigma(u_2,\Omega_2)\Sigma(u_3,\Omega_3)\Sigma^{(-)}(v_4,\Omega_4)\rangle_\beta.
\eea The position space Feynman diagrams are shown in Figure \ref{31fourpointCarrolliancorrelator}.
This correlator vanishes at zero temperature due to the kinematic constraint. However, this does not guarantee that it is still zero at finite temperature. In momentum space, we use path $\mathcal C$ and obtain 
\bs\begin{align}
    i\mathcal{C}_{1,1,1,1}&=-i\lambda n_{11}(\omega_1)n_{11}(\omega_2)n_{11}(\omega_3)n_{11}^{(-)}(\omega_4)=-i\lambda\prod_{j=1}^4 n(\omega_j),\\
    i\mathcal C_{2,2,2,2}&=+i\lambda n_{21}(\omega_1)n_{21}(\omega_2)n_{21}(\omega_3)n_{21}^{(-)}(\omega_4)=+i\lambda \prod_{j=1}^3 (1+n(\omega_j)) n(\omega_4).
\end{align}
\es The energy conservation is 
\be 
\omega_1+\omega_2+\omega_3=\omega_4
\ee and then 
\bea 
i\mathcal C_{1,1,1,1}+i\mathcal C_{2,2,2,2}=-i\lambda n(\omega_1)n(\omega_2)n(\omega_3)n(\omega_4)\left (1-e^{\beta\omega_4}\right)=+i\lambda n(\omega_1)n(\omega_2)n(\omega_3).\label{momenC}
\eea 
Note that for the path $\mathcal C'$, we find 
\bs\begin{align}
    i\mathcal C_{1,1,1,1}&=-i\lambda(1+n(\omega_1))(1+n(\omega_2))(1+n(\omega_3))(1+n(\omega_4)),\\
    i\mathcal C_{2,2,2,2}&=+i\lambda n(\omega_1)n(\omega_2)n(\omega_3)(1+n(\omega_4))
\end{align}\es and then 
\bea 
i\mathcal C_{1,1,1,1}+i\mathcal C_{2,2,2,2}=-i\lambda n(\omega_1)n(\omega_2)n(\omega_3)(1+n(\omega_4))[-1+e^{\beta(\omega_1+\omega_2+\omega_3)}]=-i\lambda n(\omega_1)n(\omega_2)n(\omega_3)e^{\beta \omega_4}.\nn\\
\eea Note that it is not the same as \eqref{momenC}. However, the discrepancy is only superficial since the bulk-to-boundary propagator in the momentum space depends on the contour. We can prove that they lead to the same correlator in position space. More precisely,   
using the Fourier transform, we get 
\bea 
&&\langle \Sigma(u_1,\Omega_1)\Sigma(u_2,\Omega_2)\Sigma(u_3,\Omega_3)\Sigma^{(-)}(v_4,\Omega_4)\rangle_\beta\nn\\&=&i\lambda\left(\frac{1}{4\pi}\right)^4\left(\int_{\mathcal C} \prod_{j=1}^4 d\omega_j\right) \delta^{(4)}(q')e^{i\omega_1u_1+i\omega_2 u_2+i\omega_3u_3-i\omega_4 v_4} n(\omega_1)n(\omega_2)n(\omega_3)\nn\\&=&-i\lambda\left(\frac{1}{4\pi}\right)^4\left(\int_{\mathcal C'} \prod_{j=1}^4 d\omega_j\right) \delta^{(4)}(q')e^{-i\omega_1u_1-i\omega_2 u_2-i\omega_3u_3+i\omega_4 v_4} n(\omega_1)n(\omega_2)n(\omega_3)e^{\beta\omega_4}\label{cor31}
\eea where the momentum $q'$
\bea 
q'=\omega_1 n_1+\omega_2 n_2+\omega_3n_3+\omega_4\bar n_4.
\eea 
Note that the second and the third line of \eqref{cor31} are related to each other by changing the variable $\omega_j$ to $-\omega_j$.
Similar to the previous discussion, we fix $z_1=0,\ z_2=z,\ z_3=1,\ z_4=0$ and then the  constraint $q'=0$ is solved by 
\bea 
\omega_1=\frac{z-1}{1+z^2}\omega_2,\quad
\omega_3=-\frac{2z}{1+z^2}\omega_2,\quad
\omega_4=\frac{z(z-1)}{1+z^2}\omega_2,\quad
\bar z=z.
\eea The integral becomes
\bea 
&&\langle \Sigma(u_1,\Omega_1)\Sigma(u_2,\Omega_2)\Sigma(u_3,\Omega_3)\Sigma^{(-)}(v_4,\Omega_4)\rangle_\beta\nn\\&=&\frac{i\lambda}{256\pi^4}\frac{1+z^2}{2}\int_{\mathcal C}d\omega_2 e^{i\omega_2\chi'}\omega_2^{-1} n(\omega_1)n(\omega_2)n(\omega_3)
\eea where 
\be 
\chi'=u_2+\frac{z-1}{1+z^2}u_1-\frac{2z}{1+z^2}u_3-\frac{z(z-1)}{1+z^2}v_4.
\ee 
Similar to the type $(2,2)$ correlator, we take time derivative and then 
\bea 
&&\langle \dot\Sigma(u_1,\Omega_1)\dot\Sigma(u_2,\Omega_2)\dot\Sigma(u_3,\Omega_3)\dot\Sigma^{(-)}(v_4,\Omega_4)\rangle_\beta\nn\\&=&iF(\lambda,z)\int_{-\infty}^\infty d\omega \omega^3 e^{i\omega\chi'}n(\alpha_1\omega)n(\omega)n(-\alpha_3\omega)\nn\\&=&iF(\lambda,z)\int_{0}^\infty d\omega \omega^3 e^{i\omega\chi'}\Big[f^{(3,1)}_0+\sum_{j=1}^3 f^{(3,1)}_j n(|\alpha_j|\omega)+\sum_{j_1<j_2}f^{(3,1)}_{j_1j_2}n(|\alpha_{j_1}|\omega)n(|\alpha_{j_2}|\omega)\nn\\&&+f^{(3,1)}_{123}n(|\alpha_{j_1}|\omega)n(|\alpha_{j_2}|\omega)n(|\alpha_{j_3}|\omega)\Big]\nn\\&&-iF(\lambda,z)\int_0^\infty d\omega \omega^3 e^{-i\omega\chi'} \Big[f^{(3,1,-)}_0+\sum_{j=1}^3 f^{(3,1,-)}_j n(|\alpha_j|\omega)+\sum_{j_1<j_2}f^{(3,1,-)}_{j_1j_2}n(|\alpha_{j_1}|\omega)n(|\alpha_{j_2}|\omega)\nn\\&&+f^{(3,1,-)}_{123}n(|\alpha_{j_1}|\omega)n(|\alpha_{j_2}|\omega)n(|\alpha_{j_3}|\omega)\Big].
\eea The constants $f_{\cdots}^{(3,1)}$ and $f_{\cdots}^{(3,1,-)}$ are fixed to be products of step functions
\bs\begin{align}
    f_0^{(3,1)}&=f_{1}^{(3,1)}=f_{3}^{(3,1)}=f_{12}^{(3,1)}=f_{13}^{(3,1)}=f_{0}^{(3,1,-)}=f_{2}^{(3,1,-)}=0,\\
    f_2^{(3,1)}&=\theta(z)\theta(1-z),\\
    f_{12}^{(3,1)}&=-\theta(z-1)+\theta(z)\theta(1-z),\\
    f_{23}^{(3,1)}&=\theta(z)\theta(1-z)-\theta(-z),\\
    f_{123}^{(3,1)}&=-f_{13}^{(3,1,-)}=-f_{123}^{(3,1,-)}=-\theta(z-1)+\theta(z)\theta(1-z)-\theta(-z),\\
    f_{1}^{(3,1,-)}&=f_{12}^{(3,1,-)}=\theta(-z),\\
    f_{3}^{(3,1,-)}&=f_{23}^{(3,1,-)}=\theta(z-1).
\end{align}\es Using the Barnes zeta function, we find
\bea 
&&\langle \dot\Sigma(u_1,\Omega_1)\dot\Sigma(u_2,\Omega_2)\dot\Sigma(u_3,\Omega_3)\dot\Sigma^{(-)}(v_4,\Omega_4)\rangle_\beta\nn\\&=&6iF(\lambda,z)\Big[\sum_{j=1}^3 f^{(3,1)}_j \zeta_1(4;\beta|\alpha_j|-i\chi';\beta|\alpha_j|)+\sum_{j_1<j_2}f_{j_1j_2}^{(3,1)}\zeta_2(4;\beta|\alpha_{j_1}|+\beta|\alpha_{j_2}|-i\chi';\beta|\alpha_{j_1}|+\beta|\alpha_{j_2}|)\nn\\&&-\sum_{j_1<j_2}f_{j_1j_2}^{(3,1,-)}\zeta_2(4;\beta|\alpha_{j_1}|+\beta|\alpha_{j_2}|+i\chi';\beta|\alpha_{j_1}|+\beta|\alpha_{j_2}|)\nn\\&&+f_{123}^{(3,1)}\zeta_3(4; \beta|\alpha_{1}|+\beta|\alpha_{2}|+\beta|\alpha_{3}|-i\chi';\beta|\alpha_{1}|+\beta|\alpha_{2}|+\beta|\alpha_{3}|)\nn\\&& -f_{123}^{(3,1,-)}\zeta_3(4; \beta|\alpha_{1}|+\beta|\alpha_{2}|+\beta|\alpha_{3}|+i\chi';\beta|\alpha_{1}|+\beta|\alpha_{2}|+\beta|\alpha_{3}|)\Big].
\eea One can obtain the high and low temperature expansion as type (2,2) correlator. We will not repeat it here. When $T=0$, the type $(3,1)$ correlator becomes exactly zero and is consistent with the previous calculation. As the temperature $T\not=0$, the type $(3,1)$ correlator is non-vanishing. 

%\paragraph{High temperature expansion.}
%\paragraph{Low temperature expansion.}
\subsection{Type $(1,3)$}
\begin{figure}
    \centering
    \usetikzlibrary{decorations.text}
    \begin{tikzpicture} [scale=0.8]
        \draw[draw,thick] (-1,5) node[above]{\footnotesize $i^+$} -- (4,0) node[right]{\footnotesize $i^0$};
        \draw[draw,thick] (4,0) -- (-1,-5) node[below]{\footnotesize $i^-$};
        \draw[draw,thick] (-1,5) -- (-1,-5);
        \node at (3.3,1.3) {\footnotesize $\mathcal{I}^+$};
        \node at (3.3,-1.3) {\footnotesize $\mathcal{I}^-$};
        \draw[draw,thick] (0,0) node[left]{\footnotesize $-i\lambda$} -- (0.25,-1.75) node[left]{\footnotesize $D_{11}$} -- (0.5,-3.5) node[below right]{\footnotesize $(v_2,\Omega_2)$};
        \draw[draw,thick] (0,0)  -- (1.5,-2.5) node[below right]{\footnotesize $(v_3,\Omega_2)$};
        \node at (0.9,-2.2) {\footnotesize $D_{11}$};
        \draw[draw,thick] (0,0) -- (2.3,-1.7) node[below right]{\footnotesize $(v_4,\Omega_3)$};
        \node at (1.5,-0.7) {\footnotesize $D_{11}$};
        \draw[draw,thick] (0,0)  -- (1.5,2.5) node[above right]{\footnotesize $(u_1,\Omega_1)$};
        \node at (1,0.8) {\footnotesize $D_{11}$};
    \end{tikzpicture}\hspace{1cm}
    \begin{tikzpicture} [scale=0.8]
        \draw[draw,thick] (-1,5) node[above]{\footnotesize $i^+$} -- (4,0) node[right]{\footnotesize $i^0$};
        \draw[draw,thick] (4,0) -- (-1,-5) node[below]{\footnotesize $i^-$};
        \draw[draw,thick] (-1,5) -- (-1,-5);
        \node at (3.3,1.3) {\footnotesize $\mathcal{I}^+$};
        \node at (3.3,-1.3) {\footnotesize $\mathcal{I}^-$};
        \draw[draw,thick] (0,0) node[left]{\footnotesize $+i\lambda$} -- (0.25,-1.75) node[left]{\footnotesize $D_{21}$} -- (0.5,-3.5) node[below right]{\footnotesize $(v_2,\Omega_2)$};
        \draw[draw,thick] (0,0)  -- (1.5,-2.5) node[below right]{\footnotesize $(v_3,\Omega_2)$};
        \node at (0.9,-2.2) {\footnotesize $D_{21}$};
        \draw[draw,thick] (0,0) -- (2.3,-1.7) node[below right]{\footnotesize $(v_4,\Omega_3)$};
        \node at (1.5,-0.7) {\footnotesize $D_{21}$};
        \draw[draw,thick] (0,0)  -- (1.5,2.5) node[above right]{\footnotesize $(u_1,\Omega_1)$};
        \node at (1,0.8) {\footnotesize $D_{21}$};
    \end{tikzpicture}
    \caption{\centering{Four-point Carrollian correlator at tree level of type (1,3) in $\Phi^4$ theory.}}
    \label{13fourpointCarrolliancorrelator}
\end{figure}
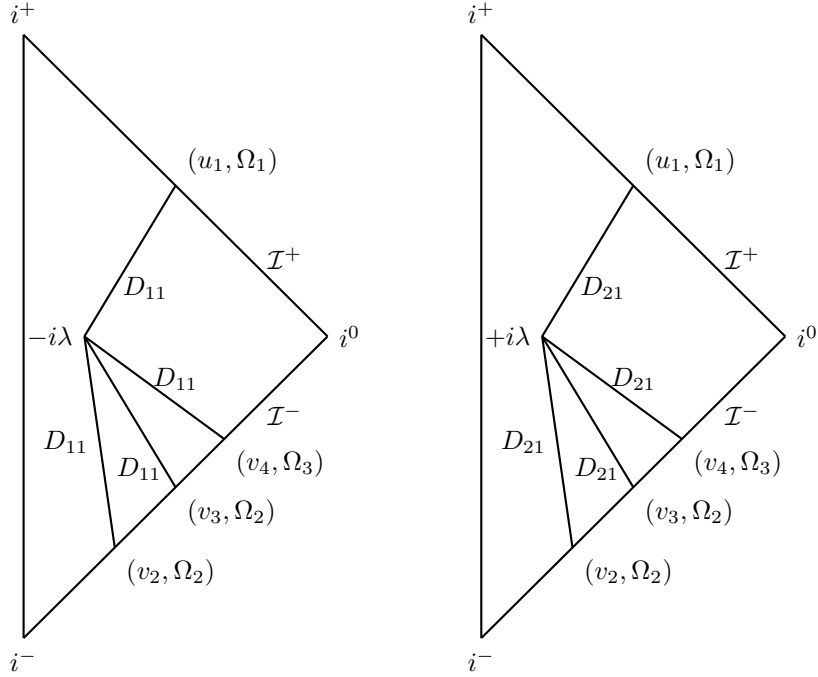

For type $(1,3)$ Carrollian correlator, the position space Feynman diagram is shown in Figure \ref{13fourpointCarrolliancorrelator}. This is the dual diagram of the type $(3,1)$ correlator. We will not repeat the computation here.
\section{Discussion}\label{dis}
In this work, we have proposed the thermal propagators and Feynman rules which are the building blocks for the
Carrollian correlators at the null infinity of finite temperature field theory. We used the real-time formalism in the derivation and thus  the degrees of freedom of the bulk fields have doubled. Interestingly, the finite temperature bulk-to-boundary propagator obeys the extended Bose-Einstein distribution for bosonic field in the position space. There are three kinds of  boundary-to-boundary propagators. The $\mathcal{I}^-$ to $\mathcal{I}^+$ propagator is always finite despite the thermal effects while the $\mathcal I^-$ to $\mathcal I^-$ or $\mathcal I^+$ to $\mathcal I^+$ propagator suffers an IR  divergence. We have derived this divergence in two different ways and they match with each other. The IR divergence is crucial to regularize the electric branch.  The bulk-to-bulk and bulk-to-boundary propagators as well as the $\mathcal{I}^-$ to $\mathcal I^+$ propagator reduce to the ones at zero temperature smoothly.  
We have also checked that one should consider an alternative limit 
\be 
r\to\infty,\quad \bar\beta=\frac{\beta}{r}\quad \text{finite}
\ee to connect the boundary-to-boundary propagators smoothly at zero temperature and finite temperature. Then we apply this formalism to compute the four-point Carrollian correlator at finite temperature. At tree level, the correlators are already fruitful compared with the zero temperature ones. At zero temperature, the  four-point correlator is  non-vanishing for $2\to 2$ scattering processes only. However, at finite temperature, there are non-trivial correlators that correspond to $1\to 3$, $2\to 2$ and $3\to 1$ scattering processes. All the correlators can be written as the summation of Barnes zeta functions. There are several points that deserve further study.  % are separated into two branches. The electric branch is finite while the magnetic branch suffers an IR  divergence. However, it is crucial to include the magnetic branch to regularize the electric branch.  %The finite temperature Feynman rules are modified compared with the zero temperature ones. At finite temperature, the momentum space Feynman rules should 

\begin{itemize}
    { 
    \item \textbf{Disappearance of the magnetic branch.}  Recently there has been growing interest in magnetic Carrollian theories. For instance,  in \cite{cotler2025soft}, it was shown that three-dimensional pure quantum gravity with zero cosmological constant can be reformulated as a magnetic Carrollian theory living on null infinity. In \cite{campoleoni2022magnetic}, magnetic Carrollian gravity  was analysed  in Hamiltonian formalism. Moreover, recent discussion on the connection between magnetic branch and soft theorem can be found in \cite{jorstad2024comment}. However, the discussion on the magnetic branch at finite temperature is scarce\footnote{One can find some comments on thermal Carrollian theories in \cite{de2023carroll}. }.  We have previously noted that in the holographic boundary-to-boundary propagator at finite temperature (\ref{Bfiniteuv})
    \bea
    B_{ab}(u,\Omega;v',\Omega')=\frac{1}{4\pi}\log\left(1-e^{-\frac{2\pi}{\beta}(u-v'-i\epsilon_{ab})}\right)\delta(\Omega-\Omega^{'\text{P}}),
    \eea
    the magnetic branch vanishes while the electric branch remains non-vanishing. %In contrast, in \cite{Bagchi:2016bcd,Bagchi:2019xfx,Duval:2014uoa,basu2018dynamical}, the magnetic branch emerges from a certain Carroll contraction.
    \iffalse 
    However, in our finite-temperature setup, it remains unclear whether the magnetic branch of boundary-to-boundary correlator encodes any information about the bulk theory. While in \cite{Bagchi:2016bcd,Bagchi:2019xfx,duval2014carroll,basu2018dynamical,bergshoeff2024carroll}, the magnetic theories and branches were constructed from the field theory on the boundary, we emphasize that our holographic analysis yields a vanishing magnetic branches in the boundary-to-boundary correlator at finite temperature.\fi  However, this does not imply the magnetic branch always disappears at finite temperature. As discussed  below (\ref{Bfiniteuv}), the finite-temperature boundary-to-boundary propagator violates the Ward identities for Lorentz boosts. Consequently, the symmetry at finite temperature becomes less restrictive, preventing the two-point function on the null boundary from being fully fixed. As a consequence, one cannot rule out the magnetic branch in more general thermal Carroll CFTs. We regard this question as highly significant for the development of magnetic  Carrollian field theories at finite temperature, and it merits further investigation. 
    
    }

 \item \textbf{Thermal Carrollian CFTs.} In thermal Carrollian CFTs, Lorentz boost symmetry is broken, while the symmetries for spacetime translation and rotation still exist. In addition, thermal Carrollian CFTs must also satisfy the KMS symmetry due to periodicity of the Euclidean time direction. In general, we expect that any finite-temperature Carrollian CFTs satisfy the symmetry for translations and rotations, together with the KMS condition.   

There are similar phenomena in thermal CFTs.
In a two-dimensional thermal CFT where the geometry is topologically a cylinder, thermal two-point functions are completely fixed by conformal symmetry and the KMS condition, and can be obtained via conformal transformation from the plane to the cylinder\cite{francesco2012conformal}. However, in dimensions $d>2$, these symmetries are not sufficient to fully determine correlation functions, even for low-point correlators such as two-point correlators. In \cite{iliesiu2018conformal}, thermal CFT data were constrained by using method from conformal bootstrap, where thermal one- and two-point functions of local operators on the plane were studied. The thermal one-point function is fixed by symmetry and dimensional analysis up to a coefficient which cannot be determined by KMS symmetry, while the OPE of thermal two-point function satisfies a nontrivial thermal crossing equation following from the KMS condition and they have used the thermal inversion formula to determine the  one-point coefficients.
Motivated by these interesting developments in thermal CFTs, it would be interesting to formulate a thermal Carrollian field theory intrinsically on a Carrollian manifold, and to uncover additional structures that are beyond the holographic description.

    \item \textbf{Divergences.} We illustrate the problem using type $(2,2)$ correlator as an example. We have shown that the four-point correlator
    \bea 
    \langle \dot\Sigma(u_1,\Omega_1)\dot \Sigma(u_2,\Omega_2)\dot \Sigma^{(-)}(v_3,\Omega_3)\dot \Sigma^{(-)}(v_4,\Omega_4)\rangle_\beta
    \eea  is finite. However, it does not imply that the original correlator $\mathcal C^{(2,2)}$ is  also finite. One may try to analytically continue the Barnes zeta function  to obtain  $\mathcal C^{(2,2)}$. However, the Barnes zeta function $\zeta_r(c+1;x;\cdots)$ suffers a pole structure for $c=0,1,2,\cdots,r-1$ which obscures the discussion. In thermal quantum field theory, it is always expected that the divergences of the correlators are only from the one at zero temperature. Therefore, it would be nice to check this point in thermal Carrollian field theory.
%Note $z=0,1,\infty$, two of the fields are collinear and the above solution is not valid. We will assume $z\not=0,1,\infty$ from now on.
\item \textbf{Pole structure and the imaginary time formalism.} One can also compute the correlators using residue theorem. 
For the four-point correlators, the conservation of four-momentum always reduces them to the summation of the following form 
\be 
\int d\omega\omega^m e^{-i\omega\chi}\prod_{j\in J} n(\alpha_j\omega).
\ee where $m$ is an integer and $J$ is a subset of $\{1,2,3,4\}$. 
\iffalse 
Therefore, the four point{\xhy four-point} correlator becomes 
\bea 
&&\langle \Sigma(u_1,\Omega_1)\Sigma(u_2,\Omega_2)\Sigma^{(-)}(v_3,\Omega_3)\Sigma^{(-)}(v_4,\Omega_4)\rangle_\beta\nn\\&=&\frac{i\lambda}{256\pi^4}\left(\prod_{j=1}^4\int_{\mathcal C} d\omega_j \right) e^{-i\omega_1 u_1-i\omega_2 u_2+i\omega_3 v_3+i\omega_4 v_4} \delta^{(4)}(q)n(\omega_1)n(\omega_2)[1+n(\omega_3)+n(\omega_4)]
\nn\\
&=&\frac{i\lambda}{256\pi^4}\frac{1+z^2}{2}\int_{\mathcal C}d\omega_2 e^{i\omega_2\chi}\omega_2^{-1} n(\omega_1)n(\omega_2)[1+n(\omega_3)+n(\omega_4)]\nn\\&=&\frac{i\lambda}{256\pi^4}\frac{1+z^2}{2}\int_{\mathcal C'}d\omega_2 e^{-i\omega_2\chi}\omega_2^{-1}[1+n(\omega_1)][(1+n(\omega_2)][1+n(\omega_3)+n(\omega_4)].
\eea The factor $\frac{2\omega_2}{1+z^2}$ is the Jacobian from changing variables and the function $\chi$ is 
\be 
\chi=u_2+\frac{z-1}{1+z^2}u_1-\frac{2z}{1+z^2}v_3-\frac{z(z-1)}{1+z^2}v_4.
\ee 
At the last step, we have changed the path $\mathcal C$ to $\mathcal C'$.
There is no step function from the integration of Dirac function at finite temperature since the integrated frequency can be negative. \fi 
Note that there are four families of poles in the complex plane of $
\omega$  which correspond to the poles of $n(\alpha_j\omega)$ with $j=1,2,3,4$ respectively
\bs\begin{align}
\omega_*^{(1)}(k)&=\frac{1+z^2}{z-1}\frac{2\pi i k}{\beta},\\
\omega_*^{(2)}(k)&=\frac{2\pi ik}{\beta},\\
\omega_*^{(3)}(k)&=\frac{1+z^2}{2z}\frac{2\pi ik}{\beta},\\
\omega_*^{(4)}(k)&=\frac{1+z^2}{z(z-1)}\frac{2\pi ik}{\beta}.
\end{align}\es We have assumed $k$  an integer.  The pole at the origin 
\be 
\omega_*=0
\ee is the common pole of the occupation numbers, which corresponds to the contribution from the modes of zero energy.  When $z$ is a rational number, there could be poles that coincide with each other. It would be interesting to understand whether the rational $z$ is special. The poles are exactly the ones that appear in the imaginary time formalism \cite{Matsubara:1955ws}. It would be rather interesting to explore the imaginary time  formalism in more details for thermal Carrollian field theory.

{In AdS/CFT, poles of the retarded Green's function in the boundary CFT correspond to the frequencies of quasi-normal modes (QNMs) of a black hole in asymptotically AdS spacetimes\cite{son2002minkowski,birmingham2002conformal,horowitz2000quasinormal}. Subject to some particular boundary conditions in asymptotically AdS black hole spacetime, quasi-normal modes are associated with the perturbations of matter or gravitational field\cite{chan1997scalar,wang2000quasinormal,govindarajan2001quasi,cardoso2001scalar,wang2002scalar,chen2009quasi,siopsis2005quasi,amado2024scalar}. In thermal CFTs, the location of the poles of the retarded Green's functions describes the linear response \cite{balasubramanian1999bulk} and  is also  associated with the process of thermalization \cite{horowitz2000quasinormal}. 
\iffalse
\begin{align}
    \tau^{(\rho\sim1)}=\ln{[\sinh^2{(\De\phi/2)}]},\quad \text{with }\De\phi=2\pi N\pm\gamma,
\end{align}
where $\rho$ is the coordinate in Penrose diagram, $N$ is an integer and $\gamma\in[0,\pi]$ the minimal angular distance. 
\fi

However, the QNMs studied in \cite{son2002minkowski,birmingham2002conformal} are obtained under the condition of purely ingoing flux at the horizon and purely outgoing flux at asymptotic infinity. In contrast, our analysis allows for both ingoing and outgoing waves at null infinity. Consequently, we have not identified a direct correspondence between the pole structure of our propagators and  quasi-normal modes, as different boundary conditions generally lead to different modes. It would be interesting to explore thermal correlators in the dual theory that have the pole structure of the QNMs of black hole. Such a connection would represent a significant development in the context of flat holography.

Regarding the pole structure of the propagators in the position space, we notice that the discontinuous surface (\ref{bulkboundarygeodesic})
\be
u+\ell\cdot x+i\beta N=0,
\ee
correspond to  poles of the bulk-to-boundary propagators (\ref{retardedbb}). Similarly, the boundary-to-boundary propagator (\ref{Bfiniteuv}) also exhibits poles
\begin{align}
    u-v'+i\beta N=0,
\end{align}
where $N$ is an integer. In fact, there exists an infinite number of complex poles. In the special case $N=0$, the pole equation describes the trajectory of a light ray travelling from past null infinity to future null infinity in the spacetime\cite{Li:2024kbo}. For $N\neq0$, however, the complex poles are related to the inverse temperature $\beta$. The appearance of infinite complex poles can be attributed to the compactness of the time direction at finite temperature. For an  asymptotically flat spacetime such as Schwarzschild spacetime, the quasi-normal modes and the singular structure of the Green's function have been studied in detail in \cite{wardell2009green}. Actually, according to the theorem of the ``Propagation of Singularities"\cite{duistermaat1994fourier,hormander2009analysis}, one expects the Green's function to be singular when its two argument points
are connected by a null geodesic. 
 
%It was found in \cite{wardell2009green} that the singularities of Green's function for a scalar field in Schwarzschild spacetime are in concordance with coordinate time on the null geodesic.

}

\iffalse We will assume $z$ is an irrational number such that these poles do not coincide. 
Using residue theorem, the integral can be rendered to the infinite series of the following form 
\bea 
F_1(x;q)=\sum_{k=1}^\infty \frac{1}{k}\frac{q^{x  k}}{q^k-1}
\eea and 
\bea 
F_2(x,y;q)=\sum_{k=1}^\infty\frac{1}{k}\frac{q^{x k}}{(q^k-1)(q^{yk}-1)}.
\eea \fi

\item \textbf{Loop corrections.} For the type $(4,0)$ correlator, it has been shown that the tree level correction vanishes due to the conservation of four-momentum and the identity 
\be 
\prod_{j=1}^4 n(\omega_j)=\prod_{j=1}^4(1+ n(\omega_j)),\quad\text{for}\quad  \sum_{j=1}^4 \omega_j=0.
\ee At zero temperature, the correlator receives no loop correction since the conservation of the energy cannot be satisfied for all outgoing modes because  their frequencies are always positive. However, at finite temperature, the conservation of the four-momentum does not imply that the type $(4,0)$ correlator is automatically vanishing (recall that there are both outgoing and incoming modes now). It would be interesting to explore this correlator in the future. As an illustration, 
the s-channel Feynman diagram of the one-loop correction of the type $(4,0)$ Carrollian correlator has been given in Figure \ref{feynoneloop40}. 
\begin{figure}
    \centering
    \usetikzlibrary{decorations.text}
    \begin{tikzpicture} [scale=0.8]
  % \label{sigmatimepath}
        \draw[draw,thick] (0,5) node[above]{\footnotesize $i^+$} -- (5,0) node[right]{\footnotesize $i^0$};
        \draw[draw,thick] (5,0) -- (0,-5) node[below]{\footnotesize $i^-$};
        \draw[draw,thick](0,5) -- (0,-5);
        \node at (2.8,2.8) {\footnotesize $\mathcal{I}^+$};
        \node at (2.8,-2.8) {\footnotesize $\mathcal{I}^-$};
        \draw[draw,thick] (0.5,4.5) node[right,yshift=0.1cm]{\footnotesize $(u_1,\Omega_1)$} -- (0.9,1.8) node[left]{\footnotesize $a$};
        \draw[draw,thick](0.9,1.8) -- (1.6,3.4) node[right,yshift=0.1cm]{\footnotesize $(u_2,\Omega_2)$};
        \fill (0.9,1.8) circle (2pt);
        \coordinate (X) at (1.5,1);
        \draw[thick, rotate around={130:(X)}] (X) ellipse (1 and 0.8);
        \fill (2.2,0.3) circle (2pt);
        \draw[draw, thick] (2.2,0.3) node[below]{\footnotesize $b$} -- (3.6,1.4) node[right,yshift=0.2cm,xshift=-0.2cm]{\footnotesize $(u_3,\Omega_3)$};
        \draw[draw,thick] (2.2,0.3) -- (4.5,0.5) node[right,yshift=0.15cm,xshift=-0.1cm]{\footnotesize $(u_4,\Omega_4)$};
    \end{tikzpicture}
    \caption{One-loop Feynman diagram at s-channel for four-point Carrollian correlator of type (4,0) in $\Phi^4$ theory.}
    \label{feynoneloop40}
\end{figure}
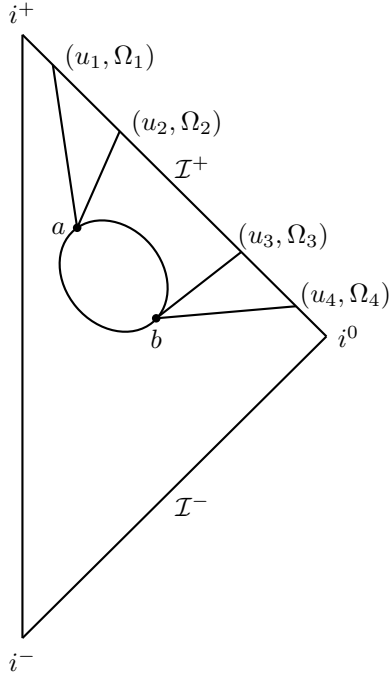
In momentum space, the s-channel correlator at one-loop is
\bs\label{schannelC}\begin{align}
    i\mathcal C^{(4,0)}_{1,1,1,1}&=n_{11}(\omega_1)n_{11}(\omega_2)n_{11}(\omega_3)n_{11}(\omega_4) I_{1,1}^{\text{one-loop}}(p_1+p_2),\\
    i\mathcal C^{(4,0)}_{2,2,2,2}&=n_{21}(\omega_1)n_{21}(\omega_2)n_{21}(\omega_3)n_{21}(\omega_4)I_{2,2}^{\text{one-loop}}(p_1+p_2),\\
    i\mathcal C^{(4,0)}_{1,1,2,2}&=-n_{11}(\omega_1)n_{11}(\omega_2)n_{21}(\omega_3)n_{21}(\omega_4)I_{1,2}^{\text{one-loop}}(p_1+p_2),\\
    i\mathcal C^{(4,0)}_{2,2,1,1}&=-n_{21}(\omega_1)n_{21}(\omega_2)n_{11}(\omega_3)n_{11}(\omega_4)I_{2,1}^{\text{one-loop}}(p_1+ p_2),
\end{align}\es 
where we have defined the one-loop integral at finite temperature
\be 
I_{ab}^{\text{one-loop}}(k)=\frac{(i\lambda)^2}{2}\prod_{j=1}^4\int \frac{d^4p}{(2\pi)^4}G_{ab}(p)G_{ab}(p+k).\ee  The summation of the four correlators \eqref{schannelC} at the s-channel is not likely to be vanishing, otherwise the momentum space propagators should satisfy rather non-trivial identity. This implies that the type $(4,0)$ may receive loop corrections. The non-vanishing correction is essential  for the extended Virasoro algebra  \cite{Liu:2022mne}
\be 
[\mathcal T_{f_1},\mathcal T_{f_2}]=-\frac{ic}{48\pi}\mathcal I_{f_1\dddot{f}_2-f_2\dddot{f}_1}+i\mathcal T_{f_1\dot{f}_2-f_2\dot{f}_1}.\label{vir}
\ee To illustrate this, we draw a null hypersurface in Figure \ref{geodesic}. 
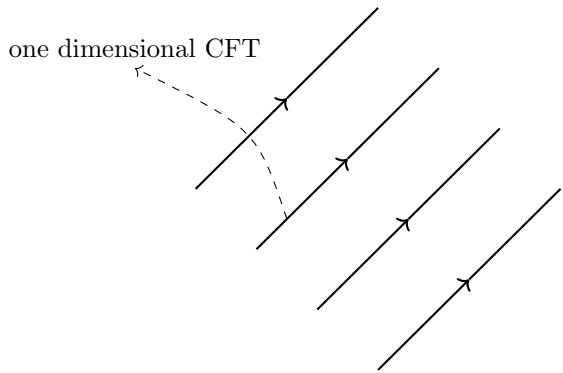
\begin{figure}
    \centering
    \usetikzlibrary{decorations.text}
    \begin{tikzpicture} [scale=0.8]
        \draw[draw,thick](1.5,1.5) -- (3,3);
        \draw[->,thick](0,0) -- (1.5,1.5);
        \draw[draw,thick](2.5,0.5) -- (4,2);
        \draw[->,thick] (1,-1) -- (2.5,0.5);
        \draw[draw,thick](5,1) -- (3.5,-0.5);
        \draw[->,thick] (2,-2) -- (3.5,-0.5);
        \draw[draw,thick](4.5,-1.5) -- (6,0);
        \draw[->,thick](3,-3) -- (4.5,-1.5);
        \draw[->,dashed] (1.5,-0.5) .. controls (1,1) ..(-1,2) node[above]{\footnotesize \text{one dimensional CFT}};
    \end{tikzpicture}
    \caption{A Carrollian manifold may be regarded as a null hypersurface that generated by null geodesics. There is an effective one-dimensional conformal field theory on each generator of a null hypersurface.}
    \label{geodesic}
\end{figure}
The algebra indicates that, roughly speaking, for each  generator of the null hypersurface, there is an effective one-dimensional conformal field theory. 
At this moment, there are several supports on this algebra. At first, as has been discussed, 
 the type $(n>2,0)$ correlator receives no loop correction due to conservation of energy at zero temperature. Therefore, the one and  two-point correlators of the flux operator 
 \be 
 \langle\mathcal T_f\rangle,\quad \langle\mathcal T_{f_1}\mathcal T_{f_2}\rangle
 \ee are not corrected since they are constructed by the composite operator $:\dot\Sigma^2:$ which can be treated as the limit 
 \be 
 :\dot\Sigma^2(u,\Omega):=\lim_{u'\to u,\quad \Omega'\to \Omega}\dot\Sigma(u,\Omega)\dot\Sigma(u',\Omega')-\langle \dot\Sigma(u,\Omega)\dot\Sigma(u',\Omega')\rangle.
 \ee This implies that the algebra \eqref{vir} is still valid after considering interactions.
 
 At finite temperature, we have not checked the algebra \eqref{vir} since the type $(2,0)$ and $(4,0)$ correlators may receive loop-corrections\footnote{At tree level, there is no correction for the type $(2,0)$ correlator and the type $(4,0)$ correlator has been shown to be vanishing. Therefore, the algebra \eqref{vir} is valid at the tree level even at finite temperature.}.
Interestingly, the two-point correlator 
\be 
\langle \dot\Sigma(u,\Omega)\dot\Sigma(u',\Omega')\rangle_\beta=\partial_u\partial_{u'}B(u,\Omega;u',\Omega')=- \frac{\pi}{4\beta^2}\frac{1}{\sinh^2\frac{\pi(u-u'-i\epsilon)}{\beta}}\delta(\Omega-\Omega')
\ee 
is exactly the same one for a primary operator with conformal weight $h=1$ at finite temperature in one dimensional conformal field theory \cite{Cardy:1984epx}. This fact is consistent with the commutator 
\be [\mathcal T_f, \dot\Sigma(u,\Omega)]=f(u,\Omega)\ddot{\Sigma}(u,\Omega)+h\dot f(u,\Omega)\dot\Sigma
\ee with $h=1$. To further check the algebra \eqref{vir}, one should at least consider the one-loop correction of the  four-point correlator in Figure
\ref{feynoneloop40} and the two-loop correction of two-point correlator in Figure \ref{twoloopsl} because both of them are of order $\mathcal{O}(\lambda^2)$. In general, an $n$-loop correction of the two-point correlator is the same order as an $(n-1)$-loop correction of the four-point correlator. It would be nice to explore the loop corrections of the Carrollian correlator at finite temperature.%Suppose the algebra \eqref{vir} to be correct,  there should be a relation between the $n$-loop amplitude and $(n-1)$-loop amplitude. This reminds us the Cutkosky rule \cite{Cutkosky:1960sp} whose finite temperature version can be found in  \cite{Weldon:1983jn}. Therefore, it would be nice to explore the loop corrections of the Carrollian correlator at finite temperature.%check the relation between the cutting rule and the validity of the algebra.
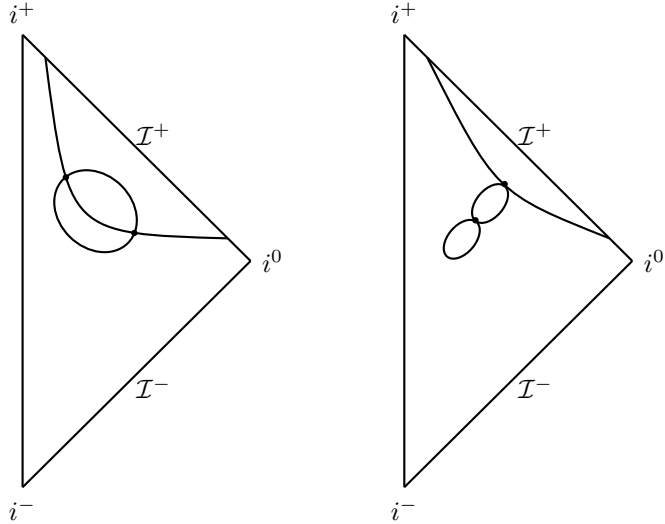
\begin{figure}
    \centering
    \usetikzlibrary{decorations.text}
    \begin{tikzpicture} [scale=0.6]
        \draw[draw,thick] (0,5) node[above]{\footnotesize $i^+$} -- (5,0) node[right]{\footnotesize $i^0$};
        \draw[draw,thick] (5,0) -- (0,-5) node[below]{\footnotesize $i^-$};
        \draw[draw,thick](0,5) -- (0,-5);
        \node at (2.88,2.8) {\footnotesize $\mathcal{I}^+$};
        \node at (2.88,-2.8) {\footnotesize $\mathcal{I}^-$};
        \fill (0.95,1.85) circle (2pt);
        \coordinate (X) at (1.6,1.1);
        \draw[thick, rotate around={135:(X)}] (X) ellipse (1 and 0.8);
        \fill (2.45,0.62) circle (2pt);
        
        \draw[draw,thick] (0.5,4.5) node[right,yshift=0.1cm]{\footnotesize $$}  .. controls (1,0.6) .. (4.5,0.5)node[right, yshift=0.15cm,xshift=-0.1cm]{\footnotesize $$};
    \end{tikzpicture}\hspace{1cm}
    \begin{tikzpicture}[scale=0.6]
        \draw[draw,thick] (0,5) node[above]{\footnotesize $i^+$} -- (5,0) node[right]{\footnotesize $i^0$};
        \draw[draw,thick] (5,0) -- (0,-5) node[below]{\footnotesize $i^-$};
        \draw[draw,thick](0,5) -- (0,-5);
        \node at (2.88,2.8) {\footnotesize $\mathcal{I}^+$};
        \node at (2.88,-2.8) {\footnotesize $\mathcal{I}^-$};
        \draw[draw,thick] (0.5,4.5)   .. controls (2,1.5) .. (4.5,0.5);
        \fill (2.2,1.7)  circle (2pt);
        \coordinate (X) at (1.88,1.27);
        \draw[thick,rotate around={-40}:(X)] (X) ellipse (0.3 and 0.5);
        \fill (1.56,0.9)  circle (2pt);
        \coordinate (Y) at (1.26,0.48);
        \draw[thick,rotate around={-40}:(Y)] (Y) ellipse (0.3 and 0.5);
    \end{tikzpicture}
    \caption{\centering{2-loop correction.}}
    \label{twoloopsl}
\end{figure}

\item \textbf{Unruh effect.} The scattering amplitude in Rindler spacetime has been explored in \cite{Li:2024kbo} in the framework of Carrollian analysis. We have been working in the Rindler vacuum such that the amplitude is much easier. To obtain the Unruh effect, one should work in Minkowski vacuum and then the field theory in the Rindler wedge is a thermal field theory. Note that the bulk-to-bulk, bulk-to-boundary and boundary-to-boundary propagators in Minkowski vacuum have been derived in \cite{Li:2024kbo},  one can use the method developed in this work to compute thermal Carrollian correlators on the  Rindler horizon.
 \item \textbf{Black holes.} Now we will comment on the application of Carrollian correlators  in black hole spacetime. In Figure \ref{fourgravitons}, we have drawn the Penrose diagram of a maximally extended Schwarzschild spacetime. This spacetime is globally hyperbolic and there are four kinds of null boundaries. As a consequence, there should be four bulk-to-boundary propagators as shown in the figure. The bulk-to-bulk propagator has been studied decades ago since the work of  \cite{Boulware:1974dm,Unruh:1976db, Hartle:1976tp}. However, the bulk-to-boundary propagators have not been explored sufficiently in the literature.  
 There are also sixteen boundary-to-boundary propagators in total and we have just shown four of them in the Penrose diagram \ref{boundarytoboundarypenrosediagram}. Using the technology developed in our work, the Carrollian correlator in an eternal black hole may be solved at tree level. However, it is expected that there are still UV divergences at loop level, similar to the one in the classic books \cite{Birrell:1982ix,DeWitt:1975ys}.
 \begin{figure}
    \centering
    \usetikzlibrary{decorations.text}
    \begin{tikzpicture} [scale=0.8]
        \draw[dashed,thick] (-3,-3) node[below]{\footnotesize $i^-$}  -- (3,3) node[above]{\footnotesize $i^+$};
        \draw[draw,thick] (3,3) -- (6,0) node[right]{\footnotesize $i^0$};
        \node at (4.8,1.8) {\footnotesize $\mathcal{I}^+$};
        \draw[dashed,thick] (-3,3) node[above] {\footnotesize $i^+$} -- (3,-3) node[below]{\footnotesize $i^-$};
        \draw[draw, thick](3,-3) -- (6,0);
        \draw[draw, thick](-3,-3) -- (-6,0) node[left]{\footnotesize $i^0$};
        \draw[draw,thick](-6,0) -- (-3,3);
        \node at(-4.8,1.8) {\footnotesize $\mathcal{I}^+$};
        \node at (-4.8,-1.8){\footnotesize $\mathcal{I}^-$};
        \node at (4.8,-1.8) {\footnotesize $\mathcal{I}^-$};
        \node at (3,0){\text{I}};
        \node at (-3,0){\text{II}};
       \draw[decorate, decoration={snake, amplitude=0.4mm, segment length=1mm}] (-3,3) -- (3,3);
       \draw[draw,thick](-3,3) -- (3,3);
       \draw[decorate, decoration={snake, amplitude=0.4mm, segment length=1mm}] (-3,-3) -- (3,-3);
       \draw[draw,thick](-3,-3) -- (3,-3);
       \draw[decorate, decoration={snake, amplitude=0.8mm,segment length=2mm}] (4,2)  .. controls (3.2,1) and (3.2,-1) .. (4,-2);
       \draw[decorate, decoration={snake, amplitude=0.8mm,segment length=2mm}] (4.8,1.2)  .. controls (4.6,0.4) and (5.4,0.2) .. (5.6,0.4);
       \draw[decorate, decoration={snake, amplitude=0.8mm,segment length=2mm}] (3.4,2.6)  .. controls (1.5,1.5) and (-1.5,1.5) .. (-3.4,2.6);
       \draw[decorate, decoration={snake, amplitude=0.8mm,segment length=2mm}] (3.5,-2.5)  .. controls (1,1) and (-1.5,1.5) .. (-4.8,1.2);
    \end{tikzpicture}
    \caption{Boundary-to-boundary propagators for graviton in a maximally extended Schwarzschild spacetime.}
    \label{boundarytoboundarypenrosediagram}
\end{figure}
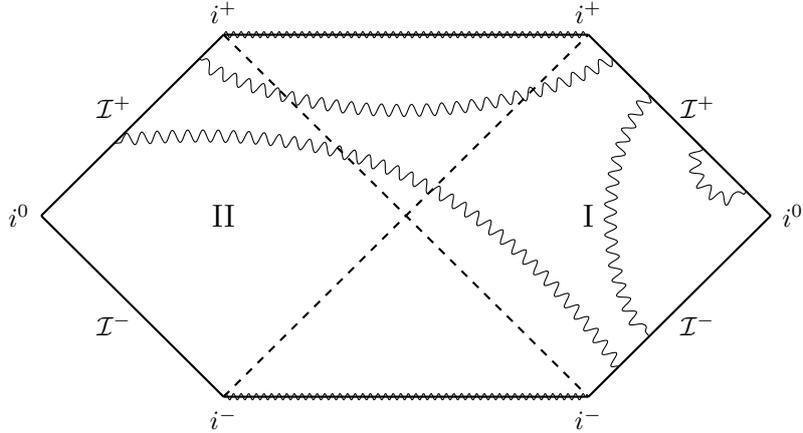
 
 Another interesting situation is that the subregion I which only contains an asymptotically flat spacetime is also globally hyperbolic, one should construct propagators from bulk to null infinity (and event horizon). We have shown several examples in Figure \ref{Schwarzschild}. In this case, the near horizon region is  approximately a Rindler spacetime \cite{Rindler:1966zz}  and the far region is asymptotically flat. One should choose suitable boundary conditions to construct these bulk-to-boundary propagators which correspond to in-equivalent vacua that are used in different situations \cite{Candelas:1980zt}, namely,  the  Boulware' vacuum,  Unruh vacuum or Hartle-Hawking vacuum. The choice of the vacuum would affect the Carrollian correlators.
 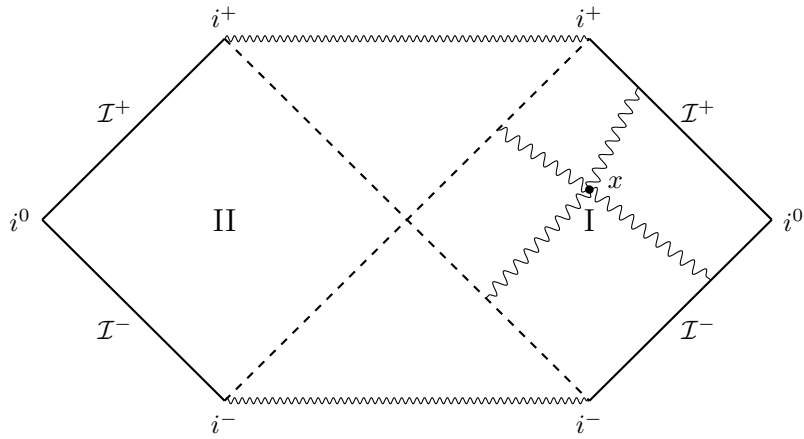
\begin{figure}
    \centering
    \usetikzlibrary{decorations.text}
    \begin{tikzpicture} [scale=0.8]
        \draw[dashed,thick] (-3,-3) node[below]{\footnotesize $i^-$}  -- (3,3) node[above]{\footnotesize $i^+$};
        \draw[draw,thick] (3,3) -- (6,0) node[right]{\footnotesize $i^0$};
        \node at (4.8,1.8) {\footnotesize $\mathcal{I}^+$};
        \draw[dashed,thick] (-3,3) node[above]{\footnotesize $i^+$} -- (3,-3) node[below]{\footnotesize $i^-$};
        \draw[draw, thick](3,-3) -- (6,0);
        \draw[draw, thick](-3,-3) -- (-6,0) node[left]{\footnotesize $i^0$};
        \draw[draw,thick](-6,0) -- (-3,3);
        \node at(-4.8,1.8) {\footnotesize $\mathcal{I}^+$};
        \node at (-4.8,-1.8){\footnotesize $\mathcal{I}^-$};
       \node at (4.8,-1.8) {\footnotesize $\mathcal{I}^-$};
        \node at (3,0){\text{I}};
        \node at (-3,0){\text{II}};
        \fill(3,0.5) circle(2pt);
        \draw[decorate, decoration={snake, amplitude=0.8mm,segment length=2mm}](3,0.5) node[right,yshift=0.1cm,xshift=0.1cm]{\footnotesize $x$}-- (3.8,2.2) node[above, xshift=0.35cm] {\footnotesize $$};
        \draw[decorate, decoration={snake, amplitude=0.8mm,segment length=2mm}](3,0.5) -- (1.5,1.5) node[left]{\footnotesize $$};

        \draw[decorate, decoration={snake, amplitude=0.8mm,segment length=2mm}](3,0.5) -- (1.3,-1.3) node[left]{\footnotesize $$};
        \draw[decorate, decoration={snake, amplitude=0.8mm,segment length=2mm}] (3,0.5) -- (5,-1) node[right]{\footnotesize $$};
       \draw[decorate, decoration={snake, amplitude=0.4mm, segment length=1mm}] (-3,3) -- (3,3);
       
       \draw[decorate, decoration={snake, amplitude=0.4mm, segment length=1mm}] (-3,-3) -- (3,-3);
       
    \end{tikzpicture}
    \caption{{Bulk-to-boundary propagators and four graviton scattering in region I of Schwarzschild solution.}}
    \label{Schwarzschild}
\end{figure}
 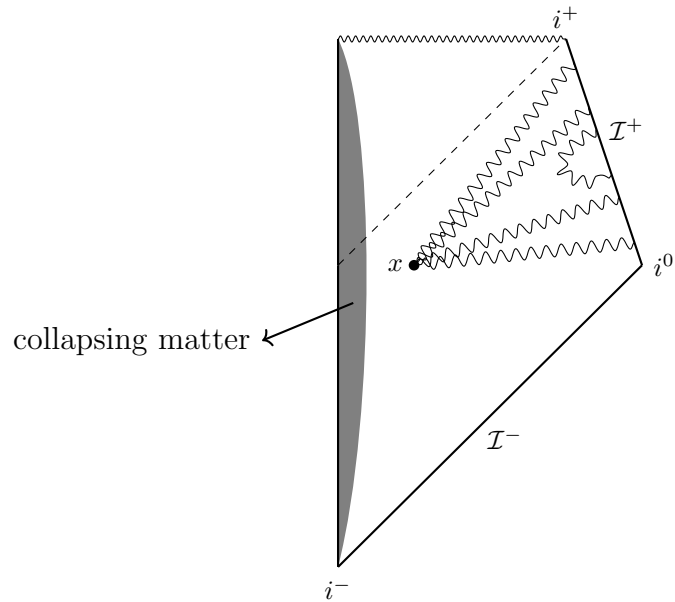
\begin{figure}
    \centering
    \usetikzlibrary{decorations.text}
    \begin{tikzpicture} [scale=1]
        \fill[gray] (0,-4)  .. controls (0.5,-2) and (0.5,2) .. (0,3);
        \draw[draw,thick] (0,0) -- (0,3);
        \draw[draw,thick] (0,0) -- (0,-4) node[below]{\footnotesize $i^-$};
        \draw[dashed](0,0) -- (3,3) node[above]{\footnotesize $i^+$};
        \draw[decorate, decoration={snake, amplitude=0.4mm, segment length=1mm}] (0,3) -- (3,3);
        \draw[draw,thick](3,3) -- (4,0)node[right]{\footnotesize $i^0$};
        \draw[draw,thick](0,-4) -- (4,0);
        \node at (3.8,1.8) {\footnotesize $\mathcal{I}^+$};
        \node at (2.2,-2.3) {\footnotesize $\mathcal{I}^-$};
        \draw[->,thick] (0.2,-0.5) -- (-1,-1) node [left] {\text{collapsing matter}};
        \draw[decorate, decoration={snake, amplitude=0.8mm,segment length=2mm}] (3.4,1.8)  .. controls (2.4,1.2) and (3.5,1) .. (3.6,1.2);
        \draw[decorate, decoration={snake, amplitude=0.8mm,segment length=2mm}] (3.1,2.7) -- (1,0) node[left] {\footnotesize $x$};
        \draw[decorate, decoration={snake, amplitude=0.8mm,segment length=2mm}] (3.3,2.1) -- (1,0);
        \draw[decorate, decoration={snake, amplitude=0.8mm,segment length=2mm}] (3.7,0.9) -- (1,0);
        \draw[decorate, decoration={snake, amplitude=0.8mm,segment length=2mm}] (3.9,0.3) -- (1,0);
        \fill(1,0) circle (2pt);
    \end{tikzpicture}
    \caption{\centering{A collapsing black hole and the two and four-point correlators for a far observer.}}
    \label{collapsingblackhole}
\end{figure}

 The most intriguing question is the Carrollian correlators through black hole collapse in astrophysics. Figure \ref{collapsingblackhole} is a Penrose diagram of spherically gravitational collapse. We have also drawn the boundary-to-boundary propagators and four-point correlators at $\mathcal{I}^+$, which  are expected to be detectable by a far observer. From Carrollian perspective, the correlators can be found by integrating out the bulk spacetime. Obviously, the black hole region could contribute to the correlators. It would be wonderful to work on this topic in the future.
\end{itemize}

\vspace{10pt}
{\noindent \bf Acknowledgments.}
The work of J.L. was supported by NSFC Grant No. 12005069. The work of H.-Y. Xiao is supported by ``the Fundamental Research Funds for the Central Universities'' with
No. YCJJ20242112.
\appendix
{
\section{Aspects of Carrollian holography}\label{Carrollian}
In this appendix we will briefly review some aspects of Carrollian holography including Carrollian symmetries and correlators. { The correlators admit an intrinsic definition on Carrollian manifolds and can be computed holographically via Carrollian amplitudes.
\subsection{Carrollian symmetries}
Carrollian symmetry originated from the ultra-relativistic contraction ($c\rightarrow0$) of the Poincar\'e group\cite{Une,Gupta1966OnAA,bacry1968possible,Henneaux:1979vn}. This Carrollian group was identified as the geometric symmetry of the Carrollian manifold  and generalized to more general groups \cite{Duval:2014uoa}. Recent studies have shown great significance of Carrollian symmetry in black hole physics\cite{penna2018near,2019CQGra..36p5002D,freidel2024carrollian,redondo2023non,ecker2023carroll}. Furthermore, a specific Carroll group is related to the BMS$_4$ group \cite{Sachs:1962wk,Bondi:1962px},
an infinite-dimensional extension of the Poincar\'e group that is important for the infrared structure of gravitational theories. 

\paragraph{Ultra-relativistic contractions.} 
Obtained from the limit $c\rightarrow0$ of the Poincar\'e group, the Carroll group is generated by the translations, the spatial rotations, and the Carrollian boosts
\begin{align}
    \mbx'=\mbx,\quad t'=t-\mathbf{b}\cdot\mbx,
\end{align}
where $\mathbf{b}$ is the boost parameter. The corresponding generators of the Carrollian algebra are $P_\mu$, $J_{ij}$ and $B_i$, $\mu=0,1,2,\cdots,d$, $i,j=1,2,\cdots,d$ which can be identified as
\begin{align}
    P_0=\p_t,\quad P_i=\p_i,\quad J_{ij}=x_i\p_j-x_j\p_i,\quad B_i=x_i\p_t,
\end{align}
with the following non-zero commutation relations
\begin{align}\label{carrollianalgebra}
    [P_i,B_j]&=\de_{ij}P_0,\quad [J_{ij},P_k]=\de_{jk}P_i-\de_{ik}P_j ,\nn\\ [J_{ij},J_{kl}]&=\de_{jk}J_{il}-\de_{ik}J_{jl}+\de_{il}J_{jk}-\de_{jl}J_{ik},\quad [J_{ij},B_k]=\de_{jk}B_i-\de_{ik}B_j .
\end{align} 
One can also consider a relativistic conformal group which has the Poincar\'e group as a subgroup with additional generators: dilatation D and special conformal transformations $K_\mu$.  
After taking the $c\rightarrow0$ limit of relativistic conformal symmetry, the dilatation and special conformal transformations $K_\mu$ are identified as
\begin{align}
    D=t\p_t+x^i\p_i,\quad K_0=x_jx^j\p_t,\quad K_i=2x_i(t\p_t+x^j\p_j)-x^jx_j\p_i,
\end{align}
and the Carrollian conformal symmetry naturally emerges. The algebra of the global part of Carrollian conformal group is generated by $\{P_\mu,J_{ij},B_i,D,K_\mu\}$ with the commutation relations (\ref{carrollianalgebra}) in additional to
\begin{align}\label{carrollianconformalalgebra}
    [D,P_i]&=-P_i,\quad [D,P_0]=-P_0,\quad [D,K_i]=K_i,\quad [D,K_0]=K_0,\nn\\ [K_0,P_i]&=-2B_i,\quad [K_i,P_0]=-2B_i,\quad [K_i,P_j]=-2\de_{ij}D-2J_{ij},\nn\\ [J_{ij},K_k]&=2(\de_{jk}K_i-\de_{ij}K_j),\quad [B_i,K_j]=\de_{ij}K_0.
\end{align}
%This global part of conformal Carroll group in $d+1$ dimension is isomorphic to the Poincar\'e group in one lower dimension\cite{Duval_2014a}. 

\paragraph{Geometric approach.}
To extend the global part of conformal Carroll group further, we turn to the geometric approach.  Considering a $d$-dimensional Carrollian manifold $\mathfrak C$ with a degenerate metric 
\be 
ds^2=\bm \gamma=\delta_{ij}dx^i dx^j,\quad i=1,2,\cdots,d
\ee associated with a null vector $\bm \chi=\partial_u$, the isometry group is generated by a vector $\bm\xi$ such that 
\be 
\mathcal L_{\bm\xi}\bm \gamma=0,\quad \mathcal L_{\bm\xi}\bm\chi=0
\ee whose solution is infinite-dimensional. One can further reduce it to a finite-dimensional Carroll group.
A much more important extension is 
the conformal Carroll group of level $k$
\begin{align}
     \text{CCarr}_k(\mathfrak C,\bm \gamma,\bm\chi)
\end{align}
which is generated by the vector $\bm\xi$ such that
\begin{align}
    \mathcal{L}_{\bm\xi}\bm \gamma=\lam\bm \gamma,\quad \mathcal{L}_{\bm\xi}\bm\chi=\mu\bm\chi,\quad \lam+k\mu=0,
\end{align}
in which $\lam$ and $\mu$ are conformal factors. Then we can get the vector $\bm\xi$ as
\begin{align}
  \bm\xi=Y^i(\bm x)\p_i+(f(\bm x)+\frac{u}{k}\partial_iY^i(\bm x))\p_u.
\end{align}
Here $f(\bm x)\in C^{\infty}(\mathbb R^d)$ and the vector $Y^A$ is a conformal Killing vector on the boundary space $\mathbb R^d$ which satisfies
\begin{align}
    \partial_iY_j+\partial_jY_i=\frac{2\gamma_{ij}}{d}\p_kY^k.
\end{align}
When $\bm\gamma$ is replaced by the metric\footnote{In spherical coordinates $\Omega=\theta^A=(\theta,\phi)$, the metric of the unit sphere is 
\be 
ds^2\equiv \gamma_{AB}d\theta^Ad\theta^B=d\theta^2+\sin^2\theta d\phi^2.
\ee In the context, we will also use the stereographic coordinates $(z,\bar z)$ 
\be 
z=\cot\frac{\theta}{2}e^{i\phi},\quad \bar z=\cot\frac{\theta}{2}e^{-i\phi},
\ee and the corresponding metric is in the form 
\be 
ds^2\equiv2\gamma_{z\bar z}dzd\bar z=\frac{4}{(1+z\bar z)^2}dz d\bar z.
\ee 
} of unit sphere $S^2$ and $k=2$, the conformal Carroll group of level 2 is isomorphic to BMS$_4$ \cite{Duval:2014uva}
\be 
 \text{CCarr}_2(\mathfrak C,\bm\gamma,\bm\chi)\simeq \text{BMS}_4.
\ee Algebraically, the original BMS group was generated by $\{L_{0,\pm1},\bar{L}_{0,\pm1},T_{r,s}\}$ where supertranslations $\{T_{r,s}\} $ are the generators of infinite-dimensional angle-dependent translations at null infinity $\scri^\pm$ and generators $\{L_{0,\pm1},\overline{L}_{0,\pm1}\}$ are a representation of the usual Lorentz group at null infinity $\scri^\pm$. 
The BMS$_4$ group can be further extended to the so called extended BMS$_4$ group by including the superrotations\cite{Barnich:2010eb}. The extended BMS$_4$ algebra is as follows,
\begin{align}
    [L_n,L_m]&=(n-m)L_{n+m},\quad [\bar{L}_n,\bar{L}_m]=(n-m)\bar{L}_{n+m},\nn\\ [L_n,T_{r,s}]&=\left(\frac{n+1}{2}-r\right)T_{n+r,s},\quad [\bar{L}_n,T_{r,s}]=\left(\frac{n+1}{2}-s\right)T_{r,n+s},\nn\\ [T_{r,s},T_{p,q}]&=0.
\end{align}
Here the superrotations $L_n$'s correspond to the global and local CKVs on the sphere at future null infinity. The global part of the extended BMS$_4$ algebra is generated by Lorentz transformations $L_0,L_{\pm 1},\overline L_0, \overline L_{\pm 1}$ and  spacetime translations $ T_{r,s}$ with $r,s=0,1$, and they together form the Poincar\'e algebra. The extended BMS$_4$ group can be generalized to a larger group which is called Carrollian diffeomorphism. 
A general Carrollian diffeomorphism is generated by the vector 
\begin{align}
    \bm\xi_{f,Z}=f(u,\Om)\p_u+Z^A(\Om)\p_A
\end{align} 
under which a scalar field from bulk reduction transforms as \cite{Liu:2023jnc}
\begin{align}
    \de_{f,Z}\Sigma(u,\Om)=f(u,\Om)\p_u\Sigma+Z^A(\Om)\nabla_A\Sigma(u,\Om)+\frac{1}{2}\nabla_CZ^C(\Om)\Sigma(u,\Om).\label{card}
\end{align} In general dimensions, the transformation law of the spinning fields under Carrollian diffeomorphism can be found in \cite{Liu:2024llk}. In this work, we only discuss the global part of the algebra since it corresponds to the Poincar\'e symmetry in the bulk.

\paragraph{Primary operators, correlators and Ward identities.}
 Next we should define primary operators on the boundary. Notice that the Lorentz algebra $\text{so}(1,3)$ is isomorphic to $\text{sl}(2,\mathbb C)$ where the latter is the global conformal algebra in two dimensions. We can utilize the knowledge of conformal field theory to define primary operators. 
{ Considering a boundary primary scalar operator $V(u,\Om)$ with conformal weight $\Delta$, the transformation law is \footnote{In our paper, the boundary is topologically $\mathbb{R}\times S^2$. Therefore, the definition of the primary field is slightly different from other works with boundary topology $\mathbb{R}\times \mathbb{R}^2$ where the transformation law of the primary field is\cite{francesco2012conformal}
\begin{align}
    V'(u',\mbx')=\left|\frac{\p \mbx'}{\p \mbx}\right|^{-\frac{\De}{2}}V(u,\mbx),\quad u\rightarrow u'=\left|\frac{\p \mbx'}{\p\mbx}\right|^{\frac{1}{2}}u,\quad \mbx\rightarrow\mbx'.
\end{align}
}
\begin{align}
    V'(u',z',\bar z')=\Gamma^{\Delta}V(u,z,\bar z)
\end{align} where
\be 
u'=\Gamma^{-1}u,\quad z'=\frac{az+b}{cz+d},\quad \bar z'=\frac{\bar a\bar z+\bar b}{\bar c\bar z+\bar d}
\ee with 
\be 
ad-bc=1,\quad a,b,c,d\in\mathbb C.
\ee The explicit form of $\Gamma$ can be found in \cite{Liu:2024nfc}.
One can check that the  infinitesimal transformation of the field $V(u,\Om)$ is equivalent to
\begin{align}
    -\de_YV(u,\Om)=\frac{1}{2}u\nabla_CY^C\dot{V}(u,\Om)+Y^A\nabla_AV(u,\Om)+\frac{\De}{2}\nabla_CY^CV(u,\Om)
\end{align} which is consistent with \eqref{card} by choosing $\Delta=1$ and  $f=\frac{1}{2}u\nabla_A Y^A$ as well as $Z^A=Y^A$. For completeness, we should also include the transformation law of the primary field under spacetime translation 
\be 
V'(u',z',\bar z')=V(u,z,\bar z),\quad u'=u+e\cdot \ell,\quad z'=z,\quad \bar z'=\bar z.
\ee where $e^\mu$ is a constant vector and $\ell^\mu$ is a null vector whose explicit form can be found in \eqref{ell}.

From the boundary perspective, the vacuum $\ket{0}$ is annihilated by all the Poincar\'e generators 
\be 
L_{n}|0\rangle=\overline L_{n}|0\rangle=T_{r,s}|0\rangle=0,\quad n=0,\pm 1\quad\text{and}\quad r,s=0,1.
\ee 
With the boundary scalar operator, the n-point Carrollian correlator can be written as
\begin{align}
    \braket{\prod_{j=1}^nV_j(u_j,\Om_j)},
\end{align}
in which we omit the vacuum $\ket{0}$. Now the boundary theory is invariant under the Poincar\'e group, as a consequence, the Carrollian correlators should satisfy the Ward identities
\begin{align}
     \braket{\prod_{j=1}^nV_j(u'_j,\Om_j)}=\braket{\prod_{j=1}^nV_j(u_j,\Om_j)}
\end{align}
for spacetime translation $u'=u-e\cdot\ell$ and
\begin{align}
    \braket{\prod_{j=1}^nV_j(u'_j,\Om'_j)}=\left(\prod_{j=1}^n\Gamma_j^{\De_j}\right)\braket{\prod_{j=1}^nV_j(u_j,\Om_j)}
\end{align}
for Lorentz transformation. One can act on the operator $V(u,\Omega)$ by all possible generators to get the descendants. As an example, we consider the previous massless scalar field $\Sigma$ at the boundary and define the $u$-descendants
\begin{align}
    V_n(u,\Om)=\left(\frac{\p}{\p u}\right)^n\Sigma(u,\Om),
\end{align}
with conformal weight
\begin{align}
    \De=1+n
\end{align} and spin 0.

Expanding the two kinds of Ward identities for $\Sigma(u,\Om)$ to  first order in the infinitesimal parameters, we reach the differential equations for the Carrollian correlators
\begin{subequations}\label{ward}
\begin{align}
    \mathcal{L}_{st}^\mu[\ell]\braket{\prod_{j=1}^n\Sigma(u_j,\Om_j)}&=0,\\\mathcal{L}_{LT}^{scalar}[Y]\braket{\prod_{j=1}^n\Sigma(u_j,\Om_j)}&=0,
\end{align}
\end{subequations}
where the differential operators read
\begin{subequations}
\begin{align}
    \mathcal{L}_{st}^\mu[\ell]&=\sum_{j=1}^n\ell^\mu_j\frac{\p}{\p u_j},\\\mathcal{L}_{LT}^{scalar}[Y]&=\sum_{j=1}^n\left(Y^A(\Om_j)\frac{\p}{\p\theta^A_j}+\frac{1}{2}\nabla\cdot Y(\Om_j)+\frac{u_j}{2}\nabla\cdot Y(\Om_j)\frac{\p}{\p u_j}\right).
\end{align}
\end{subequations}

} 

%While two-point correlation functions have been derived by taking the Carrollian limit $c\rightarrow0$ of relativistic CFT correlators\cite{Chen:2021xkw}?, 
In the following we try to deduce the two-point correlation functions from symmetries. Suppose there is a boundary scalar field $\Sigma(u,\Om)$ inserted at null infinity $\scri$. Here $(u,\Om)$ are coordinates of $\scri$ with $\Om=\theta^A=(z,\bar z)$ the stereographic coordinate. %The infinitesimal transformations of the scalar field under Carrollian diffeomorphism are\cite{Liu:2024llk}
The Ward identities for the two-point correlator
\be 
B(u_1,z_1,\bar z_1;u_2,z_2,\bar z_2)=\langle 0|\Sigma(u_1,z_1,\bar z_1)\Sigma(u_2,z_2,\bar z_2)|0\rangle
\ee can be written explicitly as
\begin{subequations}\label{Ward2}
\begin{align}
&\left(\frac{\p}{\p u_1}+\frac{\p}{\p u_2}\right)B=0, \label{timetrans}\\
        & \left(\frac{z_1+\bar{z}_1}{1+z_1\bar{z}_1}\frac{\p}{\p u_1}+\frac{z_2+\bar{z}_2}{1+z_2\bar{z}_2}\frac{\p}{\p u_2}\right)B=0, \\
       &  \left(\frac{z_1-\bar{z}_1}{1+z_1\bar{z}_1}\frac{\p}{\p u_1}+\frac{z_2-\bar{z}_2}{1+z_2\bar{z}_2}\frac{\p}{\p u_2}\right)B=0, \\
       &  \left(\frac{z_1\bar{z}_1-1}{1+z_1\bar{z}_1}\frac{\p}{\p u_1}+\frac{z_2\bar{z}_2-1}{1+z_2\bar{z}_2}\frac{\p}{\p u_2}\right)B=0,\label{spatial3}\\
   &\sum_{j=1}^2\left(\frac{u_j \left(\bar{z}_j+z_j\right)}{z_j \bar{z}_j+1}\partial_{u_j}+\frac{1}{2} \left(z_j^2-1\right)\partial_{z_j}+\frac{1}{2} \left(\bar{z}_j^2-1\right)\partial_{\bar {z}_j}+\frac{\bar{z}_j+z_j}{z_j \bar{z}_j+1}\right)B=0,\label{Lorentz1}\\
  &\sum_{j=1}^2\left(\frac{i u_j \left(\bar{z_j}-z_j\right)}{z_j \bar{z_j}+1}\partial_{u_j}-\frac{1}{2} i \left(z_j^2+1\right)\partial_{z_j}+\frac{1}{2} i \left(\bar{z_j}^2+1\right)\partial_{{\bar {z}}_j}+\frac{i \left(\bar{z}_j-z_j\right)}{z_j \bar{z}_j+1}\right)B=0,\\
 & \sum_{j=1}^2 \left(\frac{u_j \left(z_j \bar{z}_j-1\right)}{z_j \bar{z}_j+1}\partial_{u_j}-z_j\partial_{z_j}-\bar {z}_j\partial_{\bar {z}_j}+\frac{z_j \bar{z}_j-1}{z_j \bar{z}_j+1}\right)B=0,\label{Lorentz3}\\
   &\sum_{j=1}^2(iz_j\partial_{z_j}-i\bar {z}_j\partial_{\bar {z}_j})B=0,\label{rotation1}\\
   &\sum_{j=1}^2\left(\frac{1}{2} \left(z_j^2+1\right)\partial_{z_j}+\frac{1}{2} \left(\bar{z}_j^2+1\right)\partial_{\bar {z}_j}\right)B=0,\\
   &\sum_{j=1}^2\left(-\frac{1}{2} i \left(z_j^2-1\right)\partial_{z_j}+\frac{1}{2} i \left(\bar{z}_j^2-1\right)\partial_{\bar {z}_j}\right)B=0.\label{rotation3}
\end{align}
\end{subequations}
Here we have omitted the arguments in the correlator to simplify notation. There are two solution branches for the Ward identities. 
\paragraph{Magnetic branch.}
In this branch, the correlator is $u$ independent. As a consequence, only the Ward identities associated with $\text{sl}(2,\mathbb C)$ are important. One can borrow the results from 2d CFT to find 
\be B(u,\Omega;u',\Omega')=\frac{(1+z\bar{z})(1+z'\bar{z}')}{4}\frac{1}{(z-z')(\bar{z}-\bar{z}')}
\ee  up to a normalization constant. In spherical coordinates, it is 
\be 
B(u,\Omega;u',\Omega')=\frac{1}{2(1-\cos\gamma(\Omega,\Omega'))}
\ee where $\gamma(\Omega,\Omega')$ is the angle between two directions parameterized by $\Omega$ and $\Omega'$. More precisely, 
\be 
\cos\gamma(\Omega,\Omega')=\cos\theta\cos\theta'+\sin\theta\sin\theta'\cos(\phi-\phi').
\ee Note that $2(1-\cos\gamma(\Omega-\Omega'))$ is the square of the geodesic distance between $\Omega$ and $\Omega'$ on the unit sphere.

\paragraph{Electric branch.} We can also assume the correlator is $u$ dependent and then the two point function can be written as 
\be 
B(u,z,\bar z;u',z',\bar z')=\widetilde{B}(u-u')\delta(\Omega-\Omega').
\ee The correlator only depends on the difference of the time to preserve the time translation invariance. Combined with the spatial translation invariance, one can easily find that this is only possible for $\Omega=\Omega'$. This is why there is a Dirac delta function in the assumption. One can  verify that the rotation invariance is automatically satisfied and the Lorentz boost invariance leads to 
\be 
(u\partial_{u}+1)\widetilde{B}(u-u')=0\quad\Rightarrow\quad \widetilde{B}(u-u')=-\log(u-u')+\text{const.}
\ee

We have established the fact that for a primary scalar field with dimension $\Delta=1$, the general  two-point Carrollian correlator is 
\be 
B(u,\Omega;u',\Omega')=-C_E \log(u-u')\delta(\Omega-\Omega')+\frac{C_B}{1-\cos\gamma(\Omega,\Omega')} 
\ee where $C_E$ and $C_B$ are the normalization constants for the electric and magnetic branch, respectively. 
\paragraph{Remarks.}
The previous discussion can be generalized to any primary fields with conformal dimension $\Delta$ and spin $s$. We will not present them since we only focus on scalar field in this work. Further extensions to higher point correlators  are also interesting. The three-point functions can also be fixed  into several structures by solving the Ward identities (\ref{ward})\cite{Bagchi:2023cen,Salzer:2023jqv,Nguyen:2023miw,Bagchi:2023fbj}. On the other hand, the four-point functions and higher-point functions cannot be completely fixed by the global conformal Carroll group\cite{Banerjee:2018gce,Bagchi:2023cen,bagchi2017nuts,dolan2001conformal,dolan2004conformal}.

\subsection{Carrollian amplitude}\label{carrollianamplitude}
Carrollian amplitudes are massless scattering amplitudes defined in position space. In \cite{Bagchi:2023cen,Alday:2024yyj,Bagchi:2023fbj}, the Carrollian amplitudes were constructed by taking the flat limit of AdS Witten diagrams. \iffalse  In additional, Carrollian amplitudes can naturally emerge from the correlations in boundary Carrollian conformal field theories\cite{saha2022intrinsic,Bagchi:2022emh,Arcioni:2003xx,Barnich:2010eb,Bagchi:2010zz,Donnay:2022aba}.\fi However, Carrollian amplitudes can also be deduced on the foundation of bulk reduction in which the massless relativistic fields in the bulk are sent out to null infinity\cite{Liu:2024nkc,Liu:2024nfc}. The latter has also been extended to globally hyperbolic  spacetimes \cite{Li:2024kbo}. We will review the framework in this section. 

The metric of the four-dimensional Minkowski spacetime $\mathbb{R}^{1,3}$ in Cartesian coordinates $x^\mu=(t,\bm x)$ is
\begin{align}
    ds^2=\eta_{\mu\nu}dx^\mu dx^\nu=-dt^2+dx^idx^i,\quad \mu,\nu=0,1,2,3,
\end{align}
where the Minkowski matrix is $\eta_{\mu\nu}=\text{diag}(-1,+1,+1,+1)$.  Switching to the spherical coordinates $(t,r,\theta^A),\ A=1,2$, the metric can be rewritten as
\begin{align}
    ds^2=-dt^2+dr^2+r^2(d\theta^2+\sin^2{\theta}d\phi^2).
\end{align}
Given a field $\Phi(t,x)$ in the bulk, one can impose the fall-off condition
\begin{align}
    \Phi(t,\mbx)=\left\{\begin{array}{cc}
         \frac{\Sigma(u,\Om)}{r}+\mathcal{O}(r^{-2}),\quad \text{near }\scri^+  \\
         \frac{\Sigma^{(-)}(v,\Om)}{r}+\mathcal{O}(r^{-2}),\quad \text{near }\scri^- 
    \end{array}\right.
\end{align}
to obtain a boundary field $\Sigma(u,\Omega)/\Sigma^{(-)}(v,\Omega)$ at $\scri^{\pm}$. Here the coordinates $u=t-r$ and $v=t+r$ are the retarded and advanced time, respectively. The fundamental field $\Sigma(u,\Om)/\Sigma^{(-)}(v,\Om)$ encodes the propagating degree of freedom of the bulk theory and it is the leading order coefficient in the asymptotic expansion. We reinterpret it as a primary operator that is inserted at $\mathcal I^{\pm}$ with dimension $1$ and spin $0$. The bulk-to-bulk propagator that is also called Feynman propagator is defined as 
\be 
G_{\text F}(x-x')=\langle 0|\text{T}\Phi(x)\Phi(x')|0\rangle
\ee where $\text T$ denotes the time-ordering operator. Using the fall-off conditions, we can obtain two bulk-to-boundary propagators 
\bs\begin{align}
   D(u,\Omega;x')&=\langle \Sigma(u,\Omega)\Phi(x')\rangle=\lim_{r\to\infty,\quad u \ \text{finite}}r\ G_{\text{F}}(x-x'),\\
   D^{(-)}(v',\Omega';x)&=\langle \Phi(x)\Sigma^{(-)}(v',\Omega')\rangle=\lim_{r'\to\infty,\ v'\ \text{finite}} r'\ G_{\text{F}}(x-x').
\end{align}\es 
 By extrapolating the remaining bulk field to the boundary, we find three boundary-to-boundary propagators 
\bs\begin{align}
    B(u,\Omega;u',\Omega')&=\langle \Sigma(u,\Omega)\Sigma(u',\Omega')\rangle=\lim_{r'\to\infty,\quad u' \ \text{finite}} r'\ D(u,\Omega;x'),\\
    B(u,\Omega;v',\Omega')&=\langle \Sigma(u,\Omega)\Sigma^{(-)}(v',\Omega')\rangle=\lim_{r'\to\infty,\quad v' \ \text{finite}} r'\  D(u,\Omega;x')=\lim_{r\to\infty,\quad\ u\ \text{finite}}r\ D^{(-)}(v',\Omega';x),\\
    B(v,\Omega;v',\Omega')&=\langle \Sigma^{(-)}(v,\Omega)\Sigma^{(-)}(v',\Omega')\rangle=\lim_{r\to\infty,\ v\ \text{finite}}r\ D^{(-)}(v',\Omega';x).
\end{align}\es The explicit form of the propagators can be found in the paper \cite{Liu:2024nfc} where  the authors used  canonical quantization method and the primary field is written as 
\bs\begin{align} 
\Sigma(u,\Omega)&=\frac{i}{8\pi^2}\int_0^\infty d\omega ( b_{\bm p}e^{-i\omega u}+b_{\bm p}^\dagger e^{i\omega u}),\\
\Sigma^{(-)}(v,\Omega)&=-\frac{i}{8\pi^2}\int_0^\infty d\omega ( b_{\bm p^{\text P}}e^{-i\omega v }+b_{\bm p^{\text{P}}}^\dagger e^{i\omega v}).
\end{align}\es  Here $b_{\bm p}$ and $b_{\bm p}^\dagger$ are annihilation and creation operators. The superscript $\text{P}$ denotes the antipodal map, more explicitly
\be \bm p=(\omega,\theta,\phi)\quad\Rightarrow\quad  \bm p^{\text P}=(\omega,\pi-\theta,\pi+\phi).
\ee

With these fields, we can define asymptotic states from the vacuum $\ket{0}$:
\begin{align}
    \ket{\Sigma(u,\Om)}=\Sigma(u,\Om)\ket{0},\quad \ket{\Sigma^{(-)}(v,\Om^P)}=\Sigma^{(-)}(v,\Om^P)\ket{0},
\end{align}
where the spherical coordinates $\Om^P=(\pi-\theta,\pi+\phi)$ are defined as the antipodal point of $\Om=(\theta,\phi)$. Similarly, the asymptotic `multi-particle' states are
\begin{align}
    \ket{\prod_{k=1}^m\Sigma(u_k,\Om_k)}=\prod_{k=1}^m\Sigma(u_k,\Om_k)\ket{0},\quad \ket{\prod_{k=1}^n\Xi(v_k,\Om^P_k)}=\prod_{k=1}^n\Xi(v_k,\Om^P_k)\ket{0}
\end{align}
which represent the states with $m$ boundary fields inserted at future null infinity $\mathcal I^+$ and $n$ boundary fields at past null infinity $\mathcal I^-$. Then the $m\rightarrow n$ Carrollian amplitude is defined as
\begin{align}
    {}_{\text{out}}\langle \prod_{k=m+1}^{m+n}\Sigma(u_k,\Omega_k)|\prod_{k=1}^m \Sigma^{(-)}(v_k,\Omega_k^P)\rangle_{\text{in}}=\langle \prod_{k=m+1}^{m+n}\Sigma(u_k,\Omega_k)|S|\prod_{k=1}^m \Sigma^{(-)}(v_k,\Omega_k^P)\rangle
\end{align}
where $S$ is the scattering operator. The left-hand side can also be understood as $(m+n)$-point correlators with $m$ fields $\Sigma^{(-)}(v,\Om)$ inserted at $(v_1,\Om_1^P),\cdots,(v_m,\Om_m^P)$ and $n$ fields $\Sigma(u,\Om)$ inserted at $(u_{m+1},\Om_{m+1}),\cdots,(u_{m+n},\Om_{m+n})$, respectively. In the following we redefine $u_j=v_j,j=1,2,\cdots,m$, transform $\Om_j^P$ to their antipodal points $\Om_j$, and relabel the boundary fields as
\bea  
\Sigma(u,\Omega,+)=\Sigma(u,\Omega),\quad \Sigma(u,\Om,-)=\Sigma^{(-)}(v,\Om^P)
\eea  to obtain a more familiar form of Carrollian amplitudes
\begin{align}
    {}_{\text{out}}\langle \prod_{k=m+1}^{m+n}\Sigma(u_k,\Omega_k,+)|\prod_{k=1}^m \Sigma(u_k,\Omega_k,-)\rangle_{\text{in}}.
\end{align}
 Carrollian amplitude can also be reduced to the $\mathcal{M}$ matrix which is related to amputated and connected Feynman diagrams
\begin{align}\label{ca=mmatrix}
    &{}_{\text{out}}\langle \prod_{k=m+1}^{m+n}\Sigma(u_k,\Omega_k,+)|\prod_{k=1}^m \Sigma(u_k,\Omega_k,-)\rangle_{\text{in}}\Big{|}_{\text{connected and amputated}}\nn\\&=(\frac{1}{8\pi^2i})\prod_{j=1}^{m+n}\int d\om_je^{-i\sigma_j\om_ju_j}(2\pi)^4\de^{(4)}(\sum_{j=1}^{m+n}p_j)i\mathcal{M}(p_1,p_2,\cdots,p_{m+n}).
\end{align}
Here $\sigma_j=\pm1,j=1,2,\cdots,m+n$ denotes the incoming or outgoing state for each operator and $\om_j,j=1,2,\cdots,m+n$ represents the energy of each state. Interested reader can refer to \cite{Liu:2024nfc} for more details.

The Carrollian amplitude can also be written as an $(m+n)-$point correlater for the boundary Carrollian field theory
\begin{align}\label{m+ncorrelator}
     \langle\prod_{j=1}^{m+n}\Sigma_j(u_j,\Om_j,\sigma_j)\rangle={}_{\text{out}}\langle \prod_{k=m+1}^{m+n}\Sigma(u_k,\Omega_k)|\prod_{k=1}^m \Sigma(u_k,\Omega_k)\rangle_{\text{in}}.
\end{align}
Note that the Carrollian amplitude (\ref{m+ncorrelator}) is a function in the Carrollian space, we denote it as
\begin{align}
    \mathcal C_n(u_1,\Om_1,\sigma_1;\cdots;u_{m+n},\Om_{m+n},\sigma_{m+n})\equiv\langle\prod_{j=1}^{m+n}\Sigma_j(u_j,\Om_j,\sigma_j)\rangle.\label{carrco}
\end{align}

The fact that Carrollian amplitudes at $\scri$ can be connected to momentum space amplitudes via  Fourier transforms can also be found in \cite{Donnay:2022wvx}, in which the Carrollian amplitude is defined as
\begin{align}\label{CarrollianamplitudeFourier}
   \mathcal C_n(\{u_1,z_1,\bar{z}_1\}^{\ep_1},\cdots,\{u_n,&z_n,\bar{z}_n\}^{\ep_n})\nn\\&=\prod_{i=1}^n\left(\int_0^{\infty}\frac{d\om_i}{2\pi}e^{i\ep_i\om_i u_i}\right)\mathcal{A}_n(\{\om_1,z_1,\bar{z}_1\}^{\ep_1},\cdots,\{\om_n,z_n,\bar{z}_n\}^{\ep_n})
\end{align}
where $\mathcal{A}_n(\{\om_1,z_1,\bar{z}_1\}^{\ep_1},\cdots,\{\om_n,z_n,\bar{z}_n\}^{\ep_n})$ refers to the $\mathcal{S}$-matrix elements in momentum space. Here $n$ denotes the total number of particles, $\om_i>0$ the energy of  each particle, and $\ep=\pm1$ tells whether the particle is outgoing or incoming. This is in general the same as (\ref{ca=mmatrix}) except  for the coordinates chosen at null infinity. It has also been shown that the Carrollian amplitude is actually the extrapolation of the bulk Green's function to the null boundary \cite{Liu:2024nfc}. Therefore, one can also use Feynman rules in  position space to compute it.

Carrollian amplitudes can also be interpreted as Carrollian CFT correlators of operators \eqref{carrco} inserted at null infinity $\scri$. Therefore, Carrollian amplitudes are a holographic version of Carrollian correlators in the sense of flat holography. Concrete examples of two-point Carrollian amplitudes have been obtained in\cite{Donnay:2022wvx} and three-point Carrollian amplitudes can be found in\cite{Salzer:2023jqv,Nguyen:2023miw}. Moreover, the tree-level Carrollian amplitudes for gluons and gravitons have been studied systematically\cite{Mason:2023mti}.

\iffalse
The momentum of a massless particle can be parameterised as
\begin{align}
    p^\mu=\ep\om q^\mu(z,\bar{z})
\end{align}
where
\begin{align}
    q^\mu(z,\bar{z})=(1,\frac{z+\bar{z}}{1+z\bar{z}},\frac{-i(z-\bar{z})}{1+z\bar{z}},\frac{1-z\bar{z}}{1+z\bar{z}})
\end{align}
is a null vector. $\om>0$ is the energy for the particle and $\ep=+1$ or $\ep=-1$ represent the outgoing or incoming particles, respectively. An $n$-point massless scalar scattering amplitude in momentum space is denoted by $\mathcal{M}_n\left(\om_1,z_1,\bar{z}_1,\ep_1;\cdots;\om_n,z_n,\bar{z}_n,\ep_n\right)$. The Carrollian amplitude is defined as the associated position space amplitude at $\scri$ obtain via Fourier transform\cite{Donnay:2022wvx,mason2024carrollian}
\begin{align}
    C(u_1,z_1,\bar{z}_1,\ep_1;\cdots;u_n,z_n,\bar{z}_n,\ep_n)=\int_0^\infty\prod_{i=1}^n\frac{d\om_i}{2\pi}e^{-i\ep_i\om_i u_i}\mathcal{M}_n\left(\om_1,z_1,\bar{z}_1,\ep_1;\cdots;\om_n,z_n,\bar{z}_n,\ep_n\right).
\end{align}
and it can be naturally interpreted as correlation functions of Carrollian operators at null infinity $\scri$\cite{alday2025carrollian}
\begin{align}
    C(u_1,z_1,\bar{z}_1,\ep_1;\cdots;u_n,z_n,\bar{z}_n,\ep_n)=\braket{\Phi(u_1,z_1,\bar{z}_1,\ep_1)\cdots\Phi(u_n,z_n,\bar{z}_n,\ep_n)}.
\end{align}
\fi
}
}
\section{Integral representation of the propagators}\label{prop}
We will discuss the integral representation of the various propagators in frequency space and clarify their relations in this appendix. 
\paragraph{From bulk-to-bulk to bulk-to-boundary propagator.}
In this paragraph, we will reduce the integral representation of the bulk-to-bulk propagators \eqref{bulktobulkint} to bulk-to-boundary propagators. Recall the retarded coordinates defined in \eqref{labely}, we find 
\bea 
D_{11}(u,\Omega;x)&=&\lim_{r\to\infty,\ u\ \text{finite}} \frac{r}{4\pi^2|\bm x-\bm y|}\int_{-\infty}^\infty d\omega n(\omega) e^{-i\omega (x^0-y^0)}\frac{e^{i\omega|\bm x-\bm y|}-e^{-i\omega|\bm x-\bm y|}}{2i}\nn\\&=&\frac{1}{8\pi^2i}\lim_{r\to\infty,\ u\ \text{finite}}\int_{-\infty}^\infty d\omega n(\omega)e^{-i\omega x^0+i\omega y^0}[e^{i\omega r-i\omega \bm x\cdot\bm \ell}-e^{-i\omega r+i\omega \bm x\cdot\bm \ell}]\nn\\&=&\frac{1}{8\pi^2 i}\lim_{r\to\infty,\ u\ \text{finite}}\int_{-\infty}^\infty d\omega n(\omega) [e^{i\omega (v-\bar\ell\cdot x)}-e^{i\omega (u+\ell\cdot x)}].\label{D11integralmid}
\eea In the second line, we have expanded the distance $|\bm x-\bm y|$ as 
\be 
|\bm x-\bm y|=r-\bm x\cdot\bm\ell+\cdots
\ee where $\cdots$ is order $\mathcal{O}(r^{-1})$. The integral \eqref{D11integralmid} can be separated into two parts
\bea 
D_{11}(u,\Omega;x)&=&\frac{1}{8\pi^2 i}\lim_{r\to\infty,\ u\ \text{finite}}\Big[\int_{\mathcal C} d\omega n(\omega) e^{i\omega (v-\bar\ell\cdot x)}-\int_{\mathcal C}d\omega n(\omega)e^{i\omega (u+\ell\cdot x)}\Big].\label{firstnull}
\eea Note that we have deformed the real axis to the contour $\mathcal C$ in the integration to avoid the pole of  $n(\omega)$ at $\omega=0$. Otherwise the two integrals are divergent and the separation is ill defined. Now consider the first integral involving $v$. In the limit $r\to\infty$ with $u$ finite, the combination $(v -\bar\ell\cdot x)\to\infty$. We should complete the contour $\mathcal{C}$ by a half circle with large radius in the upper half plane. Using the residue theorem, each of the residue becomes zero in the limit $v\to\infty$. Therefore, we can discard the first integral in  \eqref{firstnull} and then 
\be 
D_{11}(u,\Omega;x)=-\frac{1}{8\pi^2 i}\int_{\mathcal C}d\omega n(\omega) e^{i\omega(u+\ell\cdot x)}.\label{intd11}
\ee One can also deform the real axis to the contour $\mathcal C'$ to obtain 
\bea 
D_{11}(u,\Omega;x)&=&\frac{1}{8\pi^2 i}\lim_{r\to\infty,\ u\ \text{finite}}\Big[\int_{\mathcal C'} d\omega n(\omega) e^{i\omega (v-\bar\ell\cdot x)}-\int_{\mathcal C'}d\omega n(\omega)e^{i\omega (u+\ell\cdot x)}\Big].\label{firstnull2}
\eea In this case, the first integral is non-vanishing due to the contribution of the residue at $\omega=0$
\bea 
D_{11}(u,\Omega;x)=\frac{1}{4\pi\beta}-\frac{1}{8\pi^2i}\int_{\mathcal C'}d\omega n(\omega)e^{i\omega(u+\ell\cdot x)}.\label{intd11p}
\eea The two integrals \eqref{intd11} and \eqref{intd11p} are equivalent due to the identity 
\bea 
\frac{1}{8\pi^2 i}\int_{\mathcal C'-\mathcal C}d\omega n(\omega) e^{i\omega(u+\ell\cdot x)}=\frac{1}{4\pi\beta}.
\eea

\paragraph{From bulk-to-boundary to boundary-to-boundary propagator.}
To get the integral representation of the boundary-to-boundary propagator, we use the expansion of plane wave 
\bea 
e^{i\bm p\cdot\bm x'}=4\pi\sum_{\ell,m}i^\ell j_\ell(\omega r')Y_{\ell,m}^*(\Omega) Y_{\ell,m}(\Omega')
\eea where the momentum $\bm p$ and $\bm x'$ are written in spherical coordinates
\be \bm p=(\omega,\Omega),\quad \bm x'=(r',\Omega').
\ee The large $r'$ expansion of the spherical Bessel function is 
\be 
j_{\ell}(\omega r')=\frac{\sin(\omega r'-\frac{\pi\ell}{2})}{\omega r'}=\frac{e^{i\omega r'}i^{-\ell}-e^{-i\omega r'}i^\ell}{2i\omega r'}.\label{jl}
\ee Therefore, 
\bea 
D_{11}(u,\Omega;x')&=&-\frac{1}{8\pi^2 i}\int_{\mathcal C}d\omega n(\omega) e^{i\omega u-i\omega t'+i\bm p\cdot\bm x'}\nn\\&\sim&-\frac{1}{8\pi^2i\times 2ir'}\int_{\mathcal C}d\omega \frac{n(\omega)}{\omega}e^{i\omega u}4\pi \sum_{\ell,m}[e^{-i\omega u'}-(-1)^\ell e^{-i\omega v'}]Y_{\ell,m}^*(\Omega)Y_{\ell,m}(\Omega').
\eea When $r'\to\infty$ with $v'$ finite, we have $u'\to-\infty$ and then the integral involving $u'$ vanishes due to the residue theorem. Therefore, 
\bea 
B(u,\Omega;v',\Omega')=-\frac{1}{4\pi}\int_{\mathcal C}\frac{d\omega}{\omega}n(\omega) e^{i\omega(u-v')}\delta(\Omega-\Omega^{'\text{P}}).\label{finiteB}
\eea 

One can consider an alternative limit $r'\to\infty$ with $u'$ finite. In this case, $v'\to\infty$ and then the integral involving $v'$ is non-vanishing due to the residue at the origin 
\bea 
B(u,\Omega;u',\Omega')=\frac{1}{4\pi}\int_{\mathcal C}\frac{d\omega}{\omega}n(\omega) e^{i\omega(u-u')}\delta(\Omega-\Omega')-\frac{1}{2\beta}(u-v'+\frac{i}{2}\beta)\delta(\Omega-\Omega^{'\text{P}}).
\eea The first term is the same form of the integral \eqref{finiteB} which has shown to be finite. We will denote this finite term as
\bea 
B^{\text{finite}}(u,\Omega;v',\Omega')=\frac{1}{4\pi}\int_{\mathcal C}\frac{d\omega}{\omega}n(\omega) e^{i\omega(u-u')}\delta(\Omega-\Omega').
\eea The second term still contains a term proportional to $r'$ which is divergent 
\bea 
B^{\text{div}(1)}(u,\Omega;u',\Omega')&=&-\frac{1}{2\beta}(u-u'-2r'+\frac{i}{2}\beta)\delta(\Omega-\Omega^{'\text{P}})\nn\\&=&\left(\frac{r'}{\beta}-\frac{u-u'}{2\beta}-\frac{i}{4}\right)\delta(\Omega-\Omega^{'\text{P}}).
\eea 
At the next order, \eqref{jl} should be corrected by 
\bea 
j_\ell(\omega r')&=& \frac{\sin(\omega r'-\frac{\pi\ell}{2})}{\omega r'}+\ell(\ell+1)\frac{\cos(\omega r'-\frac{\pi\ell}{2})}{2\omega^2 r'^2}+\cdots\nn\\&=&\frac{i^{-\ell}e^{i\omega r'}-i^\ell e^{-i\omega r'}}{2i\omega r'}+\ell(\ell+1)\frac{i^{-\ell}e^{i\omega r'}+i^\ell e^{-i\omega r'}}{4\omega^2r'^2}+\cdots.
\eea 
For the outgoing modes, the leading term proportional to $r'^{-1}$ is enough. However, for incoming modes, we should consider the subleading term which is proportional to $r'^{-2}$. To be more precise, the contribution of the incoming modes are
\bea 
B^{\text{div}(2)}(u,\Omega;u',\Omega')&=&-\frac{1}{8\pi^2 i}\int_{\mathcal C}d\omega n(\omega) 4\pi \left(\sum_{\ell,m}Y_{\ell,m}^*(\Omega)Y_{\ell,m}(\Omega')(-1)^{\ell}\left(\frac{-1}{2i\omega}+\frac{\ell(\ell+1)}{4\omega^2 r'}\right)\right)e^{i\omega(u-v')}\nn\\&=&\left(\frac{r'}{\beta}-\frac{u-u'}{2\beta}-\frac{i}{4}\right)\delta(\Omega-\Omega^{'\text{P}})+\left(\frac{r'}{2\beta}-\frac{u-u'}{2\beta}-\frac{i}{4}\right)\nabla^2\delta(\Omega-\Omega^{'\text{P}})
\eea 
where we have used the fact that the spherical Harmonics are the eigenfunctions of the Laplace operator on $S^2$
\be 
\nabla^2 Y_{\ell,m}(\Omega)=-\ell(\ell+1)Y_{\ell,m}(\Omega).
\ee 
Interestingly, the incoming modes of order $r'^{-2}$ will also contribute a divergent term to the boundary-to-boundary propagator. It is reasonable to consider all order contributions from the incoming modes. To solve this problem, we notice the spherical Bessel function is a linear superposition of the spherical Hankel functions
\bea 
j_\ell(\omega r')=\frac{1}{2}\left(h_\ell^{(1)}(\omega r')+ h_\ell^{(2)}(\omega r')\right)
\eea where $h_\ell^{(1)}$ is the spherical Hankel function of the first kind while $h^{(2)}_\ell$ the second kind. The $h_\ell^{(1)}$ and $h_{\ell}^{(2)}$ correspond to outgoing  and incoming modes, respectively.
The spherical Hankel function of the second kind has the asymptotic expansion near $r'\to\infty$
\bea 
h_{\ell}^{(2)}(\omega r')=-e^{-i\omega r'} i^\ell \sum_{k=0}^\infty \frac{H_k(-\ell(\ell+1))}{\omega^{k+1}r'^{k+1}}
\eea where 
\be H_k(x)=\frac{i^{k-1}}{2^k k!}\prod_{m=1}^k (x+m(m-1)).
\ee Therefore, the divergent part is 
\bea 
B^{\text{div}}(u,\Omega;u',\Omega')&=&-\frac{i}{4\pi}\int_{\mathcal C}d\omega n(\omega)e^{i\omega(u-v')}\sum_{k=0}^\infty \frac{H_k(\nabla^2)}{\omega^{k+1}r'^k}\delta(\Omega-\Omega^{'\text{P}}).
\eea Since $v'\to\infty$, we can use the residue theorem and only the soft mode $\omega=0$ has non-vanishing contribution. To extract the residue for $r'\to\infty$, we define the function 
\be 
g(\omega;r')=n(\omega)\omega^{-1-k}e^{i\omega(u-v')}r'^{-k}=n(\omega)\omega^{-1-k}e^{i\omega(u-u'-2r')}r'^{-k}.
\ee We expand it for large $r'$, the non-vanishing terms are 
\be 
g(\omega;r')\sim n(\omega)\omega^{-1-k}r'^{-k}\sum_{j=k}^\infty \frac{(i\omega(u-u' -2r'))^j}{j!}. 
\ee When $j>k+1$, there will be no residue for the function $g(\omega;r')$ at $\omega=0$ in the large $r'$ limit. Therefore, the relevant terms are $j=k$ or $j=k+1$
\be 
g(\omega;r')\sim n(\omega)\omega^{-1}\frac{i^k (-2)^k}{k!}+n(\omega)\frac{i^{k+1}(-2)^{k+1}r'}{(k+1)!}+n(\omega)\frac{i^{k+1} (k+1)(-2)^k}{(k+1)!}(u-u').
\ee Since the residue 
\bea 
\text{Res}_{\omega=0}n(\omega)=\frac{1}{\beta},\quad \text{Res}_{\omega=0}n(\omega)\omega^{-1}=-\frac{1}{2},
\eea we get 
\bea 
B^{\text{div}}(u,\Omega;u',\Omega')=-\frac{1}{2}\sum_{k=0}^\infty[\frac{r'}{\beta}\frac{i^{k+1}(-2)^{k+1}}{(k+1)!}+\frac{i^{k+1}(-2)^k}{k!}(\frac{u-u'}{\beta}+\frac{i}{2})] H_k(\nabla^2)\delta(\Omega-\Omega^{'\text{P}}).
\eea The series  of $k$ can be summarized to a closed form 
\bea 
B^{\text{div}}(u,\Omega;u',\Omega')=\Big[{}_2F_1(\Delta_+,\Delta_-,2;1)\frac{r'}{\beta}-\frac{1}{2}(\frac{u-u'}{\beta}+\frac{i}{2}){}_2F_1(\Delta_+,\Delta_-,1;1)\Big]\delta(\Omega-\Omega^{'\text{P}}),
\eea where the operators $\Delta_+$ and $\Delta_-$ are
\bea \Delta_{\pm}=\frac{1}{2}(1\pm\sqrt{1-4\nabla^2}).
\eea The hypergeometric function ${}_2F_1(a,b,c;x)$ can be expanded as an infinite series
\bea 
{}_2F_1(a,b,c;x)=\sum_{n=0}^\infty \frac{(a)_n(b)_n}{(c)_n n!}x^n
\eea where the Pochhammer symbol  is 
\be (a)_n=a(a-1)\cdots (a-n+1).
\ee 
 The operators $\Delta_{\pm}$ can act on the spherical Harmonic function with the eigenvalues
\bs\begin{align}
    \Delta_+ Y_{\ell,m}(\Omega)&=(\ell+1)Y_{\ell,m}(\Omega),\\
    \Delta_-Y_{\ell,m}(\Omega)&=-\ell Y_{\ell,m}(\Omega).
\end{align}\es 
The Legendre polynomial can be defined by the Hypergeometric function by \cite{1963cma..book.....W}
\be 
P_\ell(z)={}_2F_1(\ell+1,-\ell,1;\frac{1-z}{2}),
\ee 
then the operator ${}_2F_1(\Delta_+,\Delta_-,1;1)$ can act on the Dirac delta function formally 
\bea 
{}_2F_1(\Delta_+,\Delta_-,1;1)\delta(\Omega-\Omega^{'\text{P}})&=&{}_2F_1(\Delta_+,\Delta_-,1;1)\sum_{\ell,m}(-1)^\ell Y_{\ell,m}(\Omega)Y_{\ell,m}^*(\Omega')\nn\\&=&\sum_{\ell,m}(-1)^\ell P_\ell(-1)Y_{\ell,m}(\Omega)Y_{\ell,m}^*(\Omega')\nn\\&=&\sum_{\ell,m}P_\ell(1)Y_{\ell,m}(\Omega)Y_{\ell,m}^*(\Omega')\nn\\&=&\sum_{\ell,m}Y_{\ell,m}(\Omega)Y_{\ell,m}^*(\Omega')\nn\\&=&\delta(\Omega-\Omega').\label{F21}
\eea We have used the completeness relation of the spherical Harmonic function 
\be 
\sum_{\ell,m}Y_{\ell,m}(\Omega)Y_{\ell,m}^*(\Omega')=\delta(\Omega-\Omega')
\ee and the parity of the Legendre polynomial
\be P_\ell(-z)=(-1)^\ell P_\ell(z)
\ee as well as the special value 
\be 
P_\ell(1)=1.
\ee 
The operator ${}_2F_1(\Delta_+,\Delta_-,1;1)$ is rather interesting since it transforms the Dirac delta function $\delta(\Omega-\Omega^{'\text{P}})$ to an alternative Dirac delta function $\delta(\Omega-\Omega')$. One can extend the operator to 
${}_2F_1(\Delta_+,\Delta_-,1;\frac{1-z}{2})$ such that 
\bea 
{}_2F_1(\Delta_+,\Delta_-,1;\frac{1-z}{2})\delta(\Omega-\Omega')=\frac{1}{2\pi}\delta(z-\cos\gamma(\Omega-\Omega')).
\eea 
%identity \eqref{F21} is rather amusing since 

Similarly, one can also define the associated Legendre function through the Hypergeometric function
\be 
P_\ell^m(z)=\frac{1}{\Gamma(1-m)}\left(\frac{1+z}{1-z}\right)^{m/2}{}_2F_1(\ell+1,-\ell,1-m;\frac{1-z}{2}).
\ee Therefore, we find 
\bea 
{}_2F_1(\Delta_+,\Delta_-,2;z)\delta(\Omega-\Omega^{'\text{P}})&=&{}_2F_1(\Delta_+,\Delta_-,2;z)\sum_{\ell,m}(-1)^\ell Y_{\ell,m}(\Omega)Y_{\ell,m}^*(\Omega')\nn\\&=&\sqrt{\frac{1-z}{z}}\sum_{\ell,m}P_\ell^{-1}(1-2z)(-1)^\ell Y_{\ell,m}(\Omega)Y^*_{\ell,m}(\Omega').
\eea As $z\to 1$, we have the limit
\be 
\lim_{z\to 1}\sqrt{\frac{1-z}{z}}P_\ell^{-1}(1-2z)=\delta_{\ell,0}.
\ee Therefore, we obtain 
\bea 
{}_2F_1(\Delta_+,\Delta_-,2;1)\delta(\Omega-\Omega^{'\text{P}})=\frac{1}{4\pi}.\label{delta4}
\eea 
We can also use the identity 
\bea 
{}_2F_1(a,1-a,2;1)=\frac{\sin a\pi}{\pi a(1-a)}
\eea and then 
\bea 
{}_2F_1(\Delta_+,\Delta_-,2,1)\delta(\Omega-\Omega^{'\text{P}})&=&\sum_{\ell,m}\frac{\sin(\ell+1)\pi}{\pi (\ell+1)(-\ell)}(-1)^\ell Y_{\ell,m}(\Omega)Y^*_{\ell,m}(\Omega').
\eea Only the $\ell=0$ mode has non-trivial contribution, and we can use the limit 
\be \lim_{x\to 0} \frac{\sin\pi x}{\pi x}=1
\ee to obtain 
\be 
{}_2F_1(\Delta_+,\Delta_-,2,1)\delta(\Omega-\Omega^{'\text{P}})=\frac{1}{4\pi}.
\ee The above result is consistent with \eqref{delta4}. Therefore, the divergent part of the boundary-to-boundary propagator becomes
\bea 
B^{\text{div}}(u,\Omega;u',\Omega')&=&-\frac{1}{2}(\frac{u-u'}{\beta}+\frac{i}{2})\delta(\Omega-\Omega')+\frac{r'}{4\pi\beta}.
\eea 
Note that the first term should be understood as \bea 
-\frac{1}{2}(\frac{u-u'}{\beta}+\frac{i}{2})=-\frac{1}{4\pi}\log e^{\frac{2\pi}{\beta}(u-u'+\frac{i}{2}\beta)}
\eea to satisfy the KMS symmetry. 
The total boundary-to-boundary propagator is 
\bea 
B(u,\Omega;u',\Omega')&=&\frac{1}{4\pi}\int_{\mathcal C}\frac{d\omega}{\omega}n(\omega)e^{i\omega(u-u')}\delta(\Omega-\Omega')-\frac{1}{2}(\frac{u-u'}{\beta}+\frac{i}{2})\delta(\Omega-\Omega')+\frac{r'}{4\pi\beta}\nn\\&=&-\frac{1}{4\pi}\log \left(1-e^{\frac{2\pi}{\beta}(u-u'-i\epsilon)}\right)\delta(\Omega-\Omega')+\frac{r'}{4\pi\beta}\nn\\&=&\frac{1}{4\pi}\int_{\mathcal C}\frac{d\omega}{\omega}(1+n(\omega))e^{-i\omega(u-u'-i\epsilon)}\delta(\Omega-\Omega')+\frac{r'}{4\pi\beta}.\label{divBu}
\eea 

To remove the divergence, we should consider the derivative of the boundary field with respect to time. For example,  
\bs\begin{align}
\partial_u B(u,\Omega;u',\Omega')&=-\frac{i}{4\pi}\int_{\mathcal C}d\omega (1+n(\omega))e^{-i\omega(u-u'-i\epsilon)}\delta(\Omega-\Omega'),\\
\partial_{u'} B(u,\Omega;u',\Omega')&=\frac{i}{4\pi}\int_{\mathcal C}d\omega (1+n(\omega))e^{-i\omega(u-u'-i\epsilon)}\delta(\Omega-\Omega')=-\partial_u B(u,\Omega;u',\Omega').
\end{align}\es The divergent part of  \eqref{divBu} is contributed by the soft mode, its effect has been removed by considering the time derivative.
 
\section{Step function}\label{step}
The step function is defined as 
\be 
\theta(x)=\left\{\begin{array}{cc}1,&x>0,\\ 0,&x<0\end{array}\right.
\ee  
and  related to the Dirac delta function through differentiation
\be 
\delta(x)=\frac{d\theta(x)}{dx}.
\ee 
The summation of the step functions whose arguments are opposite is equal to one
\be 
\theta(x)+\theta(-x)=1.
\ee
One can also construct the sign function through their difference
\be
s(x)=\theta(x)-\theta(-x).
\ee  
When the argument is a product, we have
\bs\begin{align}  
\theta(x y)&=\theta(x)\theta(y)+\theta(-x)\theta(-y),\\
s(x y)&=s(x)s(y).
\end{align}
\es The non-vanishing products 
$\theta(\pm \alpha_1)\theta(\pm\alpha_2)\theta(\pm\alpha_3)\theta(\pm \alpha_4)$ are
\bs\label{identheta}\begin{align}
 &\theta(\alpha_1)\theta(\alpha_2)\theta(\alpha_3)\theta(\alpha_4)=\theta(z-1),\\
  & \theta(-\alpha_1)\theta(\alpha_2)\theta(-\alpha_3)\theta(\alpha_4)=\theta(-z).
\end{align}\es 

\iffalse 
\bs\label{identheta}\begin{align}
    &\theta(\alpha_1)\theta(\alpha_2)\theta(\alpha_3)\theta(\alpha_4)=\theta(z-1),\\ & \theta(-\alpha_1)\theta(-\alpha_2)\theta(-\alpha_3)\theta(-\alpha_4)=0,\\
    &\theta(\alpha_1)\theta(\alpha_2)\theta(-\alpha_3)\theta(-\alpha_4)=0,\\
   & \theta(-\alpha_1)\theta(-\alpha_2)\theta(\alpha_3)\theta(\alpha_4)=0,\\
   & \theta(\alpha_1)\theta(-\alpha_2)\theta(\alpha_3)\theta(-\alpha_4)=0,\\
    & \theta(\alpha_1)\theta(-\alpha_2)\theta(-\alpha_3)\theta(\alpha_4)=0,\\
     & \theta(-\alpha_1)\theta(\alpha_2)\theta(\alpha_3)\theta(-\alpha_4)=0,\\
     & \theta(-\alpha_1)\theta(\alpha_2)\theta(-\alpha_3)\theta(\alpha_4)=\theta(-z),\\
   &\theta(\alpha_1)\theta(\alpha_2)\theta(\alpha_3)\theta(-\alpha_4)=0,\\
    &\theta(\alpha_1)\theta(\alpha_2)\theta(-\alpha_3)\theta(\alpha_4)=0,\\
     &\theta(\alpha_1)\theta(-\alpha_2)\theta(\alpha_3)\theta(\alpha_4)=0,\\
      &\theta(-\alpha_1)\theta(\alpha_2)\theta(\alpha_3)\theta(\alpha_4)=0,\\
       &\theta(-\alpha_1)\theta(-\alpha_2)\theta(-\alpha_3)\theta(\alpha_4)=0,\\
    &\theta(-\alpha_1)\theta(-\alpha_2)\theta(\alpha_3)\theta(-\alpha_4)=0,\\
     &\theta(-\alpha_1)\theta(\alpha_2)\theta(-\alpha_3)\theta(-\alpha_4)=0,\\
      &\theta(\alpha_1)\theta(-\alpha_2)\theta(-\alpha_3)\theta(-\alpha_4)=0.
\end{align}\es \fi 
The functions $f$  can be constructed from the step function and the sign function 
\bs\begin{align}
    f_0&=\theta(\alpha_1\omega)\theta(\alpha_2\omega)\theta(\alpha_3\omega)\theta(\alpha_4\omega)-\theta(\alpha_1\omega)\theta(\alpha_2\omega)\theta(-\alpha_3\omega)\theta(-\alpha_4\omega),\\
    f_1&=s(\alpha_1\omega)\theta(\alpha_2\omega)\theta(\alpha_3\omega)\theta(\alpha_4\omega)-s(\alpha_1\omega)\theta(\alpha_2\omega)\theta(-\alpha_3\omega)\theta(-\alpha_4\omega),\\
    f_2&=\theta(\alpha_1\omega)s(\alpha_2\omega)\theta(\alpha_3\omega)\theta(\alpha_4\omega)-\theta(\alpha_1\omega)s(\alpha_2\omega)\theta(-\alpha_3\omega)\theta(-\alpha_4\omega),\\
    f_3&=\theta(\alpha_1\omega)\theta(\alpha_2\omega)s(\alpha_3\omega)\theta(\alpha_4\omega)+\theta(\alpha_1\omega)\theta(\alpha_2\omega)s(\alpha_3\omega)\theta(-\alpha_4\omega)=\theta(\alpha_1\omega)\theta(\alpha_2\omega)s(\alpha_3\omega),\\
    f_{4}&=\theta(\alpha_1\omega)\theta(\alpha_2\omega)\theta(\alpha_3\omega)s(\alpha_4\omega)+\theta(\alpha_1\omega)\theta(\alpha_2\omega)\theta(-\alpha_3\omega)s(\alpha_4\omega)=\theta(\alpha_1\omega)\theta(\alpha_2\omega)s(\alpha_4\omega),\\
    f_{12}&=s(\alpha_1\omega)s(\alpha_2\omega)\theta(\alpha_3\omega)\theta(\alpha_4\omega)-s(\alpha_1\omega)s(\alpha_2\omega)\theta(-\alpha_3\omega)\theta(-\alpha_4\omega),\\
    f_{13}&=s(\alpha_1\omega)\theta(\alpha_2\omega)s(\alpha_3\omega)\theta(\alpha_4\omega)+s(\alpha_1\omega)\theta(\alpha_2\omega)s(\alpha_3\omega)\theta(-\alpha_4\omega)=s(\alpha_1\omega)\theta(a_2\omega)s(\alpha_3\omega),\\
    f_{14}&=s(\alpha_1\omega)\theta(\alpha_2\omega)\theta(\alpha_3\omega)s(\alpha_4\omega)+s(\alpha_1\omega)\theta(\alpha_2\omega)\theta(-\alpha_3\omega)s(\alpha_4\omega)=s(\alpha_1\omega)\theta(a_2\omega)s(\alpha_4\omega),\\
    f_{23}&=\theta(\alpha_1\omega)s(\alpha_2\omega)s(\alpha_3\omega)\theta(\alpha_4\omega)+\theta(\alpha_1\omega)s(\alpha_2\omega)s(\alpha_3\omega)\theta(-\alpha_4\omega)=\theta(\alpha_1\omega)s(\alpha_2\omega)s(\alpha_3\omega),\\
    f_{24}&=\theta(\alpha_1\omega)s(\alpha_2\omega)\theta(\alpha_3\omega)s(\alpha_4\omega)+\theta(\alpha_1\omega)s(\alpha_2\omega)\theta(-\alpha_3\omega)s(\alpha_4\omega)=\theta(\alpha_1\omega)s(\alpha_2\omega)s(\alpha_4\omega),\\
    f_{34}&=\theta(\alpha_1\omega)\theta(\alpha_2\omega)s(\alpha_3\omega)s(\alpha_4\omega)-\theta(\alpha_1\omega)\theta(\alpha_2\omega)s(\alpha_3\omega)s(\alpha_4\omega)=0,\\
    f_{123}&=s(\alpha_1\omega)s(\alpha_2\omega)s(\alpha_3\omega)\theta(\alpha_4\omega)+s(\alpha_1\omega)s(\alpha_2\omega)s(\alpha_3\omega)\theta(-\alpha_4\omega)=s(\alpha_1\omega)s(\alpha_2\omega)s(\alpha_3\omega),\\
    f_{124}&=s(\alpha_1\omega)s(\alpha_2\omega)\theta(\alpha_3\omega)s(\alpha_4\omega)+s(\alpha_1\omega)s(\alpha_2\omega)\theta(-\alpha_3\omega)s(\alpha_4\omega)=s(\alpha_1 \omega)s(\alpha_2 \omega)s(\alpha_4 \omega),\\
    f_{134}&=s(\alpha_1\omega)\theta(\alpha_2\omega)s(\alpha_3\omega)s(\alpha_4\omega)-s(\alpha_1\omega)\theta(\alpha_2\omega)s(\alpha_3\omega)s(\alpha_4\omega)=0,\\
    f_{234}&=\theta(\alpha_1\omega)s(\alpha_2\omega)s(\alpha_3\omega)s(\alpha_4\omega)-\theta(\alpha_1\omega)s(\alpha_2\omega)s(\alpha_3\omega)s(\alpha_4\omega)=0,\\
    f_{1234}&=s(\alpha_1\omega)s(\alpha_2\omega)s(\alpha_3\omega)s(\alpha_4\omega)-s(\alpha_1\omega)s(\alpha_2\omega)s(\alpha_3\omega)s(\alpha_4\omega)=0.
\end{align}\es 
\section{Barnes zeta function}\label{Barnes}
Barnes zeta function is direct generalization of Riemann zeta function  with multiple variables \cite{1901RSPTA.196..265B}. The property of Barnes zeta function can be found in \cite{Ruijsenaars2000OnBM,komori2010barnesmultiplezetafunctionsramanujans} and it has been reviewed in the book \cite{barneszeta}. Let the real part of $x$ and $w_i,\ i=1,2,\cdots,r$  be positive numbers, the Dirichlet series of the Barnes zeta function is
\bea 
\zeta_r(c;x;w_1,w_2,\cdots,w_r)=\sum_{m_1=0}^\infty \sum_{m_2=0}^\infty\cdots\sum_{m_r=0}^\infty (x+m_1w_1+m_2w_2+\cdots+m_rw_r)^{-c}.
\eea The series is convergent for $\text{Re}\ c>r$. For $r=1$, the series is proportional to the
Hurwitz zeta function
\be 
\zeta(c;x)=\sum_{m=0}^\infty(x+m)^{-c}.
\ee To be more precise, we have 
\bea 
\zeta_1(c;x;w)=\sum_{m=0}^\infty (x+m w)^{-c}=w^{-c}\zeta(c;\frac{x}{w}).
\eea
Consider the scaling transformation $w_i\to \lambda\omega_i,\ \lambda>0$, we have 
\be 
\zeta_r(c;x;\lambda w_1,\lambda w_2,\cdots,\lambda w_n)=\lambda^{-c}\zeta_r(c;\frac{x}{\lambda};w_1,w_2,\cdots,w_r).\label{scaling}
\ee 

By using the integral representation of the Gamma function 
\be 
\Gamma(s)=\int_0^\infty dx e^{-x}x^{s-1},\quad \text{Re}\ s>1, 
\ee the Barnes zeta function can be expressed as the following integral 
\bea 
\zeta_r(c;x;w_1,w_2,\cdots,w_r)&=&\frac{1}{\Gamma(c)}\int_0^\infty dt\ t^{c-1}\frac{e^{-xt}}{\prod_{j=1}^r (1-e^{-w_j t})}\nn\\&=&\frac{1}{\Gamma(c)}\int_0^\infty dt t^{ c-1}\frac{e^{(w_1+w_2+\cdots+w_r-x)t}}{\prod_{j=1}^r (e^{w_j t}-1)}.
\eea Using contour integrals, one can extend the Barnes zeta function to the whole $c$ plane except for the points $c=1,2,\cdots,r$ where the function is singular. Using the definition of the occupation number, we find 
\bea 
\zeta_r(c;x;w_1,w_2,\cdots,w_r)=\frac{1}{\Gamma(c)}\int_0^\infty dt\ t^{c-1} e^{-xt}\prod_{j=1}^r (1+n(w_j/\beta)).
\eea 
The generating function of the Bernoulli numbers is 
\be 
\frac{t}{e^t-1}=\sum_{m=0}^\infty 
B_m\frac{t^m}{m!},\label{bernoulli2}
\ee where the first few Bernoulli numbers are 
\be 
B_0=1,\quad B_1=-\frac{1}{2},\quad B_2=\frac{1}{12},\quad B_3=-\frac{1}{720},\cdots.
\ee
They are the special value of the Bernoulli polynomials $B_n(x)$ 
\be 
B_n=B_n(0).
\ee The Bernoulli polynomials are defined through the generating function 
\bea 
\frac{t e^{x t}}{e^t-1}=\sum_{n=0}^\infty \frac{B_n(x)}{n!}t^n.
\eea 
We set $x=1$ and then 
\bea 
\frac{t e^t}{e^t-1}=t+\frac{t}{e^t-1}=t+\sum_{n=0}^\infty \frac{B_n(0)}{n!} t^n=\sum_{n=0}^\infty \frac{B_n(1)}{n!}t^n.
\eea Therefore, 
\be 
B_n(1)=B_n(0)+\delta_{n,1}.
\ee 
In general, $n$-th Bernoulli polynomial is a polynomial of degree $n$.
By rescaling $t\to w t$, we find 
\be 
\frac{t e^{xw t}}{e^{w t}-1}=\sum_{n=0}^\infty \frac{B_{n}(x)}{n!} w^{n-1} t^n.\label{bernoulli}
\ee 
Now we compute the product of the generating function \eqref{bernoulli} and define  generalized Bernoulli numbers
\bea 
\frac{t^r e^{(\sum_{j=1}^r w_j)x t}}{\prod_{j=1}^r (e^{w_j t}-1)}=\sum_{n=0}^\infty\frac{ B_{r,n}(x;w_1,w_2,\cdots,w_r)}{n!}t^n
\eea where 
\bea 
B_{r,n}(x;w_1,w_2,\cdots,w_r)=\sum_{m_1+m_2+\cdots+m_r=n} \frac{n!}{m_1!m_2!\cdots m_r!}\prod_{j=1}^r B_{m_j}(x)w_j^{m_j-1}.
\eea When $x=1$ and $n=0$, the special value is
\bea 
B_{r,0}(1;w_1,w_2,\cdots,w_r)=\left(B_0(1)\right)^3 w_1^{-1}w_2^{-1}w_3^{-1}=\frac{1}{w_1w_2w_3}.
\eea 
Therefore, we can find another series for the Barnes zeta function
\bea 
\zeta_r(c;x;w_1,w_2,\cdots,w_r)&=&\frac{1}{\Gamma(c)}\int_0^\infty dt\ t^{c-r-1}e^{-x t}\sum_{n=0}^\infty \frac{B_{r,n}(1;w_1,w_2,\cdots,w_r)}{n!}t^n\nn\\&=&\frac{1}{\Gamma(c)}\sum_{n=0}^\infty \frac{\Gamma(c-r+n)B_{r,n}(1;w_1,w_2,\cdots,w_r)}{n!x^{c-r+n}}.\label{seriesB}%\nn\\&=&\sum_{m_1=0}^\infty\sum_{m_2=0}^\infty \cdots\sum_{m_r=0}^\infty \frac{B_{m_1}B_{m_2}\cdots B_{m_r}}{m_1!m_2!\cdots m_r!}w_1^{m_1}w_2^{m_2}\cdots w_r^{m_r}t^{m_1+m_2+\cdots+m_r}\nn\\&=&\sum_{m_1=0}^\infty\sum_{m_2=0}^\infty\cdots\sum_{m_r=0}^\infty \frac{B_{m_1m_2\cdots m_r}(c;w_1,w_2,\cdots,w_r)}{x^{c-r+(m_1+m_2+\cdots+m_r)}}.
\eea The last step requires $\text{Re}\ x>0$. \iffalse We may also use compute the products of \eqref{bernoulli2} to define an alternative multi-variable Bernoulli polynomials
\bea 
\frac{t^r}{\prod_{j=1}^r (e^{w_j t}-1)}=\sum_{n=0}^\infty \frac{\mathcal B_{r,n}(w_1,w_2,\cdots,w_r)}{n!} t^n
\eea 
where 
\bea 
\mathcal B_{r,n}(w_1,w_2,\cdots,w_r)=\sum_{m_1+m_2+\cdots+m_r=n} \frac{n!}{m_1!m_2!\cdots m_r!}\prod_{j=1}^r B_{m_j}w_j^{m_j-1}.
\eea 
Therefore, the Barnes zeta function becomes
\bea 
\zeta_r(c;x;w_1,w_2,\cdots,w_r)=\frac{1}{\Gamma(c)}\int_0^\infty t^{c-r-1}e^{}
\eea

\bea 
B_{m_1m_2\cdots m_r}(c;w_1,w_2,\cdots,w_r)=\frac{\Gamma(c-r+m_1+m_2+\cdots+m_r)}{\Gamma(c)}\frac{B_{m_1}B_{m_2}\cdots B_{m_r}}{m_1!m_2!\cdots m_r!}w_1^{m_1}w_2^{m_2}\cdots w_r^{m_r}.
\eea  \fi

In the context, we are interested in the following integrals 
\bea 
I(c;\chi;b_1,b_2,\cdots,b_r)=\int_0^\infty d\omega \omega^c e^{-i\omega\chi} \prod_j n(b_j \omega)
\eea where $b_1,b_2,\cdots$ are positive numbers. Using the geometric series
\be 
n(b_j\omega)=\sum_{m=1}^\infty e^{-m\beta b_j\omega}=\sum_{m=0}^\infty e^{-(m+1)\beta b_j\omega},
\ee we obtain
\bea 
&&I(c;\chi;b_1,b_2,\cdots,b_r)\nn\\&=&\int_0^\infty d\omega \omega^c e^{-i\omega \chi} \sum_{m_1=0}^\infty\sum_{m_2=0}^\infty\cdots\sum_{m_r=0}^\infty e^{-\sum_{j=1}^r(m_j+1)b_j\beta\omega}\nn\\&=&\sum_{m_1=0}^\infty\sum_{m_2=0}^\infty\cdots\sum_{m_r=0}^\infty\int_0^\infty d\omega \omega^c e^{-(\sum_{j=1}^r (m_j+1)b_j \beta\omega+i\chi)}\nn\\&=&\Gamma(c+1)\sum_{m_1=0}^\infty\sum_{m_2=0}^\infty\cdots\sum_{m_r=0}^\infty (i\chi+\sum_{j=1}^r (m_j+1){ \beta}b_j)^{-c-1}\nn\\&=&\Gamma(c+1)\zeta_r(c+1;\beta(b_1+b_2+\cdots+b_r)+i\chi;\beta b_1,\beta b_2,\cdots,\beta b_r).
\eea 
In the low temperature expansion, we use the scaling law \eqref{scaling} to obtain 
\bea 
&&I(c;\chi;b_1,b_2,\cdots,b_r)=\Gamma(c+1)\beta^{-c-1}\zeta_r(c+1;b_1+b_2+\cdots+b_r+\frac{i\chi}{\beta};b_1,b_2,\cdots,b_r).\label{hightem}
\eea  
\iffalse 
In the high temperature expansion, we need the series expansion \eqref{seriesB}
\bea 
&&I(c;\chi;b_1,b_2,\cdots,b_r)\nn\\&=&\Gamma(c+1)\zeta_r(c+1;\beta(b_1+b_2+\cdots+b_r)+i\chi;\beta b_1,\beta b_2,\cdots,\beta b_r)\nn\\&=&\sum_{n=0}^\infty \frac{\Gamma(c+1-r+n)B_{r,n}(1;\beta b_1,\beta b_2,\cdots,\beta b_r)}{n!\left(\beta(b_1+b_2+\cdots+b_r)+i\chi\right)^{c+1-r+n}}.
\eea 
We choose $c=r=3$, then 
\bea 
I(3;\chi;b_1,b_2,\cdots,b_r)=\sum_{n=0}^\infty\frac{ B_{3,n}(1;\beta b_1,\beta b_2,\beta b_3)}{(\beta(b_1+b_2+b_3)+i\chi)^{n+1}}
\eea The leading order term is 
\bea 
I(3;\chi;b_1,b_2,b_3)=\frac{B_{3,0}(1;\beta b_1,\beta b_2,\beta b_3)}{\beta(b_1+b_2+b_3)+i\chi}=B_{3,0}(1;\beta b_1,\beta b_2,\beta b_3)\Big[\frac{1}{\beta(b_1+b_2+b_3)}-i\pi\delta(\chi)\Big].
\eea 
\fi 
To obtain high temperature limit, we can use the expansion of the occupation number by using Bernoulli numbers 
\bea 
n(\omega)=\sum_{n=0}^\infty \frac{B_n(0)}{n!}(\beta\omega)^{n-1},\quad 1+n(\omega)=\sum_{n=0}^\infty \frac{B_n(1)}{n!}(\beta \omega)^{n-1}.
\eea 
This can be understood as a high temperature expansion of the occupation number. 
Therefore, we define the following polynomials 
\bea 
\prod_{j=1}^{r_1} (1+n(b_j\omega))\prod_{k=r_1+1}^{r_1+r_2} n(b_k\omega)=(\beta\omega)^{-r_1-r_2}b_1^{-1}\cdots b_{r_1+r_2}^{-1}\sum_{n=0}^\infty \frac{P_{r_1,r_2,n}(b_1,b_2,\cdots,b_{r_1+r_2})}{n!}(\beta\omega)^{n}.\label{expansionn}
\eea 
In particular, we have 
\bea 
P_{1,0,n}(b)=B_n(1)b^n,\quad P_{0,1,n}(b)=B_n(0)b^n.
\eea In general, the polynomial $P$
is 
\bea 
P_{r_1,r_2,n}(b_1,b_2,\cdots,b_{r_1+r_2})=\sum'_{m_1,m_2,\cdots,m_{r_1+r_2}}\frac{n!}{m_1!m_2!\cdots m_{r_1+r_2}!}\left(\prod_{j=1}^{r_1}B_{m_j}(1)\right)\left(\prod_{j=r_1+1}^{r_1+r_2}B_{m_j}(0)\right)\left( \prod_{i=1}^{r_1+r_2} b_i^{m_i}\right).\nn\\
\eea The summation is over all possible non-negative integers of $m_1,m_2,\cdots,m_{r_1+r_2}$ with summation fixed to $n$
\bea 
\sum'_{m_1,m_2,\cdots,m_{r_1+r_2}}(\cdots)=\sum_{m_1=0}^\infty \sum_{m_2=0}^\infty\cdots\sum_{m_{r_1+r_2}=0}^\infty  \delta_{n,m_1+m_2+\cdots+m_{r_1+r_2}}(\cdots).
\eea 
The polynomial $P$  is homogeneous with
\bea 
P_{r_1,r_2,n}(\lambda b_1,\lambda b_2,\cdots,\lambda  b_{r_1+r_2})=\lambda^n P_{r_1,r_2,n}(b_1,b_2,\cdots,b_{r_1+r_2}).
\eea 

When $n=0$, we always have 
\be 
P_{r_1,r_2,0}(b_1,b_2,\cdots,b_{r_1+r_2})=1.
\ee

\bibliography{refs}

\providecommand{\href}[2]{#2}\begingroup\raggedright\begin{thebibliography}{100}

\bibitem{Bekenstein:1973ur}
J.~D. Bekenstein, ``{Black holes and entropy},'' {\em Phys. Rev. D} {\bf 7}
  (1973) 2333--2346.

\bibitem{Hawking:1975vcx}
S.~W. Hawking, ``{Particle Creation by Black Holes},'' {\em Commun. Math.
  Phys.} {\bf 43} (1975) 199--220. [Erratum: Commun.Math.Phys. 46, 206 (1976)].

\bibitem{Bardeen:1973gs}
J.~M. Bardeen, B.~Carter, and S.~W. Hawking, ``{The Four laws of black hole
  mechanics},'' {\em Commun. Math. Phys.} {\bf 31} (1973) 161--170.

\bibitem{Unruh:1976db}
W.~G. Unruh, ``{Notes on black hole evaporation},'' {\em Phys. Rev. D} {\bf 14}
  (1976) 870.

\bibitem{1993gr.qc....10026T}
G.~{'t Hooft}, ``{Dimensional Reduction in Quantum Gravity},'' {\em arXiv
  e-prints} (Oct., 1993) gr--qc/9310026,
  \href{http://www.arXiv.org/abs/gr-qc/9310026}{{\tt gr-qc/9310026}}.

\bibitem{Susskind:1994vu}
L.~Susskind, ``{The World as a hologram},'' {\em J. Math. Phys.} {\bf 36}
  (1995) 6377--6396, \href{http://www.arXiv.org/abs/hep-th/9409089}{{\tt
  hep-th/9409089}}.

\bibitem{Maldacena:1997re}
J.~M. Maldacena, ``{The Large N limit of superconformal field theories and
  supergravity},'' {\em Adv. Theor. Math. Phys.} {\bf 2} (1998) 231--252,
  \href{http://www.arXiv.org/abs/hep-th/9711200}{{\tt hep-th/9711200}}.

\bibitem{Polchinski:1999ry}
J.~Polchinski, ``{S matrices from AdS space-time},''
  \href{http://www.arXiv.org/abs/hep-th/9901076}{{\tt hep-th/9901076}}.

\bibitem{Susskind:1998vk}
L.~Susskind, ``{Holography in the flat space limit},'' {\em AIP Conf. Proc.}
  {\bf 493} (1999), no.~1, 98--112,
  \href{http://www.arXiv.org/abs/hep-th/9901079}{{\tt hep-th/9901079}}.

\bibitem{Giddings:1999jq}
S.~B. Giddings, ``{Flat space scattering and bulk locality in the AdS / CFT
  correspondence},'' {\em Phys. Rev. D} {\bf 61} (2000) 106008,
  \href{http://www.arXiv.org/abs/hep-th/9907129}{{\tt hep-th/9907129}}.

\bibitem{deBoer:2003vf}
J.~de~Boer and S.~N. Solodukhin, ``{A Holographic reduction of Minkowski
  space-time},'' {\em Nucl. Phys. B} {\bf 665} (2003) 545--593,
  \href{http://www.arXiv.org/abs/hep-th/0303006}{{\tt hep-th/0303006}}.

\bibitem{Arcioni:2003xx}
G.~Arcioni and C.~Dappiaggi, ``{Exploring the holographic principle in
  asymptotically flat space-times via the BMS group},'' {\em Nucl. Phys. B}
  {\bf 674} (2003) 553--592,
  \href{http://www.arXiv.org/abs/hep-th/0306142}{{\tt hep-th/0306142}}.

\bibitem{Arcioni:2003td}
G.~Arcioni and C.~Dappiaggi, ``{Holography in asymptotically flat space-times
  and the BMS group},'' {\em Class. Quant. Grav.} {\bf 21} (2004) 5655,
  \href{http://www.arXiv.org/abs/hep-th/0312186}{{\tt hep-th/0312186}}.

\bibitem{Mann:2005yr}
R.~B. Mann and D.~Marolf, ``{Holographic renormalization of asymptotically flat
  spacetimes},'' {\em Class. Quant. Grav.} {\bf 23} (2006) 2927--2950,
  \href{http://www.arXiv.org/abs/hep-th/0511096}{{\tt hep-th/0511096}}.

\bibitem{Pasterski:2016qvg}
S.~Pasterski, S.-H. Shao, and A.~Strominger, ``{Flat Space Amplitudes and
  Conformal Symmetry of the Celestial Sphere},'' {\em Phys. Rev. D} {\bf 96}
  (2017), no.~6, 065026, \href{http://www.arXiv.org/abs/1701.00049}{{\tt
  1701.00049}}.

\bibitem{Pasterski:2017kqt}
S.~Pasterski and S.-H. Shao, ``{Conformal basis for flat space amplitudes},''
  {\em Phys. Rev. D} {\bf 96} (2017), no.~6, 065022,
  \href{http://www.arXiv.org/abs/1705.01027}{{\tt 1705.01027}}.

\bibitem{Pasterski:2017ylz}
S.~Pasterski, S.-H. Shao, and A.~Strominger, ``{Gluon Amplitudes as 2d
  Conformal Correlators},'' {\em Phys. Rev. D} {\bf 96} (2017), no.~8, 085006,
  \href{http://www.arXiv.org/abs/1706.03917}{{\tt 1706.03917}}.

\bibitem{Donnay:2022aba}
L.~Donnay, A.~Fiorucci, Y.~Herfray, and R.~Ruzziconi, ``{Carrollian Perspective
  on Celestial Holography},'' {\em Phys. Rev. Lett.} {\bf 129} (2022), no.~7,
  071602, \href{http://www.arXiv.org/abs/2202.04702}{{\tt 2202.04702}}.

\bibitem{Bagchi:2022emh}
A.~Bagchi, S.~Banerjee, R.~Basu, and S.~Dutta, ``{Scattering Amplitudes:
  Celestial and Carrollian},'' {\em Phys. Rev. Lett.} {\bf 128} (2022), no.~24,
  241601, \href{http://www.arXiv.org/abs/2202.08438}{{\tt 2202.08438}}.

\bibitem{Une}
J.~M. L\'evy-Leblond, ``{Une nouvelle limite non-relativiste du groupe de
  Poincar\'e},'' {\em Ann. Inst. H Poincar\'e} {\bf 3} (1965), no.~1, 1--12.

\bibitem{Gupta1966OnAA}
N.~Gupta, ``On an analogue of the galilei group,'' {\em Nuovo Cimento Della
  Societa Italiana Di Fisica A-nuclei Particles and Fields} {\bf 44} (1966)
  512--517.

\bibitem{Duval:2014uva}
C.~Duval, G.~W. Gibbons, and P.~A. Horvathy, ``{Conformal Carroll groups and
  BMS symmetry},'' {\em Class. Quant. Grav.} {\bf 31} (2014) 092001,
  \href{http://www.arXiv.org/abs/1402.5894}{{\tt 1402.5894}}.

\bibitem{Duval:2014lpa}
C.~Duval, G.~W. Gibbons, and P.~A. Horvathy, ``{Conformal Carroll groups},''
  {\em J. Phys. A} {\bf 47} (2014), no.~33, 335204,
  \href{http://www.arXiv.org/abs/1403.4213}{{\tt 1403.4213}}.

\bibitem{Duval:2014uoa}
C.~Duval, G.~W. Gibbons, P.~A. Horvathy, and P.~M. Zhang, ``{Carroll versus
  Newton and Galilei: two dual non-Einsteinian concepts of time},'' {\em Class.
  Quant. Grav.} {\bf 31} (2014) 085016,
  \href{http://www.arXiv.org/abs/1402.0657}{{\tt 1402.0657}}.

\bibitem{Barnich:2010eb}
G.~Barnich and C.~Troessaert, ``{Aspects of the BMS/CFT correspondence},'' {\em
  JHEP} {\bf 05} (2010) 062, \href{http://www.arXiv.org/abs/1001.1541}{{\tt
  1001.1541}}.

\bibitem{Sachs:1962wk}
R.~K. Sachs, ``{Gravitational waves in general relativity. 8. Waves in
  asymptotically flat space-times},'' {\em Proc. Roy. Soc. Lond. A} {\bf 270}
  (1962) 103--126.

\bibitem{Campiglia:2014yka}
M.~Campiglia and A.~Laddha, ``{Asymptotic symmetries and subleading soft
  graviton theorem},'' {\em Phys. Rev. D} {\bf 90} (2014), no.~12, 124028,
  \href{http://www.arXiv.org/abs/1408.2228}{{\tt 1408.2228}}.

\bibitem{Campiglia:2015yka}
M.~Campiglia and A.~Laddha, ``{New symmetries for the Gravitational
  S-matrix},'' {\em JHEP} {\bf 04} (2015) 076,
  \href{http://www.arXiv.org/abs/1502.02318}{{\tt 1502.02318}}.

\bibitem{1978JMP....19.1542A}
A.~{Ashtekar} and R.~O. {Hansen}, ``{A unified treatment of null and spatial
  infinity in general relativity. I. Universal structure, asymptotic
  symmetries, and conserved quantities at spatial infinity.},'' {\em Journal of
  Mathematical Physics} {\bf 19} (July, 1978) 1542--1566.

\bibitem{Ashtekar:1981sf}
A.~Ashtekar, ``{Asymptotic Quantization of the Gravitational Field},'' {\em
  Phys. Rev. Lett.} {\bf 46} (1981) 573--576.

\bibitem{Ashtekar:1981bq}
A.~Ashtekar and M.~Streubel, ``{Symplectic Geometry of Radiative Modes and
  Conserved Quantities at Null Infinity},'' {\em Proc. Roy. Soc. Lond. A} {\bf
  376} (1981) 585--607.

\bibitem{Ashtekar:1987tt}
A.~Ashtekar, {\em {Asymptotic Quantization: Based on 1984 Naples Lectures}}.
\newblock Bibliopolis, 1987.

\bibitem{Liu:2022mne}
W.-B. Liu and J.~Long, ``{Symmetry group at future null infinity: Scalar
  theory},'' {\em Phys. Rev. D} {\bf 107} (2023), no.~12, 126002,
  \href{http://www.arXiv.org/abs/2210.00516}{{\tt 2210.00516}}.

\bibitem{Liu:2023qtr}
W.-B. Liu and J.~Long, ``{Symmetry group at future null infinity II: Vector
  theory},'' {\em JHEP} {\bf 07} (2023) 152,
  \href{http://www.arXiv.org/abs/2304.08347}{{\tt 2304.08347}}.

\bibitem{Liu:2023gwa}
W.-B. Liu and J.~Long, ``{Symmetry group at future null infinity III:
  Gravitational theory},'' {\em JHEP} {\bf 10} (2023) 117,
  \href{http://www.arXiv.org/abs/2307.01068}{{\tt 2307.01068}}.

\bibitem{Li:2023xrr}
A.~Li, W.-B. Liu, J.~Long, and R.-Z. Yu, ``{Quantum flux operators for
  Carrollian diffeomorphism in general dimensions},'' {\em JHEP} {\bf 11}
  (2023) 140, \href{http://www.arXiv.org/abs/2309.16572}{{\tt 2309.16572}}.

\bibitem{Liu:2024nkc}
W.-B. Liu and J.~Long, ``{Holographic dictionary from bulk reduction},'' {\em
  Phys. Rev. D} {\bf 109} (2024), no.~6, L061901,
  \href{http://www.arXiv.org/abs/2401.11223}{{\tt 2401.11223}}.

\bibitem{Liu:2024rvz}
W.-B. Liu, J.~Long, and X.-H. Zhou, ``{Electromagnetic helicity flux operators
  in higher dimensions},'' \href{http://www.arXiv.org/abs/2407.20077}{{\tt
  2407.20077}}.

\bibitem{Guo:2024qzv}
S.-M. Guo, W.-B. Liu, and J.~Long, ``{Quantum flux operators in the fermionic
  theory and their supersymmetric extension},''
  \href{http://www.arXiv.org/abs/2412.20829}{{\tt 2412.20829}}.

\bibitem{Liu:2023jnc}
W.-B. Liu, J.~Long, and X.-H. Zhou, ``{Quantum flux operators in higher spin
  theories},'' {\em Phys. Rev. D} {\bf 109} (2024), no.~8, 086012,
  \href{http://www.arXiv.org/abs/2311.11361}{{\tt 2311.11361}}.

\bibitem{long2025gravitational}
J.~Long and R.-Z. Yu, ``Gravitational helicity flux density from two-body
  systems,'' {\em Classical and Quantum Gravity} {\bf 42} (2025), no.~4,
  045005.

\bibitem{Bagchi:2019xfx}
A.~Bagchi, A.~Mehra, and P.~Nandi, ``{Field Theories with Conformal Carrollian
  Symmetry},'' {\em JHEP} {\bf 05} (2019) 108,
  \href{http://www.arXiv.org/abs/1901.10147}{{\tt 1901.10147}}.

\bibitem{Baiguera:2022lsw}
S.~Baiguera, G.~Oling, W.~Sybesma, and B.~T. S\o{}gaard, ``{Conformal Carroll
  scalars with boosts},'' {\em SciPost Phys.} {\bf 14} (2023), no.~4, 086,
  \href{http://www.arXiv.org/abs/2207.03468}{{\tt 2207.03468}}.

\bibitem{saha2022intrinsic}
A.~Saha, ``Intrinsic approach to 1+ 1d carrollian conformal field theory,''
  {\em Journal of High Energy Physics} {\bf 2022} (2022), no.~12, 1--54.

\bibitem{Ciambelli:2019lap}
L.~Ciambelli, R.~G. Leigh, C.~Marteau, and P.~M. Petropoulos, ``{Carroll
  Structures, Null Geometry and Conformal Isometries},'' {\em Phys. Rev. D}
  {\bf 100} (2019), no.~4, 046010,
  \href{http://www.arXiv.org/abs/1905.02221}{{\tt 1905.02221}}.

\bibitem{Bagchi:2022eav}
A.~Bagchi, A.~Banerjee, S.~Dutta, K.~S. Kolekar, and P.~Sharma, ``{Carroll
  covariant scalar fields in two dimensions},''
  \href{http://www.arXiv.org/abs/2203.13197}{{\tt 2203.13197}}.

\bibitem{de2023carroll}
J.~de~Boer, J.~Hartong, N.~A. Obers, W.~Sybesma, and S.~Vandoren, ``Carroll
  stories,'' {\em Journal of High Energy Physics} {\bf 2023} (2023), no.~9,
  1--59.

\bibitem{gupta2021constructing}
N.~Gupta and N.~V. Suryanarayana, ``Constructing carrollian cfts,'' {\em
  Journal of High Energy Physics} {\bf 2021} (2021), no.~3, 1--23.

\bibitem{chen2023constructing}
B.~Chen, R.~Liu, H.~Sun, and Y.-f. Zheng, ``Constructing carrollian field
  theories from null reduction,'' {\em Journal of High Energy Physics} {\bf
  2023} (2023), no.~11, 1--52.

\bibitem{sharma2025studies}
A.~Sharma, ``Studies on carrollian quantum field theories,'' {\em arXiv
  preprint arXiv:2502.00487} (2025).

\bibitem{bekaert2024holographic}
X.~Bekaert, A.~Campoleoni, and S.~Pekar, ``Holographic carrollian conformal
  scalars,'' {\em Journal of High Energy Physics} {\bf 2024} (2024), no.~5,
  1--40.

\bibitem{banerjee2023one}
K.~Banerjee, R.~Basu, B.~Krishnan, S.~Maulik, A.~Mehra, and A.~Ray, ``One-loop
  quantum effects in carroll scalars,'' {\em Physical Review D} {\bf 108}
  (2023), no.~8, 085022.

\bibitem{ciambelli2024dynamics}
L.~Ciambelli, ``Dynamics of carrollian scalar fields,'' {\em Classical and
  Quantum Gravity} {\bf 41} (2024), no.~16, 165011.

\bibitem{bagchi2023magic}
A.~Bagchi, A.~Banerjee, R.~Basu, M.~Islam, and S.~Mondal, ``Magic fermions:
  Carroll and flat bands,'' {\em Journal of High Energy Physics} {\bf 2023}
  (2023), no.~3, 1--39.

\bibitem{banerjee2023carroll}
A.~Banerjee, S.~Dutta, and S.~Mondal, ``Carroll fermions in two dimensions,''
  {\em Physical Review D} {\bf 107} (2023), no.~12, 125020.

\bibitem{bergshoeff2024carroll}
E.~A. Bergshoeff, A.~Campoleoni, A.~Fontanella, L.~Mele, and J.~Rosseel,
  ``Carroll fermions,'' {\em SciPost Physics} {\bf 16} (2024), no.~6, 153.

\bibitem{hao2024bms}
P.-X. Hao, W.~Song, Z.~Xiao, and X.~Xie, ``Bms-invariant free fermion models,''
  {\em Physical Review D} {\bf 109} (2024), no.~2, 025002.

\bibitem{yu2023free}
Z.-f. Yu and B.~Chen, ``Free field realization of the bms ising model,'' {\em
  Journal of High Energy Physics} {\bf 2023} (2023), no.~8, 1--47.

\bibitem{ara2024flat}
N.~Ara, A.~Banerjee, R.~Basu, and B.~Krishnan, ``Flat bands and compact
  localised states: A carrollian roadmap,'' {\em arXiv preprint
  arXiv:2412.18965} (2024).

\bibitem{ekiz2025quantization}
E.~Ekiz, E.~O. Kahya, and U.~Zorba, ``Quantization of carrollian fermions,''
  {\em Physical Review D} {\bf 111} (2025), no.~10, 105019.

\bibitem{islam2023carrollian}
M.~Islam, ``Carrollian yang-mills theory,'' {\em Journal of High Energy
  Physics} {\bf 2023} (2023), no.~5, 1--39.

\bibitem{barducci2019vector}
A.~Barducci, R.~Casalbuoni, and J.~Gomis, ``Vector susy models with carroll or
  galilei invariance,'' {\em Physical Review D} {\bf 99} (2019), no.~4, 045016.

\bibitem{Rivera-Betancour:2022lkc}
D.~Rivera-Betancour and M.~Vilatte, ``{Revisiting the Carrollian scalar
  field},'' {\em Phys. Rev. D} {\bf 106} (2022), no.~8, 085004,
  \href{http://www.arXiv.org/abs/2207.01647}{{\tt 2207.01647}}.

\bibitem{Henneaux:2021yzg}
M.~Henneaux and P.~Salgado-Rebolledo, ``{Carroll contractions of
  Lorentz-invariant theories},'' {\em JHEP} {\bf 11} (2021) 180,
  \href{http://www.arXiv.org/abs/2109.06708}{{\tt 2109.06708}}.

\bibitem{Bagchi:2016bcd}
A.~Bagchi, R.~Basu, A.~Kakkar, and A.~Mehra, ``{Flat Holography: Aspects of the
  dual field theory},'' {\em JHEP} {\bf 12} (2016) 147,
  \href{http://www.arXiv.org/abs/1609.06203}{{\tt 1609.06203}}.

\bibitem{Bagchi:2019clu}
A.~Bagchi, R.~Basu, A.~Mehra, and P.~Nandi, ``{Field Theories on Null
  Manifolds},'' {\em JHEP} {\bf 02} (2020) 141,
  \href{http://www.arXiv.org/abs/1912.09388}{{\tt 1912.09388}}.

\bibitem{basu2018dynamical}
R.~Basu and U.~N. Chowdhury, ``Dynamical structure of carrollian
  electrodynamics,'' {\em Journal of High Energy Physics} {\bf 2018} (2018),
  no.~4, 1--23.

\bibitem{kraus2025carrollian}
P.~Kraus and R.~M. Myers, ``Carrollian partition functions and the flat limit
  of ads,'' {\em Journal of High Energy Physics} {\bf 2025} (2025), no.~1,
  1--54.

\bibitem{Bagchi:2023fbj}
A.~Bagchi, P.~Dhivakar, and S.~Dutta, ``{AdS Witten diagrams to Carrollian
  correlators},'' {\em JHEP} {\bf 04} (2023) 135,
  \href{http://www.arXiv.org/abs/2303.07388}{{\tt 2303.07388}}.

\bibitem{Donnay:2022wvx}
L.~Donnay, A.~Fiorucci, Y.~Herfray, and R.~Ruzziconi, ``{Bridging Carrollian
  and celestial holography},'' {\em Phys. Rev. D} {\bf 107} (2023), no.~12,
  126027, \href{http://www.arXiv.org/abs/2212.12553}{{\tt 2212.12553}}.

\bibitem{Salzer:2023jqv}
J.~Salzer, ``{An embedding space approach to Carrollian CFT correlators for
  flat space holography},'' {\em JHEP} {\bf 10} (2023) 084,
  \href{http://www.arXiv.org/abs/2304.08292}{{\tt 2304.08292}}.

\bibitem{Nguyen:2023miw}
K.~Nguyen, ``{Carrollian conformal correlators and massless scattering
  amplitudes},'' {\em JHEP} {\bf 01} (2024) 076,
  \href{http://www.arXiv.org/abs/2311.09869}{{\tt 2311.09869}}.

\bibitem{Liu:2024nfc}
W.-B. Liu, J.~Long, and X.-Q. Ye, ``{Feynman rules and loop structure of
  Carrollian amplitudes},'' {\em JHEP} {\bf 05} (2024) 213,
  \href{http://www.arXiv.org/abs/2402.04120}{{\tt 2402.04120}}.

\bibitem{Mason:2023mti}
L.~Mason, R.~Ruzziconi, and A.~Yelleshpur~Srikant, ``{Carrollian Amplitudes and
  Celestial Symmetries},'' \href{http://www.arXiv.org/abs/2312.10138}{{\tt
  2312.10138}}.

\bibitem{Ruzziconi:2024zkr}
R.~Ruzziconi, S.~Stieberger, T.~R. Taylor, and B.~Zhu, ``{Differential
  Equations for Carrollian Amplitudes},''
  \href{http://www.arXiv.org/abs/2407.04789}{{\tt 2407.04789}}.

\bibitem{Liu:2024llk}
W.-B. Liu, J.~Long, H.-Y. Xiao, and J.-L. Yang, ``{On the definition of
  Carrollian amplitudes in general dimensions},''
  \href{http://www.arXiv.org/abs/2407.20816}{{\tt 2407.20816}}.

\bibitem{Li:2024kbo}
A.~Li, J.~Long, and J.-L. Yang, ``{Carrollian propagator and amplitude in
  Rindler spacetime},'' \href{http://www.arXiv.org/abs/2410.20372}{{\tt
  2410.20372}}.

\bibitem{Witten:1998qj}
E.~Witten, ``{Anti-de Sitter space and holography},'' {\em Adv. Theor. Math.
  Phys.} {\bf 2} (1998) 253--291,
  \href{http://www.arXiv.org/abs/hep-th/9802150}{{\tt hep-th/9802150}}.

\bibitem{Alday:2024yyj}
L.~F. Alday, M.~Nocchi, R.~Ruzziconi, and A.~Yelleshpur~Srikant, ``{Carrollian
  Amplitudes from Holographic Correlators},''
  \href{http://www.arXiv.org/abs/2406.19343}{{\tt 2406.19343}}.

\bibitem{Chou:1984es}
K.-c. Chou, Z.-b. Su, B.-l. Hao, and L.~Yu, ``{Equilibrium and Nonequilibrium
  Formalisms Made Unified},'' {\em Phys. Rept.} {\bf 118} (1985) 1--131.

\bibitem{landsman1987real}
N.~P. Landsman and C.~G. Van~Weert, ``Real-and imaginary-time field theory at
  finite temperature and density,'' {\em Physics Reports} {\bf 145} (1987),
  no.~3-4, 141--249.

\bibitem{Son:2002sd}
D.~T. Son and A.~O. Starinets, ``{Minkowski space correlators in AdS / CFT
  correspondence: Recipe and applications},'' {\em JHEP} {\bf 09} (2002) 042,
  \href{http://www.arXiv.org/abs/hep-th/0205051}{{\tt hep-th/0205051}}.

\bibitem{Herzog:2002pc}
C.~P. Herzog and D.~T. Son, ``{Schwinger-Keldysh propagators from AdS/CFT
  correspondence},'' {\em JHEP} {\bf 03} (2003) 046,
  \href{http://www.arXiv.org/abs/hep-th/0212072}{{\tt hep-th/0212072}}.

\bibitem{le2000thermal}
M.~Le~Bellac, {\em Thermal field theory}.
\newblock Cambridge university press, 2000.

\bibitem{hollik2014quantum}
W.~Hollik, ``Quantum field theory and the standard model,'' 2014.

\bibitem{rammer2011quantum}
J.~Rammer, ``Quantum field theory of non-equilibrium states,'' {\em Quantum
  Field Theory of Non-equilibrium States} (2011).

\bibitem{Umezawa:1982nv}
H.~Umezawa, H.~Matsumoto, and M.~Tachiki, {\em {THERMO FIELD DYNAMICS AND
  CONDENSED STATES}}.
\newblock 1982.

\bibitem{Schwinger:1960qe}
J.~S. Schwinger, ``{Brownian motion of a quantum oscillator},'' {\em J. Math.
  Phys.} {\bf 2} (1961) 407--432.

\bibitem{Keldysh:1964ud}
L.~V. Keldysh, ``{Diagram technique for nonequilibrium processes},'' {\em Zh.
  Eksp. Teor. Fiz.} {\bf 47} (1964) 1515--1527.

\bibitem{chu1994unified}
H.~Chu and H.~Umezawa, ``A unified formalism of thermal quantum field theory,''
  {\em International Journal of Modern Physics A} {\bf 9} (1994), no.~14,
  2363--2409.

\bibitem{Jordan:1986ug}
R.~D. Jordan, ``{Effective Field Equations for Expectation Values},'' {\em
  Phys. Rev. D} {\bf 33} (1986) 444--454.

\bibitem{chen2023higher}
B.~Chen, R.~Liu, and Y.-f. Zheng, ``On higher-dimensional carrollian and
  galilean conformal field theories,'' {\em SciPost Physics} {\bf 14} (2023),
  no.~5, 088.

\bibitem{laine2016basics}
M.~Laine and A.~Vuorinen, ``Basics of thermal field theory,'' {\em Lect. Notes
  Phys} {\bf 925} (2016), no.~1, 1701--01554.

\bibitem{marchetto2023broken}
E.~Marchetto, A.~Miscioscia, and E.~Pomoni, ``Broken (super) conformal ward
  identities at finite temperature,'' {\em Journal of High Energy Physics} {\bf
  2023} (2023), no.~12, 1--48.

\bibitem{Kubo:1957mj}
R.~Kubo, ``{Statistical mechanical theory of irreversible processes. 1. General
  theory and simple applications in magnetic and conduction problems},'' {\em
  J. Phys. Soc. Jap.} {\bf 12} (1957) 570--586.

\bibitem{Martin:1959jp}
P.~C. Martin and J.~S. Schwinger, ``{Theory of many particle systems. 1.},''
  {\em Phys. Rev.} {\bf 115} (1959) 1342--1373.

\bibitem{Lindell1993DeltaFE}
I.~V. Lindell, ``Delta function expansions, complex delta functions and the
  steepest descent method,'' {\em American Journal of Physics} {\bf 61} (1993)
  438--442.

\bibitem{1996tah..book.....P}
A.~D. {Poularikas}, {\em {The transforms and applications handbook}}.
\newblock 1996.

\bibitem{Brewster2016GeneralizedDF}
R.~A. Brewster and J.~D. Franson, ``Generalized delta functions and their use
  in quantum optics,'' {\em Journal of Mathematical Physics} {\bf 59} (2016)
  012102.

\bibitem{cotler2025soft}
J.~Cotler, K.~Jensen, S.~Prohazka, M.~Riegler, and J.~Salzer, ``Soft gravitons
  in three dimensions,'' {\em Journal of High Energy Physics} {\bf 2025}
  (2025), no.~7, 1--40.

\bibitem{campoleoni2022magnetic}
A.~Campoleoni, M.~Henneaux, S.~Pekar, A.~P{\'e}rez, and P.~Salgado-Rebolledo,
  ``Magnetic carrollian gravity from the carroll algebra,'' {\em Journal of
  High Energy Physics} {\bf 2022} (2022), no.~9, 1--22.

\bibitem{jorstad2024comment}
E.~J{\o}rstad and S.~Pasterski, ``A comment on boundary correlators: Soft
  omissions and the massless s-matrix,'' {\em arXiv preprint arXiv:2410.20296}
  (2024).

\bibitem{francesco2012conformal}
P.~Francesco, P.~Mathieu, and D.~S{\'e}n{\'e}chal, {\em Conformal field
  theory}.
\newblock Springer Science \& Business Media, 2012.

\bibitem{iliesiu2018conformal}
L.~Iliesiu, M.~Kolo{\u{g}}lu, R.~Mahajan, E.~Perlmutter, and D.~Simmons-Duffin,
  ``The conformal bootstrap at finite temperature,'' {\em Journal of High
  Energy Physics} {\bf 2018} (2018), no.~10, 1--71.

\bibitem{Matsubara:1955ws}
T.~Matsubara, ``{A New approach to quantum statistical mechanics},'' {\em Prog.
  Theor. Phys.} {\bf 14} (1955) 351--378.

\bibitem{son2002minkowski}
D.~T. Son and A.~O. Starinets, ``Minkowski-space correlators in ads/cft
  correspondence: Recipe and applications,'' {\em Journal of High Energy
  Physics} {\bf 2002} (2002), no.~09, 042.

\bibitem{birmingham2002conformal}
D.~Birmingham, I.~Sachs, and S.~N. Solodukhin, ``Conformal field theory
  interpretation of black hole quasinormal modes,'' {\em Physical review
  letters} {\bf 88} (2002), no.~15, 151301.

\bibitem{horowitz2000quasinormal}
G.~T. Horowitz and V.~E. Hubeny, ``Quasinormal modes of ads black holes and the
  approach to thermal equilibrium,'' {\em Physical Review D} {\bf 62} (2000),
  no.~2, 024027.

\bibitem{chan1997scalar}
J.~Chan and R.~B. Mann, ``Scalar wave falloff in asymptotically anti--de sitter
  backgrounds,'' {\em Physical Review D} {\bf 55} (1997), no.~12, 7546.

\bibitem{wang2000quasinormal}
B.~Wang, C.-Y. Lin, and E.~Abdalla, ``Quasinormal modes of
  reissner-nordstr{\"o}m anti-de sitter black holes,'' {\em Physics Letters B}
  {\bf 481} (2000), no.~1, 79--88.

\bibitem{govindarajan2001quasi}
T.~Govindarajan and V.~Suneeta, ``Quasi-normal modes of adsblack holes: a
  superpotential approach,'' {\em Classical and Quantum Gravity} {\bf 18}
  (2001), no.~2, 265.

\bibitem{cardoso2001scalar}
V.~Cardoso and J.~P. Lemos, ``Scalar, electromagnetic, and weyl perturbations
  of btz black holes: Quasinormal modes,'' {\em Physical Review D} {\bf 63}
  (2001), no.~12, 124015.

\bibitem{wang2002scalar}
B.~Wang, E.~Abdalla, and R.~B. Mann, ``Scalar wave propagation in topological
  black hole backgrounds,'' {\em Physical Review D} {\bf 65} (2002), no.~8,
  084006.

\bibitem{chen2009quasi}
B.~Chen and Z.-b. Xu, ``Quasi-normal modes of warped black holes and warped
  ads/cft correspondence,'' {\em Journal of High Energy Physics} {\bf 2009}
  (2009), no.~11, 091.

\bibitem{siopsis2005quasi}
G.~Siopsis, ``On quasi-normal modes and the ads5/cft4 correspondence,'' {\em
  Nuclear physics B} {\bf 715} (2005), no.~1-2, 483--498.

\bibitem{amado2024scalar}
J.~B. Amado and B.~Gwak, ``Scalar quasi-normal modes of accelerating
  kerr-newman-ads black holes,'' {\em Journal of High Energy Physics} {\bf
  2024} (2024), no.~2, 1--29.

\bibitem{balasubramanian1999bulk}
V.~Balasubramanian, P.~Kraus, and A.~Lawrence, ``Bulk versus boundary dynamics
  in anti--de sitter spacetime,'' {\em Physical Review D} {\bf 59} (1999),
  no.~4, 046003.

\bibitem{wardell2009green}
B.~Wardell, ``Green functions and radiation reaction from a spacetime
  perspective,'' {\em arXiv preprint arXiv:0910.2634} (2009).

\bibitem{duistermaat1994fourier}
J.~J. Duistermaat and L.~H{\"o}rmander, ``Fourier integral operators. ii,'' in
  {\em Mathematics Past and Present Fourier Integral Operators}, pp.~129--215.
\newblock Springer, 1994.

\bibitem{hormander2009analysis}
L.~H{\"o}rmander, {\em The analysis of linear partial differential operators
  IV: Fourier integral operators}.
\newblock Springer, 2009.

\bibitem{Cardy:1984epx}
J.~L. Cardy, ``{Conformal invariance and universality in finite-size
  scaling},'' {\em J. Phys. A} {\bf 17} (1984), no.~7, L385.

\bibitem{Boulware:1974dm}
D.~G. Boulware, ``{Quantum Field Theory in Schwarzschild and Rindler Spaces},''
  {\em Phys. Rev. D} {\bf 11} (1975) 1404.

\bibitem{Hartle:1976tp}
J.~B. Hartle and S.~W. Hawking, ``{Path Integral Derivation of Black Hole
  Radiance},'' {\em Phys. Rev. D} {\bf 13} (1976) 2188--2203.

\bibitem{Birrell:1982ix}
N.~D. Birrell and P.~C.~W. Davies, {\em {Quantum Fields in Curved Space}}.
\newblock Cambridge Monographs on Mathematical Physics. Cambridge Univ. Press,
  Cambridge, UK, 2, 1984.

\bibitem{DeWitt:1975ys}
B.~S. DeWitt, ``{Quantum Field Theory in Curved Space-Time},'' {\em Phys.
  Rept.} {\bf 19} (1975) 295--357.

\bibitem{Rindler:1966zz}
W.~Rindler, ``{Kruskal Space and the Uniformly Accelerated Frame},'' {\em Am.
  J. Phys.} {\bf 34} (1966) 1174.

\bibitem{Candelas:1980zt}
P.~Candelas, ``{Vacuum Polarization in Schwarzschild Space-Time},'' {\em Phys.
  Rev. D} {\bf 21} (1980) 2185--2202.

\bibitem{bacry1968possible}
H.~Bacry and J.-M. L{\'e}vy-Leblond, ``Possible kinematics,'' {\em Journal of
  Mathematical Physics} {\bf 9} (1968), no.~10, 1605--1614.

\bibitem{Henneaux:1979vn}
M.~Henneaux, ``{Geometry of Zero Signature Space-times},'' {\em Bull. Soc.
  Math. Belg.} {\bf 31} (1979) 47--63.

\bibitem{penna2018near}
R.~F. Penna, ``Near-horizon carroll symmetry and black hole love numbers,''
  {\em arXiv preprint arXiv:1812.05643} (2018).

\bibitem{2019CQGra..36p5002D}
L.~{Donnay} and C.~{Marteau}, ``{Carrollian physics at the black hole
  horizon},'' {\em Classical and Quantum Gravity} {\bf 36} (Aug., 2019) 165002,
  \href{http://www.arXiv.org/abs/1903.09654}{{\tt 1903.09654}}.

\bibitem{freidel2024carrollian}
L.~Freidel and P.~Jai-akson, ``Carrollian hydrodynamics and symplectic
  structure on stretched horizons,'' {\em Journal of High Energy Physics} {\bf
  2024} (2024), no.~5, 1--43.

\bibitem{redondo2023non}
J.~Redondo-Yuste and L.~Lehner, ``Non-linear black hole dynamics and carrollian
  fluids,'' {\em Journal of High Energy Physics} {\bf 2023} (2023), no.~2,
  1--28.

\bibitem{ecker2023carroll}
F.~Ecker, D.~Grumiller, J.~Hartong, A.~P{\'e}rez, S.~Prohazka, and R.~Troncoso,
  ``Carroll black holes,'' {\em SciPost Physics} {\bf 15} (2023), no.~6, 245.

\bibitem{Bondi:1962px}
H.~Bondi, M.~G.~J. van~der Burg, and A.~W.~K. Metzner, ``{Gravitational waves
  in general relativity. 7. Waves from axisymmetric isolated systems},'' {\em
  Proc. Roy. Soc. Lond. A} {\bf 269} (1962) 21--52.

\bibitem{Bagchi:2023cen}
A.~Bagchi, P.~Dhivakar, and S.~Dutta, ``{Holography in Flat Spacetimes: the
  case for Carroll},'' \href{http://www.arXiv.org/abs/2311.11246}{{\tt
  2311.11246}}.

\bibitem{Banerjee:2018gce}
S.~Banerjee, ``{Null Infinity and Unitary Representation of The Poincare
  Group},'' {\em JHEP} {\bf 01} (2019) 205,
  \href{http://www.arXiv.org/abs/1801.10171}{{\tt 1801.10171}}.

\bibitem{bagchi2017nuts}
A.~Bagchi, M.~Gary, {\em et al.}, ``The nuts and bolts of the bms bootstrap,''
  {\em Classical and Quantum Gravity} {\bf 34} (2017), no.~17, 174002.

\bibitem{dolan2001conformal}
F.~A. Dolan and H.~Osborn, ``Conformal four point functions and the operator
  product expansion,'' {\em Nuclear Physics B} {\bf 599} (2001), no.~1-2,
  459--496.

\bibitem{dolan2004conformal}
F.~A. Dolan and H.~Osborn, ``Conformal partial waves and the operator product
  expansion,'' {\em Nuclear Physics B} {\bf 678} (2004), no.~1-2, 491--507.

\bibitem{1963cma..book.....W}
E.~T. {Whittaker} and G.~N. {Watson}, {\em {A course of modern analysis}}.
\newblock 1963.

\bibitem{1901RSPTA.196..265B}
E.~W. {Barnes}, ``{The Theory of the Double Gamma Function},'' {\em
  Philosophical Transactions of the Royal Society of London Series A} {\bf 196}
  (Jan., 1901) 265--387.

\bibitem{Ruijsenaars2000OnBM}
S.~Ruijsenaars, ``On barnes' multiple zeta and gamma functions,'' {\em Advances
  in Mathematics} {\bf 156} (2000) 107--132.

\bibitem{komori2010barnesmultiplezetafunctionsramanujans}
Y.~Komori, K.~Matsumoto, and H.~Tsumura, ``Barnes multiple zeta-functions,
  ramanujan's formula, and relevant series involving hyperbolic functions,''
  \href{http://www.arXiv.org/abs/1006.3336}{{\tt 1006.3336}}.

\bibitem{barneszeta}
M.~K. Tsuneo~Arakawa, Tomoyoshi~Ibukiyama, {\em Bernoulli Numbers and Zeta
  Functions}.
\newblock Springer Tokyo, 2014.

\end{thebibliography}\endgroup
\end{document}